\documentclass[utf8]{FrontiersinHarvard} %

\usepackage{url,hyperref,lineno,subcaption}
\usepackage{amsmath,amssymb,mathtools}
\usepackage{graphicx}
\usepackage{tabularx,xltabular,makecell,booktabs,tablefootnote}
\usepackage{pifont}
\usepackage[super]{nth}
\newcommand{\xmark}{\ding{55}}%

\usepackage[onehalfspacing]{setspace}
\usepackage{capt-of}%
\usepackage{tcolorbox}

\usepackage{tikz,forest}
\usetikzlibrary{external}
\usetikzlibrary{positioning}
\usetikzlibrary{shapes,backgrounds}
\usetikzlibrary{trees}
\useforestlibrary{edges}

\def\keyFont{\fontsize{8}{11}\helveticabold }
\def\firstAuthorLast{Hayes {et~al.}} %
\def\Authors{Ben Hayes\,$^{1,\ast}$, Jordie Shier\,$^{1,\ast}$, Gy\"orgy Fazekas\,$^{1}$, Andrew McPherson\,$^{1,2}$, Charalampos Saitis\,$^{1}$}
\def\Address{$^{1}$Centre for Digital Music, School of Electronic Engineering and Computer Science, Queen Mary University of London, United Kingdom\\
$^{2}$Dyson School of Design Engineering, Imperial College London, United Kingdom }
\def\corrAuthor{Jordie Shier, Ben Hayes, Queen Mary University of London, 327 Mile End Rd, Bethnal Green, London E1 4NS, UK}

\def\corrEmail{\{j.m.shier,b.j.hayes\}@qmul.ac.uk}

\begin{document}
\onecolumn
\firstpage{1}

\title {A Review of Differentiable Digital Signal Processing for Music \& Speech Synthesis} 

\author[\firstAuthorLast ]{\Authors} %
\address{} %
\correspondance{} %

\extraAuth{}%

\maketitle

\begin{abstract}

\section{}
The term ``differentiable digital signal processing'' describes a family of techniques in which loss function gradients are backpropagated through digital signal processors, facilitating their integration into neural networks. 
This article surveys the literature on differentiable audio signal processing, focusing on its use in music \& speech synthesis.
We catalogue applications to tasks including music performance rendering, sound matching, and voice transformation, discussing the motivations for and implications of the use of this methodology.
This is accompanied by an overview of digital signal processing operations that have been implemented differentiably.
Finally, we highlight open challenges, including optimisation pathologies, robustness to real-world conditions, and design trade-offs, and discuss directions for future research.

\tiny
 \keyFont{ \section{Keywords:} digital signal processing, machine learning, audio synthesis, automatic differentiation, neural networks}
\end{abstract}

\begin{tcolorbox}[title=Citing this Article]
This is version 1 of this pre-print manuscript. It should be cited as follows.
\\
\\
\textit{Hayes, B., Shier, J., Fazekas, G., McPherson, A., Saitis, C., 2023. A Review of Differentiable Digital Signal Processing for Music \& Speech Synthesis. (Under Review) Frontiers in Signal Processing.}
\end{tcolorbox}

\section{Introduction}\label{sec:introduction}

\begin{figure}[t]
    \centering
    \includegraphics{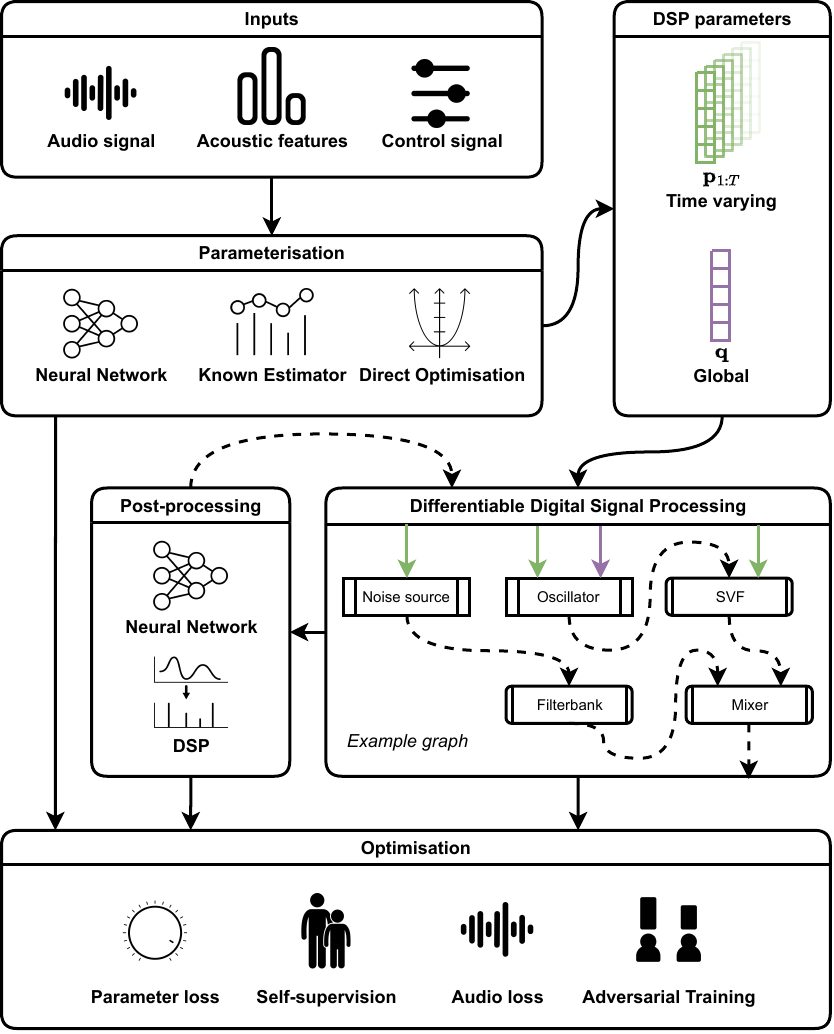}
    \caption{A high level overview of the general structure of a typical DDSP synthesis system. Not every depicted component is present in every system, however we find this structure broadly encompasses the work we have surveyed.}
    \label{fig:ddsp-overview}
\end{figure}

Audio synthesis, the artificial production of sound, has been an active field of research for over a century. 
Early inventions included entirely new categories of musical instrument~\citep{cahill_art_1897,ssergejewitsch_method_1928} and the first machines for artificial speech production~\citep{dudley_vocoder_1939,dudley_speaking_1950}, while the latter half of the \nth{20} century saw a proliferation of research into digital methods for sound synthesis, built on advances in signal processing~\citep{smith_physical_2010,keller_fundamentals_1994} and numerical methods~\citep{bilbao_numerical_2009}.
Applications of audio synthesis have since come to permeate daily life, from music~\citep{holmes_electronic_2008}, through synthetic voices, to the sound design in films, TV shows, video games, and even the cockpits of cars~\citep{dupre_spatial_2021}.

In recent years, the field has undergone something of a technological revolution.
The publication of WaveNet~\citep{oord_wavenet_2016}, an autoregressive neural network which produced a quantised audio signal sample-by-sample, first illustrated that deep learning might be a viable methodology for audio synthesis.
Over the following years, new methods for neural audio synthesis abounded, from refinements to WaveNet~\citep{oord_parallel_2018} to the application of entirely different classes of generative model~\citep{donahue_adversarial_2019,chen_wavegrad_2020,kumar_melgan_2019,kong_hifi-gan_2020}, with the majority of work focusing on speech~\citep{tan_survey_2021} and music~\citep{huzaifah_deep_2020} synthesis.

Nonetheless, modelling audio signals remained challenging.
Upsampling layers, crucial components of workhorse architectures such as generative adversarial networks~\citep{goodfellow_generative_2014} and autoencoders, were found to cause undesirable signal artifacts~\citep{pons_upsampling_2021}.
Similarly, frame-based estimation of audio signals was also found to be more challenging than might na\"ively be assumed, due to the difficulty of ensuring phase coherence between successive frames, where frame lengths are independent of the frequencies contained in a signal~\citep{engel_gansynth_2019}.

Aiming to address such issues, one line of research explored the integration of domain knowledge from speech synthesis and signal processing into neural networks.
Whilst some methods combined the outputs of classical techniques with neural networks~\citep{valin_lpcnet_2019}, others integrated them by expressing the signal processing elements differentiably~\citep{wang_neural_2019-2,juvela_gelp_2019}. 
This was crystalised in the work of \cite{engel_ddsp_2020}, who introduced the terminology \textit{differentiable digital signal processing} (DDSP).
In particular, Engel et al. suggested that some difficulties in neural audio synthesis could be explained by certain biases induced by the underlying models.
The proposed advantage of DDSP was thus to gain a domain-appropriate inductive bias by incorporating a known signal model to the neural network.
Implementing the signal model differentiably allowed loss gradients to be backpropagated through its parameters, in a manner similar to differentiable rendering~\citep{kato_differentiable_2020}. 

In subsequent years, DDSP was applied to tasks including music performance synthesis~\citep{wu_midi-ddsp_2022,jonason_control-synthesis_2020}, instrument modelling~\citep{renault_differentiable_2022}, synthesiser sound matching~\citep{masuda_synthesizer_2021}, speech synthesis and voice transformation~\citep{choi_nansy_2022}, singing voice synthesis and conversion~\citep{yu_singing_2023,nercessian_differentiable_2023}, sound-effect generation~\citep{barahona-rios_noisebandnet_2023,hagiwara_modeling_2022}, and more.
The technology has also been deployed in a number of publicly available software instruments and real-time tools.\footnote{These include Google Magenta's DDSP-VST (\url{https://magenta.tensorflow.org/ddsp-vst}), Bytedance's Mawf (\url{https://mawf.io/}), Neutone Inc.'s Neutone (\url{https://neutone.space/}), ACIDS-IRCAM's \textit{ddsp\~} (\url{https://github.com/acids-ircam/ddsp_pytorch}) and Aalborg University's JUCE implementation (\url{https://github.com/SMC704/juce-ddsp}). Accessed 21st August 2023.}
Fig. \ref{fig:ddsp-overview} illustrates the general structure of a typical DDSP synthesis system and we list included papers in Table~\ref{tab:papers}.

Differentiable signal processing has also been applied in tasks related to audio engineering, such as audio effect modelling~\citep{carson_differentiable_2023,kuznetsov_differentiable_2020,lee_differentiable_2022}, automatic mixing and intelligent music production~\citep{steinmetz_style_2022,martinez_ramirez_differentiable_2021}, and filter design~\citep{colonel_direct_2022-1}.
Whilst many innovations from this work have found use in synthesis, and vice versa, we do not set out to comprehensively review these tasks areas.
Instead, we address this work where it is pertinent to our discussion of differentiable audio synthesis, and refer readers to the works of \citet{ramirez_deep_2020}, \citet{moffat_approaches_2019}, \citet{de_man_intelligent_2019}, for reviews of the relevant background, and to the work of \citet{steinmetz_deep_2022} for a summary of the state of differentiable signal processing in this field.

In light of the growing body of literature on DDSP, we hope for this article to benefit two groups.
The first consists of those unfamiliar with the details of DDSP, but who wish to gain an overview of the field.
We hope that some may even discover applications of these techniques in their work.
Indeed, we note that DDSP-based audio synthesis has already found use in more diverse tasks, including pitch estimation~\citep{engel_self-supervised_2020}, source separation~\citep{schulze-forster_unsupervised_2023}, and articulatory parameter estimation~\citep{sudholt_vocal_2023}.
Through discussion of the current understanding of the limitations of DDSP, we also endeavour to allow this group to better assess the suitability of these techniques for their work.
The second group consists of those already working with DDSP, who wish to gain a broader awareness of the field.
For this group, we endeavour in particular to provide a streamlined account of relevant technical contributions, and to pair this with a discussion of valuable future research directions.

The terms \textit{differentiable digital signal processing} and \textit{DDSP} have been ascribed various meanings in the literature.
For the sake of clarity, whilst also wishing to acknowledge the contributions of \cite{engel_ddsp_2020}, we therefore adopt the following disambiguation in this article:

\begin{enumerate}
    \item We use the general term \textit{differentiable digital signal processing} and the acronym \textit{DDSP} to describe the technique of implementing digitial signal processing operations using automatic differentiation software.
    \item To refer to Engel, et al.'s Python library, we use the term \textit{the DDSP library}.
    \item We refer to the differentiable spectral modelling synthesiser and neural network controller introduced by \cite{engel_ddsp_2020}, like other work, in terms of their specific contributions, e.g. \textit{Engel, et al.'s differentiable spectral-modelling synthesiser}.
\end{enumerate}

{\tiny
\begin{xltabular}{\textwidth}{llX}
\toprule
\textbf{Authors} & \textbf{Year} & \textbf{Contributions} \\
\midrule
\multicolumn{3}{c}{\textbf{Speech Synthesis}} \\\midrule
\citeauthor{valin_lpcnet_2019} & \citeyear{valin_lpcnet_2019} & LPC integration to WaveRNN. \\
\citeauthor{wang_neural_2019-1} & \citeyear{wang_neural_2019-1} & Neural source-filter (NSF). \\
\citeauthor{juvela_gelp_2019} & \citeyear{juvela_gelp_2019} & Differentiable LPC. \\
\citeauthor{wang_neural_2019} & \citeyear{wang_neural_2019} & Differentiable sinc FIR design. \\
\citeauthor{wang_neural_2019-2} & \citeyear{wang_neural_2019-2} & Further NSF models. \\
\citeauthor{wang_using_2020} & \citeyear{wang_using_2020} & Cyclic noise source for NSF. \\
\citeauthor{mv_sfnet_2020} & \citeyear{mv_sfnet_2020} & Fully differentiable source-filter model. \\
\citeauthor{tian_featherwave_2020} & \citeyear{tian_featherwave_2020} & Multi-band LPC. \\
\citeauthor{vipperla_bunched_2020} & \citeyear{vipperla_bunched_2020} & Bunched LPC. \\
\citeauthor{fabbro_speech_2020} & \citeyear{fabbro_speech_2020} & Differentiable harmonic-plus-noise for speech. \\
\citeauthor{nercessian_end--end_2021} & \citeyear{nercessian_end--end_2021} & Harmonic-plus-noise for voice conversion. \\
\citeauthor{subramani_end--end_2022} & \citeyear{subramani_end--end_2022} & Differentiable LPC estimation. \\
\citeauthor{choi_nansy_2022} & \citeyear{choi_nansy_2022} & Hybrid model with self-supervised disentanglement. \\
\citeauthor{kaneko_istftnet_2022} & \citeyear{kaneko_istftnet_2022} & Differentiable ISTFT-based vocoder. \\
\citeauthor{webber_autovocoder_2023} & \citeyear{webber_autovocoder_2023} & Differentiable ISTFT-based vocoder. \\
\citeauthor{sudholt_vocal_2023} & \citeyear{sudholt_vocal_2023} & Differentiable digital waveguide (Kelly-Lochbaum). \\
\citeauthor{song_dspgan_2023} & \citeyear{song_dspgan_2023} & Hybrid model combining DDSP with GAN vocoder. \\\midrule
\multicolumn{3}{c}{\textbf{Music Synthesis}} \\\midrule
\citeauthor{engel_ddsp_2020} & \citeyear{engel_ddsp_2020} & DDSP library; differentiable spectral modelling synthesiser. \\
\citeauthor{zhao_transferring_2020} & \citeyear{zhao_transferring_2020} & NSF applied to musical instrument synthesis. \\
\citeauthor{michelashvili_hierarchical_2020} & \citeyear{michelashvili_hierarchical_2020} & Hierarchical NSF model. \\
\citeauthor{castellon_towards_2020} & \citeyear{castellon_towards_2020} & DDSP-based performance rendering. \\
\citeauthor{jonason_control-synthesis_2020} & \citeyear{jonason_control-synthesis_2020} & DDSP-based performance rendering. \\
\citeauthor{caillon_rave_2021} & \citeyear{caillon_rave_2021} & Hybrid real-time audio generative model. \\
\citeauthor{carney_tone_2021} & \citeyear{carney_tone_2021} & Efficient in-browser DDSP implementation; numerically stable TF.js kernels. \\
\citeauthor{hayes_neural_2021} & \citeyear{hayes_neural_2021} & Differentiable waveshaping synthesiser. \\
\citeauthor{masuda_synthesizer_2021} & \citeyear{masuda_synthesizer_2021} & Differentiable subtractive synthesiser. \\
\citeauthor{caspe_ddx7_2022} & \citeyear{caspe_ddx7_2022} & Differentiable FM synthesiser. \\
\citeauthor{diaz_rigid-body_2022} & \citeyear{diaz_rigid-body_2022} & Differentiable modal synthesiser. \\
\citeauthor{shan_differentiable_2022} & \citeyear{shan_differentiable_2022} & Differentiable wavetable synthesiser. \\
\citeauthor{kawamura_differentiable_2022} & \citeyear{kawamura_differentiable_2022} & DDSP-based mixture model for synthesis parameter estimation. \\
\citeauthor{renault_differentiable_2022} & \citeyear{renault_differentiable_2022} & Differentiable piano model; explicit inharmonicity modelling. \\
\citeauthor{wu_midi-ddsp_2022} & \citeyear{wu_midi-ddsp_2022} & DDSP-based performance modelling. \\
\citeauthor{masuda_improving_2023} & \citeyear{masuda_improving_2023} & Semi-supervised hybrid training; differentiable ADSR. \\
\citeauthor{ye_nas-fm_2023} & \citeyear{ye_nas-fm_2023} & Neural architecture search over differentiable FM synthesisers. \\\midrule
\multicolumn{3}{c}{\textbf{Singing Voice Synthesis}} \\\midrule
\citeauthor{alonso_latent_2021} & \citeyear{alonso_latent_2021} & Experiments on autoencoder from \citet{engel_ddsp_2020} for singing voice synthesis.  \\
\citeauthor{wu_ddsp-based_2022} & \citeyear{wu_ddsp-based_2022} & Differentiable subtractive singing voice synthesiser. \\
\citeauthor{guo_improving_2022} & \citeyear{guo_improving_2022} & Differentiable filtering of sine excitation for adversarial SVC. \\
\citeauthor{yoshimura_embedding_2023} & \citeyear{yoshimura_embedding_2023} & Differentiable mel cepstral synthesis filter.\footnote{The title and abstract of the paper reference speech synthesis, but the method is evaluated solely on a singing voice synthesis task.} \\
\citeauthor{nercessian_differentiable_2023} & \citeyear{nercessian_differentiable_2023} & Differentiable WORLD vocoder. \\
\citeauthor{yu_singing_2023} & \citeyear{yu_singing_2023} & Glottal-flow wavetable; efficient all-pole IIR training algorithm and implementation. \\\midrule
\multicolumn{3}{c}{\textbf{Other}} \\\midrule
\citeauthor{shynk_adaptive_1989} & \citeyear{shynk_adaptive_1989} & Gradient-based IIR optimisation. \\
\citeauthor{back_fir_1991} & \citeyear{back_fir_1991} & Efficient gradient-based training algorithms for FIR and IIR based neural networks. \\
\citeauthor{campolucci_-line_1995} & \citeyear{campolucci_-line_1995} & Approximate online learning algorithms for IIR networks. \\
\citeauthor{bhattacharya_optimization_2020} & \citeyear{bhattacharya_optimization_2020} & Differentiable IIR filters with instantaneous backpropogation through time. \\
\citeauthor{kuznetsov_differentiable_2020} & \citeyear{kuznetsov_differentiable_2020} & Differentiable IIR filters with truncated backpropogation through time. \\
\citeauthor{nercessian_neural_2020} & \citeyear{nercessian_neural_2020} & Differentiable IIR filters via frequency sampling. \\
\citeauthor{engel_self-supervised_2020} & \citeyear{engel_self-supervised_2020} & Differentiable additive sinusoidal model; self-supervised hybrid training. \\
\citeauthor{turian_im_2020} & \citeyear{turian_im_2020} & Experiments on differentiable frequency estimation pathologies. \\
\citeauthor{turian_one_2021} & \citeyear{turian_one_2021} & Differentiable modular synthesiser; billion sound dataset. \\
\citeauthor{martinez_ramirez_differentiable_2021} & \citeyear{martinez_ramirez_differentiable_2021} & Stochastic approximation of black-box signal processor gradients. \\
\citeauthor{nercessian_lightweight_2021} & \citeyear{nercessian_lightweight_2021} & Differentiable hyperconditioned IIR filters; stability preserving activations. \\
\citeauthor{colonel_direct_2022-1} & \citeyear{colonel_direct_2022-1} &  Random polynomial sampling for differentiable IIR self-supervision. \\
\citeauthor{hagiwara_modeling_2022} & \citeyear{hagiwara_modeling_2022} & Experiments on animal vocal sound modelling via DDSP. \\
\citeauthor{lee_differentiable_2022} & \citeyear{lee_differentiable_2022} & Differentiable artificial reverberation. \\
\citeauthor{steinmetz_style_2022} & \citeyear{steinmetz_style_2022} & Neural proxies for DDSP; evaluation of gradient estimation methods. \\
\citeauthor{barahona-rios_noisebandnet_2023} & \citeyear{barahona-rios_noisebandnet_2023} & Sound effect synthesis with differentiable multiband noise synthesiser. \\
\citeauthor{carson_differentiable_2023} & \citeyear{carson_differentiable_2023} & Differentiable grey-box phaser model; differentiable LFO estimation. \\
\citeauthor{hayes_sinusoidal_2023} & \citeyear{hayes_sinusoidal_2023} & Differentiable frequency estimation; surrogate model for sinusoidal oscillator. \\
\citeauthor{schulze-forster_unsupervised_2023} & \citeyear{schulze-forster_unsupervised_2023} & Unsupervised source separation with differentiable source models. \\
\bottomrule
\caption{A summary of DDSP synthesis papers reviewed in compiling this article. Papers are grouped by major application area. Those which are applied to more than one area are grouped with their primary application.}\label{tab:papers}
\end{xltabular}
}
\section{Applications \& Tasks}\label{sec:applications}
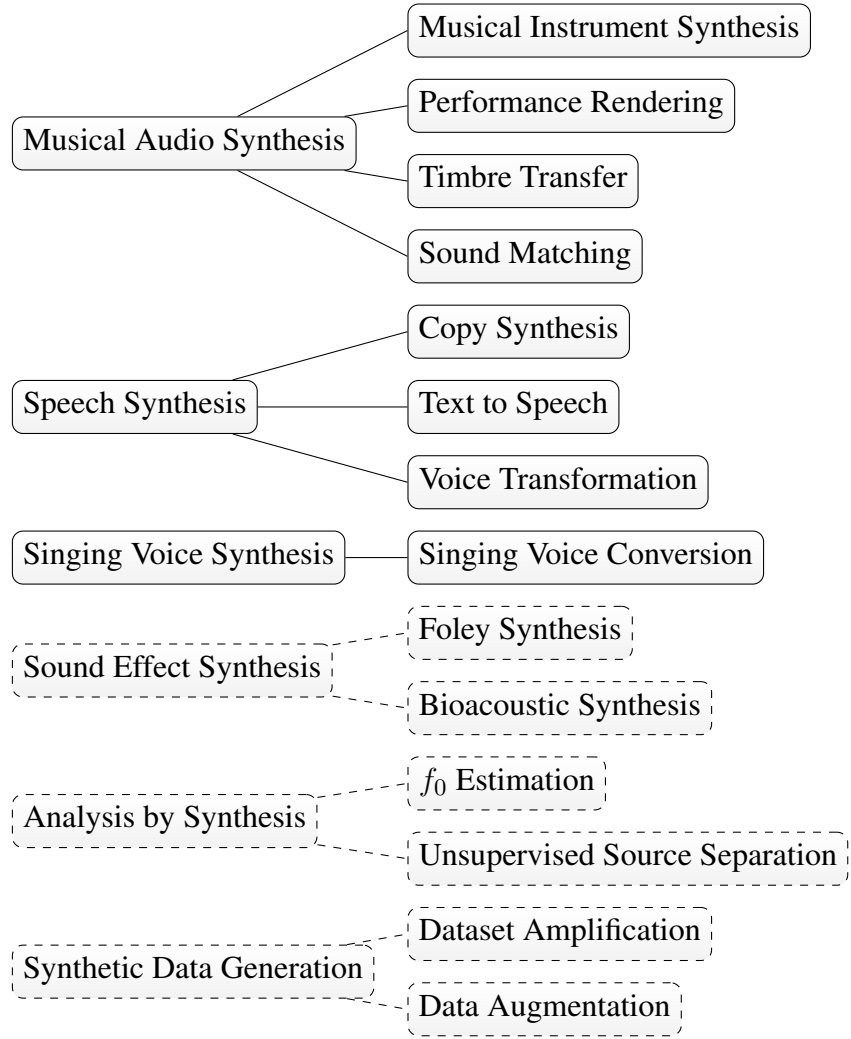
\begin{figure}
\centering
\begin{forest}
    for tree={
        grow' = east,
        draw,
        rounded corners,
        align=left,
        top color=white,
        bottom color=gray!20,
        edge={-},
        anchor=base west,
        child anchor=west,
        tier/.pgfmath=level(),
        fit=tight,
    },
    [Sound synthesis,phantom
    [Musical Audio Synthesis
        [Musical Instrument Synthesis]
        [Performance Rendering]
        [Timbre Transfer]
        [Sound Matching] 
    ]
    [Speech Synthesis
        [Copy Synthesis]
        [Text to Speech]
        [Voice Transformation]
    ]
    [Singing Voice Synthesis
        [Singing Voice Conversion]
    ]
    [Sound Effect Synthesis,dashed
        [Foley Synthesis,dashed,edge=dashed]
        [Bioacoustic Synthesis,dashed,edge=dashed]
    ]
    [Analysis by Synthesis,dashed
        [$f_0$ Estimation,dashed,edge=dashed]
        [Unsupervised Source Separation,dashed,edge=dashed]
    ]
    [Synthetic Data Generation,dashed
        [Dataset Amplification,dashed,edge=dashed]
        [Data Augmentation,dashed,edge=dashed]
    ]
    ] 
\end{forest}
\caption{A taxonomy of audio synthesis tasks to which DDSP has been applied. Further discussion on each is presented in section \ref{sec:applications}. Dashed nodes indicate audio synthesis applications of DDSP found in the literature, but which are beyond the scope of this review.}
\label{fig:task-taxonomy}
\end{figure}
In this section, we survey the tasks and application areas in which DDSP-based audio synthesis has been used, focusing on two goals.
Firstly, we aim to provide sufficient background on historical approaches to contextualize the discussion on applications of differentiable signal processing to audio synthesis.
Secondly, we seek to help practitioners in the task areas listed below, and in related fields, answer the question, ``why would I use differentiable digital signal processing?''

\subsection{Musical Audio Synthesis}\label{sec:musaud}

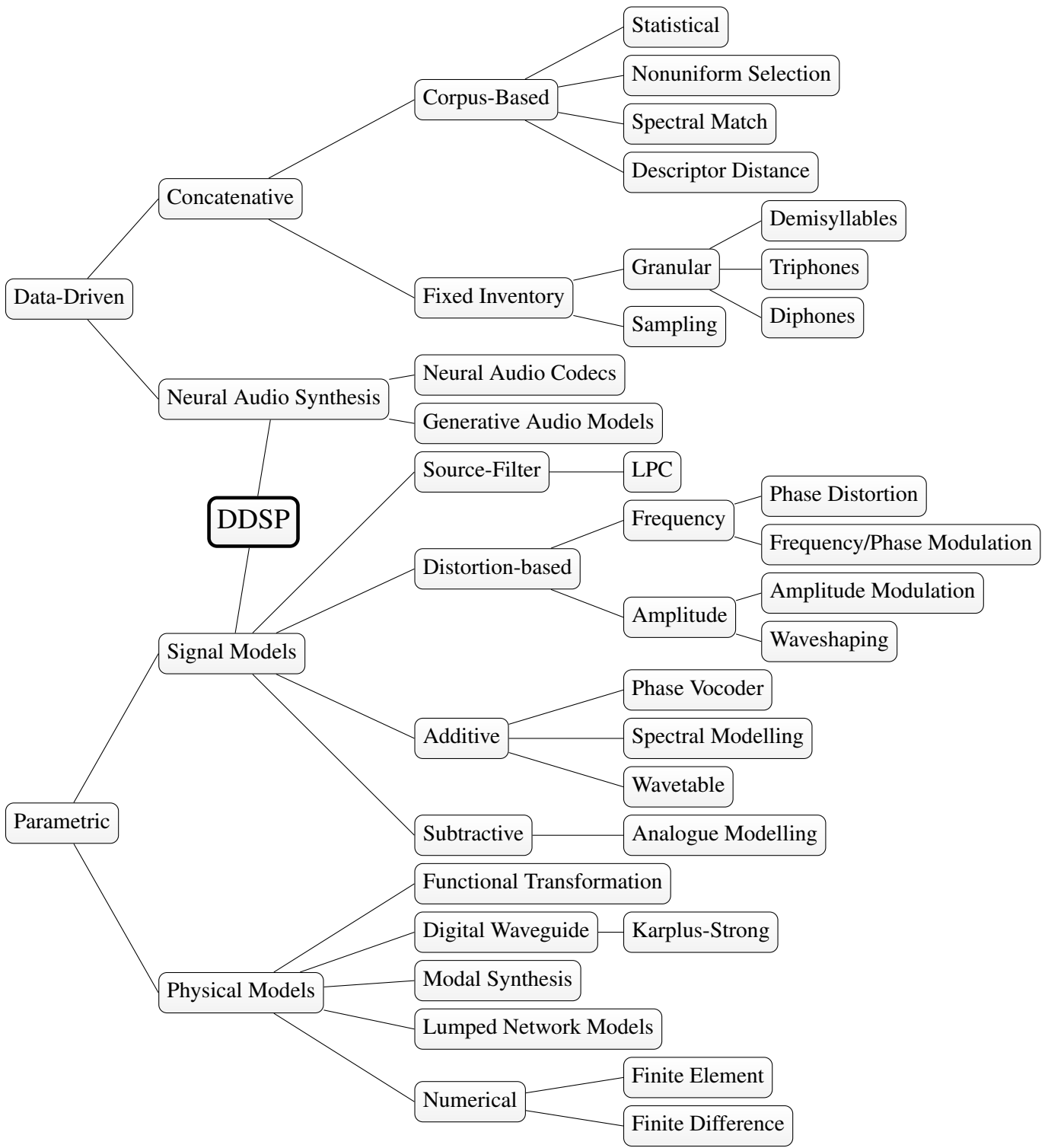
\begin{figure}[h!]
    \centering
\begin{forest}
    for tree={
        grow = east,
        draw,
        rounded corners,
        align=left,
        top color=white,
        bottom color=gray!20,
        edge={-},
        font=\small,
        s sep=0.3em,
        fit=tight,
        anchor=base west,
        child anchor=west,
        tier/.pgfmath=level(),
    },
    [Sound Synthesis,phantom
        [Parametric,name=param
            [Physical Models
                [Numerical
                    [Finite Difference]
                    [Finite Element]
                ]
                [
                    Lumped Network Models
                ]
                [
                    Modal Synthesis
                ]
                [Digital Waveguide
                    [Karplus-Strong]
                ]
                [Functional Transformation]
            ]
            [Signal Models,name=sig
                [Subtractive,name=sub
                    [Analogue Modelling]
                ]
                [Additive,name=add
                    [Wavetable,name=wt]
                    [Spectral Modelling]
                    [Phase Vocoder]
                ]
                [Distortion-based
                    [Amplitude
                        [Waveshaping]
                        [Amplitude Modulation]
                    ]
                    [Frequency
                        [Frequency/Phase Modulation]
                        [Phase Distortion]
                    ]
                ]
                [Source-Filter
                    [LPC]
                ]
            ]
        ]
        [,phantom
        [,phantom,minimum height=8em]
        [DDSP,tier=dds,name=ddsp,no edge,xshift=2em,font=\Large,line width=0.6mm]
        ]
        [Data-Driven
            [Neural Audio Synthesis,name=nas,tier=dds
                [Generative Audio Models]
                [Neural Audio Codecs]
            ]
            [Concatenative
                [Fixed Inventory 
                    [Sampling]
                    [Granular
                        [Diphones]
                        [Triphones]
                        [Demisyllables]
                    ]
                ]
                [Corpus-Based 
                    [Descriptor Distance]
                    [Spectral Match]
                    [Nonuniform Selection]
                    [Statistical]
                ]
            ]
        ]
    ]
\draw (sig) to (ddsp);
\draw (nas) to (ddsp);
\end{forest}
    \caption{A high level taxonomy of popular sound synthesis methods, based on the classifications of \cite{schwarz_corpus-based_2007} and \cite{bilbao_numerical_2009}.
    DDSP methods (bold) offer a combination of data-driven and parametric characteristics.
    This diagram is illustrative of high level relationships, and is not intended to exhaustively catalogue all audio synthesis techniques.}
    \label{fig:synth-tree}
\end{figure}

Synthesisers play an integral role in modern music creation, offering musicians nuanced control over musical timbre~\cite{holmes_electronic_2008}.
Applications of audio synthesis are diverse, ranging from faithful digital emulation of acoustic musical instruments to the creation of unique and novel sounds.
\cite{schwarz_corpus-based_2007} proposed a division of techniques for musical audio synthesis into \textbf{parametric} -- including signal models, such as spectral modelling synthesis~\citep{serra_spectral_1990}, and physical models, such as digital waveguides~\citep{smith_physical_1992} -- and \textbf{concatenative} families -- which segment, reassemble, and align samples from corpora of pre-recorded audio~\citep{schwarz_concatenative_2006}.\footnote{Concatenative methods continue to underpin the dominant professional tools for realistically simulating musical instruments. For example, the EastWest instrument libraries. \url{https://www.soundsonline.com/hollywood-solo-series}. Accessed 10th August 2023.}
We propose an updated version of this classification in Fig.~\ref{fig:synth-tree}, accommodating developments in neural audio synthesis and DDSP.

Neural audio synthesis, the production of audio with neural networks, began to attract considerable research attention  after sample-level audio generation was demonstrated to be feasible by the autoregressive WaveNet \citep{oord_wavenet_2016} and recurrent SampleRNN \cite{mehri_samplernn_2017}.
Adaptation to musical audio quickly followed \citep{engel_neural_2017} and, subsequently, further generative architectures were applied to synthesis of musical sounds \citep{hantrakul_fast_2019,esling_bridging_2018,engel_gansynth_2018}. 
The specific capabilities of these models enabled entirely new techniques, including interpolation or ``morphing'' between sounds \citep{engel_neural_2017,esling_bridging_2018,cakir_musical_2018}, in which intermediate latent representations can be manipulation to produce hybrid sounds; stochastically sampling new sounds \citep{engel_gansynth_2018,rouard_crash_2021}; multi-timbral synthesis \citep{engel_neural_2017,defossez_sing_2018,hawthorne_multi-instrument_2022}; timbre transfer \citep{huang_timbretron_2019,caillon_rave_2021,jain_att_2020}; and more.

\forestset{
    music/.style={
        top color=white,
        bottom color=green!20
    },
    speech/.style={
        top color=white,
        bottom color=blue!20
    },
    musspeech/.style={
        top color=green!20,
        bottom color=blue!20,
    } 
}

Compared to other domains, music has particularly stringent requirements for audio synthesisers.
Real-time inference is a necessity for integration of synthesisers into digital musical instruments, where action-sound latencies above 10ms are likely to be disruptive~\citep{jack_action-sound_2018}.
This has previously been challenging to address with generative audio models~\citep{huzaifah_deep_2020}, particularly at the high sample rates demanded by musical applications.
Expressive control over generation is also necessary in order to provide meaningful interfaces for musicians~\citep{devis_continuous_2023}.
The reliability of this control is also crucial, yet often challenging.
Pitch coherence, for example, is a known issue with GAN-based audio generation~\citep{song_dspgan_2023}.
Further, the comparative scarcity of high quality musical training data further compounds these issues for generative models.

\subsubsection{Musical Instrument Synthesis}

In response to the challenges of neural music synthesis, \citet{engel_ddsp_2020} implemented a differentiable \textit{spectral modelling} synthesiser~\citep{serra_spectral_1990} and effectively replaced its parameter estimation algorithm with a recurrent neural network.
Specifically, the oscillator bank was constrained to harmonic frequencies, an inductive bias which enabled monophonic modelling of instruments with predominantly harmonic spectra, including violin performances from the MusOpen library\footnote{\url{https://musopen.org/}. Accessed 26th August 2023.} and instruments from the NSynth dataset \citep{engel_neural_2017}.
The resulting model convincingly reproduced certain instrument sounds from as little as 13 minutes of training data.
It also allowed, through its low dimensional control representation, similarly convincing timbre transfer between instruments~\citep{carney_tone_2021}.

Building on this success, a number of subsequent works applied various synthesis methods to monophonic and harmonic instrument synthesis including waveshaping synthesis \citep{hayes_neural_2021}, frequency modulation synthesis \citep{caspe_ddx7_2022, ye_nas-fm_2023}, and wavetable synthesis \citep{shan_differentiable_2022}. 
Concurrently, \citet{zhao_transferring_2020} applied the a neural source-filter model to the task of musical instrument synthesis.

These works approach musical audio synthesis as the task of modelling time-varying harmonics and optionally filtered noise, utilizing input loudness and fundamental frequency signals as conditioning.
\citet{hayes_neural_2021} and \citet{shan_differentiable_2022} focus on improving computational efficiency, demonstrating how learned waveshapers and wavetables, respectively, can reduce the inference-time cost.
\citet{caspe_ddx7_2022} and \citet{ye_nas-fm_2023} focus on control and interpretability of the resulting synthesiser.
In contrast to the dense parameter space of an additive synthesier, they apply FM synthesis which, despite its complex parameter interrelationships, facilitates user intervention post-training due to its vastly smaller parameter count.
\citet{ye_nas-fm_2023} built upon the work of \citet{caspe_ddx7_2022}, further tailoring the FM synthesiser to a target instrument through neural architecture search \citep{ren_comprehensive_2022} over modulation routing.

Two common characteristics of the works discussed in this section are a reliance on pitch tracking, and the assumption of predominantly harmonic spectra.
The reasons for this reliance are discussed further in section~\ref{sec:unconstrained-additive}.
Consequently, adaptation of such methods to polyphonic, unpitched, and non-harmonic instruments, or to modelling inharmonicity induced by stiffness and nonlinear acoustic properties, is a challenging task.
Nonetheless, researchers have begun to explore solutions.
\citet{renault_differentiable_2022}, for example, proposed a method for polyphonic piano synthesis, introducing an extended pitch to model string resonance after note releases, explicit inharmonicity modeling based on piano tuning, and detuning to replicate partial interactions on piano strings.
\citet{diaz_rigid-body_2022} presented a differentiable signal model of inharmonic percussive sounds using a bank of resonant IIR filters, which they trained to match the frequency responses produced by modal decomposition using the finite element method.
This formulation was able to converge on the highly inharmonic resonant frequencies produced by excitation of arbitrarily shaped rigid bodies, with varying material parameters.

\subsubsection{Timbre Transfer}

Building on the success of neural style transfer in the image domain~\citep{gatys_neural_2015}, timbre transfer emerged as a task in musical audio synthesis.
\citet{dai_music_2018-1} define timbre transfer as the task of altering ``timbre information in a meaningful way while preserving the hidden content of performance control.''
They highlight that it requires the disentanglement of timbre and performance, giving the example of replicating a trumpet performance such that it sounds like it was played on a flute while maintaining the original musical expression.
Specific examples of timbre transfer using generative models include \citet{bitton_modulated_2018} and \citet{huang_timbretron_2019}.

Timbre transfer has been explored a number of times using DDSP.
\citet{engel_ddsp_2020}'s differentiable spectral modelling synthesiser and the associated $f_0$ and loudness control signals naturally lend themselves to the task, effectively providing a low dimensional representation of a musical performance while the timbre of a particular instrument is encoded in the network's weights.
During inference, $f_0$ and loudness signals from any instrument can be used as inputs, in a many-to-one fashion.
A similar task formulation was explored by \citet{michelashvili_hierarchical_2020}, \cite{carney_tone_2021}, \cite{hayes_neural_2021}, and \citet{caspe_ddx7_2022}.

We note that this formulation of timbre transfer necessarily implies a disentangled definition for timbre, a concept that has historically resisted precise definition~\citep{krumhansl_why_1989}.
The assumption that pitch and loudness are physical attributes that can be disentangled from a sound, leaving only its timbre, is congruent with the ANSI definition, which states that timbre is ``that attribute of auditory sensation which enables a listener to judge that two nonidentical sounds, similarly presented and having the same loudness and pitch, are dissimilar,"~\citep{ansi_psychoacoustic_1960}.
However, this definition has attracted significant criticism.\footnote{See \citep{siedenburg_perceptual_2019} for a review of literature on perceptual representations of timbre}
In particular, \citet{siedenburg_four_2017} emphasize that timbre is a perceptual attribute rather than a physical one, and that an instrument does not have a singular timbre.
Instead, its timbre depends on and covaries with multiple factors including pitch, playing effort, articulation, technique, and more.

\subsubsection{Performance Rendering}
Performance rendering systems seek to map from a symbolic musical representation to audio of that musical piece such that the musical attributes are not only correctly reflected, but expressive elements are also captured.  \cite{castellon_towards_2020, jonason_control-synthesis_2020, wu_ddsp-based_2022} augmented Engel et al.’s differentiable spectral-modelling synthesiser with a parameter generation frontend that received MIDI for performance rendering. \citet{castellon_towards_2020} and \citet{jonason_control-synthesis_2020} used recurrent neural networks to create mappings from MIDI to time-varying pitch and loudness controls. \citet{wu_ddsp-based_2022} presented a hierarchical generative system to map first from MIDI to expressive performance attributes (articulation, vibrato, timbre, etc), and then to synthesis controls.

\subsubsection{Sound Matching}
Synthesizer sound matching, also referred to as automatic synthesizer programming, aims to find synthesiser parameters that mostly closely match a target sound.
Historical approaches include genetic algorithms \citep{horner_machine_1993}, while deep learning has more recently gained popularity~\citep{yee-king_automatic_2018,barkan_inversynth_2019}.

\citet{masuda_synthesizer_2021} proposed to approach sound matching with a differentiable audio synthesiser, in contrast to previous deep learning methods which used a parameter loss function. 
In later work \citep{masuda_improving_2023}, they extended their differentiable synthesiser introduced a self-supervised training scheme blending parameter and audio losses.

\subsection{Speech Synthesis}

Artificial generation of human speech has long fascinated researchers, with its inception tracing back to \citet{dudley_vocoder_1939} at Bell Telephone Laboratories, who introduced \textit{The Vocoder}, a term now widely adopted \citep{dudley_speaking_1950}. Subsequent research has led to a substantial body of work on speech synthesis, driven by escalating demands for diverse and high-quality solutions across applications such as smartphone interfaces, translation devices, and screen readers \citep{tamamori_speaker-dependent_2017}.

In this review, two subtasks of speech synthesis are particularly pertinent, namely text-to-speech (TTS) and voice transformation.
TTS involves converting text into speech, a process often synonymous with the term speech synthesis itself \citep{tan_survey_2021}.
Voice transformation, in contrast, focuses on altering roperties of an existing speech signal, including as voice identity and mood \citep{stylianou_voice_2009}.
A central component in both tasks is the vocoder, responsible for generating speech waveforms from acoustic features.

Classical DSP-based methods for speech synthesis can be broadly split into three categories: articulatory \citep{shadle_prospects_2001, birkholz_modeling_2013}, source-filter/formant \citep{seeviour_automatic_1976}, and concatenative \citep{khan_concatenative_2016}. 
This aligns with the parametric and concatentivative distinction due to \cite{schwarz_corpus-based_2007}, discussed in section~\ref{sec:musaud}.
Specifically, articulatory and source-filter/formant synthesis are parametric methods.
Moreover, articulatory synthesis techniques are related to physical modeling approaches, as they strive to directly replicate physical movements in the human vocal tract.

Unit-selection techniques \citep{hunt_unit_1996}, a subset of concatenative synthesis, use a database of audio segmented into speech units (e.g., phonemes) and an algorithm to sequence units to produce speech.
This approach marked a shift from knowledge-based to data-driven synthesis and was considered state-of-the-art at the time. 
However, the quality of synthesis in unit-selection techniques was dependent on the database size \citep{zen_statistical_2009}. Subsequent developments in machine learning gave rise to statistical parametric speech synthesis (SPSS) systems. Unlike unit selection, SPSS systems removed the need to retain speech audio for synthesis, focusing instead on developing models, such as hidden Markov models, to predict parameters for parametric speech synthesizers \citep{zen_statistical_2009}.

As neural audio synthesis became feasible, neural vocoders such as the autoregressive WaveNet~\cite{oord_wavenet_2016} quickly became state-of-the-art for audio quality.
However, WaveNet's sequential generation was prohibitively costly, motivating the development of more efficient neural vocoders.
Approaches included improvements to neural autoregssion, including cached dilated convolutions \citep{ramachandran_fast_2017} and optimized recurrent neural networks \citep{kalchbrenner_efficient_2018}, and enabled parallel generation using flow-based models \citep{oord_parallel_2018, prenger_waveglow_2019}, denoising diffusion probabilistic models (DDPMs) \citep{kong_diffwave_2021}, and generative adversarial networks (GANs) \citep{kumar_melgan_2019, kong_hifi-gan_2020}. 
GAN-based vocoders in particular have become a common benchmark for neural vocoders, owing to their high-quality synthesis, fast inference time, and generally lightweight networks \citet{matsubara_comparison_2022, song_dspgan_2023}.
Over time, neural vocoders have shifted to focus conditioning via acoustic features \citet{tamamori_speaker-dependent_2017}, with mel-spectrograms most widely used \citep{tan_survey_2021}.

\subsubsection{DDSP-based Vocoders}

Computational efficiency is a central consideration in neural vocoders, as inference time is crucial in many application areas.
Before explicitly DDSP-based models, researchers began integrating DSP knowledge into networks.
\citet{jin_fftnet_2018} made the connection between dilated convolutions and wavelet analysis, which involves iterative filtering and downsampling steps, and proposed a network structure based on the Cooley-Tukey Fast Fourier Transform (FFT) \citep{cooley_algorithm_1965}, capable of real-time synthesis.
Similarly sub-band coding through pseudo-mirror quadrature filters (PQMF) was applied to enable greater parallelisation of WaveRNN-based models~\citep{yu_durian_2020}.

Later work saw the integration of models for speech production.
The source-filter model of voice production has proven particularly fruitful --- LPCNet \citep{valin_lpcnet_2019} augmented WaveRNN \citep{kalchbrenner_efficient_2018} with explicit linear prediction coefficient (LPC) \citep{atal_speech_1971} calculation.%
This allowed a reduction in model complexity, enabling real-time inference on a single core of an Apple A8 mobile CPU.
Further efficiency gains were made subsequently through multi-band linear prediction \citep{tian_featherwave_2020} and sample ``bunching'' \citep{vipperla_bunched_2020}.
\citet{subramani_end--end_2022} later observed that the direct calculation of LPCs limited LPCNet to acoustic features for which explicit formulas were known. 
To alleviate this issue, they proposed to backpropogate gradients through LPCs, enabling their estimation by a neural network.

Despite improved efficiency, \citet{juvela_gelp_2019} noted that LPCNet's autoregression is nonetheless a bottleneck.
To address this, they proposed GAN-Excited Linear Prediction (GELP) which produced the residual signal with a GAN signal, with explicit computation of LPCs from acoustic features, thus limiting autoregression to the synthesis filter and parallelising excitation.

LPCNet and GELP both incorporated DSP-based filters into a source-filter enhanced neural vocoder.
Conversely, \citet{wang_neural_2019-2} proposed the \textit{neural source filter} model which used DSP-based model for the excitation signal (i.e., harmonics plus noise) and a learned neural network filter.
Through ablations, they demonstrated the benefit of using such sinusoidal excitation of the neural filters.
Subsequent improvements to their neural source filter (NSF) method included the addition of a differentiable maximum voice frequency crossover filter~\citep{wang_neural_2019}, and a quasi-periodic cyclic noise excitation~\citep{wang_using_2020}.

Combining these approaches, \citet{mv_sfnet_2020} and \citet{liu_neural_2020} proposed to use differentiable implementations of DSP-based excitation and filtering, and parameterise these with a neural network. This allowed audio sample rate operations to be offloaded to efficient DSP implementations, while the parameter estimation networks could operate at frame level.

Eschewing the source-filter approach entirely,
\citet{kaneko_istftnet_2022} demonstrated that the last several layers in a neural vocoder such as HiFi-GAN \citep{kong_hifi-gan_2020} could be replaced with an inverse short-time Fourier transform (ISTFT).
The number of replaced layers can be tuned to balance efficiency and generation quality.
\citet{webber_autovocoder_2023} also used a differentiable ISTFT, reporting a real-time factor of over $100\times$ for speech at 22.05kHz on a high-end CPU. 
In contrast to a GAN, Webber et al. learned a compressed latent representation and by training a denoising autoencoder.
Nonetheless, \citet{watts_puffin_2023} noted that neural vocoders which run in real-time on a CPU often rely on powerful CPUs, limiting use in low-resource environments.
In particular, they highlight Alternative and Augmentative Communication (AAC) devices, which are used by people with speech and communication disabilities~\citep{murray_augmentative_2009}. 
Watts et al. proposed a method using the ISTFT and pitch-synchronous overlap add (PSOLA), with lightweight neural networks operating at rates below the audio sample rate.

\paragraph{Audio Quality and Robustness}

Maintaining audio quality while remaining robust to unseen input features is a critical challenge in neural vocoder research.
The ``holy grail'' is thus, perhaps, a universal vocoder capable of generating unseen speakers, styles, and languages without fine-tuning \citep{song_dspgan_2023}.
A limiting factor, however, is the mismatch between the training input distribution, and the output distribtuion of a TTS or voice transformation model.
\citet{choi_nansy_2022} suggest that end-to-end training may help, but this is often prohibitively costly.
Neural vocoders are thus typically trained from intermediate acoustic feature representations, of which mel-spectrograms are most widely used owing to their compactness \citet{webber_autovocoder_2023}. 
Even inverting mel-spectrograms, however, is non-trivial~\citep{kaneko_istftnet_2022} involving three distinct inverse problems: (i) recovery of the original-scale magnitude spectrogram; (ii) phase reconstruction; and (iii) frequency-to-time conversion. 

\citet{song_dspgan_2023} highlight that GAN vocoders struggle with periodicity, especially during prolonged vocalisations, and attribute this to unstable parallel generation.
This is compounded by over-smoothing of acoustic features output by TTS or voice transformation models.
They note that DSP vocoders can be more robust and less prone to pitch and phase errors, and thus propose to use a pre-trained neural homomorphic vocoder~\citep{liu_neural_2020} to generate mel-spectrograms for a GAN vocoder, which they confirmed experimentally to improve generalisation to unseen speakers.
Even in DSP-based models, however, design choices can influence robustness.
\citet{wang_using_2020} observed that the choice of source signal interacts with the speaker's gender: sinusoidal excitation performs better for female voices than for male voices, while a cyclic noise signal improves performance on male voices.
Additionally, they note the sine-based source signals may lead to artefacts during less periodic, expressive vocalisations such as creaky or breathy voices.

\paragraph{Control}

DDSP-based methods can also facilitate control over speech synthesis.
\citet{fabbro_speech_2020} distinguish between two categories of method: those that necessitate control and those that offer optional control.
The authors advocate for the latter, arguing that disentangling inputs into components --- namely pitch, loudness, rhythm, and timbre --- provides greater flexibility, proposing a method that enables this.
These disentangled factors are then utilized to drive a differentiable harmonic plus noise synthesizer, although the authors note that there is room for improvement in the quality of the synthesis.

\citet{choi_nansy_2022} also decompose the voice into four aspects: pitch, amplitude, linguistic features, and timbre.
They identify a that control parameters are often entangled in a mid-level representation or latent space, in existing neural vocoders, restricting control and limiting models' potential as co-creation tools.
To address this, in contrast to the single network of \citet{fabbro_speech_2020}, they combine dedicated modules for disentangling controls.
They evaluate their methodology through a range of downstream tasks, using a modified parallel WaveGAN model \citep{yamamoto_parallel_2020} with sinusoidal and noise conditioning, along with timbre and linguistic embeddings.
Reconstruction was found to be nearly identical in a copy synthesis task.

\subsubsection{Text-to-Speech Synthesis}

Text-to-speech (TTS) is the task of synthesising intelligible speech from text, and has received considerable attention due to its numerous commercial applications.
\citet{tan_survey_2021} provide a comprehensive review of the topic, including references to reviews on classical methods and historical perspectives.
While most research on DDSP audio synthesis focuses on vocoding, some studies have also assessed its application in TTS systems \citep{juvela_gelp_2019, liu_neural_2020, wang_neural_2019, choi_nansy_2022, song_dspgan_2023}.

The train-test mismatch problem is particularly salient in TTS, where acoustic features must be estimated from text rather than directly extracted from authentic audio, as in copy synthesis. This discrepancy can degrade synthesis quality.
Additionally, various TTS tasks add further complexity.
Single-speaker TTS models are trained and evaluated on one individual \citep{juvela_gelp_2019, liu_neural_2020, wang_neural_2019}.
In contrast, multi-speaker and multi-lingual TTS demand a vocoder robust to diverse acoustic features \citep{song_dspgan_2023}.
Expressive speech styles, such as reciting poetry or conveying emotion, have also been investigated \citep{song_dspgan_2023}.

\subsubsection{Voice Transformation}

An application of speech synthesis systems is the ability to modify and transform the voice.
Voice transformation is an umbrella term used to refer to a modification that is made to a speech signal that alters one or more aspects of the voice while keeping the linguistic content intact \cite{stylianou_voice_2009}.
Voice conversion (VC) is a subtask of voice transformation that seeks to modify a speech signal such that a utterance from a source speaker sounds like it was spoken by a target speaker.
Voice conversion is a longstanding research task \citep{childers_voice_1985} that has continued to receive significant attention in recent years, demonstrated by the biannual voice conversion challenges operating in 2016, 2018, and 2020.
An overview of the field is provided by \citep{mohammadi_overview_2017} and more recent applications of deep learning towards VC is reviewed by \citep{sisman_overview_2021}.

\citet{nercessian_end--end_2021} incorporated a differentiable harmonic-plus-noise synthesiser \citep{engel_ddsp_2020} to a end-to-end VC model, augmenting it with convolutional pre- and post-nets to further shape the generated signal. 
This formulation allowed end-to-end training with perceptually informed loss functions, as opposed to requiring autoregression.
Nercessian also argued that such ``oscillator driven networks'' are better equipped to produce coherent phase and follow pitch conditioning. 

\citet{choi_nansy_2022} explored zero-shot voice conversion with their NANSY++ model, which was facilitated by the distentangled intermediate representations.
Their approach was to replace the timbre embedding with that of a target speaker, while also transforming pitch conditioning for the sinusoidal generator to match the target.

In voice designing or speaker generation \citep{stanton_speaker_2022}, the goal is to provide a method to modify certain characteristics of a speaker and generate a completely unique voice.
Creation of new voice identifies has application for a number of downstream applications including audiobooks and speech-based assistants.
\citet{choi_nansy_2022} fit normalising flow models, conditioned on age and gender attributes, to generate synthesis control parameters for this purpose.

\subsection{Singing voice synthesis}

Singing voice synthesis (SVS) aims to generate realistic singing audio from a symbolic music representation and lyrics, a task that blends speech and musical instrument synthesis.
The musical context demands an emphasis on pitch and timing accuracy~\citep{saino_hmm-based_2006}, as well as ornamentation through dynamic pitch and loudness contours, as well as higher audio resolutions typical of musical recordings (i.e., 44.1kHz CD quality vs 16kHz or 24kHz as often used for speech), incurring additional computational complexity \citep{chen_hifisinger_2020}.
Applications of SVS include performance rendering from scores when an appropriate vocalist is  unavailable, modifying or correcting existing performances, and recreating performances in the likeness of singers \citep{rodet_synthesis_2002}. 

SVS methods originated in the 1960s, evolving from speech synthesis systems, and early methods can be similarly coarsely categorised into \textit{waveform} and \textit{concatenative} methods~\citep{rodet_synthesis_2002}.
A historical perspective is provided by \citet{cook_singing_1996}, and statistical methods were later introduced by~\citet{saino_hmm-based_2006}.

Early deep learning approaches to SVS included a simple feed-forward network~\citep{nishimura_singing_2016} and a WaveNet-based autoregressive approach~\cite{blaauw_neural_2017-1}, which showed improvements over then state-of-the-art concatenative synthesisers. 
Reviews by \citet{gomez_deep_2018} and \citet{cho_survey_2021} highlight challenges and opportunities in deep learning for SVS, of which five closely matched the motivations we observe across tasks for DDSP-based approaches: (i) audio artifacts; (ii) data scarcity; (iii) efficiency; (iv) explainability; (v) control.

Deep learning-based SVS systems, while achieving high quality, depend on large-scale data-driven training, exacerbated by the diverse vocal expressions in musical singing \citep{yu_singing_2023}.
The scarcity of large annotated singing datasets was noted by \citet{gomez_deep_2018} and \citet{cho_survey_2021}, contrasting with advancements in TTS research from open datasets like Hi-Fi TTS \citep{bakhturina_hi-fi_2021} and LJSpeech \citep{ito_lj_2017}.
This need for extensive data, including for specific vocal techniques like growls and rough voice \citep{gomez_deep_2018}, motivated systems like NANSY++ \citep{choi_nansy_2022}, which supported high-quality resynthesis with a fraction of the OpenCPOP dataset \citep{wang_opencpop_2022}, although it is unclear to what extent this is due to differentiable components as opposed to pre-training.
\citet{wu_ddsp-based_2022} also found that the differentiable vocoder SawSing performed well in resource-limited training.

\citet{gomez_deep_2018} and \citet{cho_survey_2021} note that the black-box nature of deep learning limits analytical understanding of learned mappings and the ability to gain domain knowledge from trained SVS systems. 
Gomez et al. acknowledged that this weakens the link between acoustics research and engineering, and foreshadowed DDSP-like innovation, hypothesising that ``transparent'' algorithms might restore it.
Indeed, explainability has motivated numerous DDSP-based SVS systems \citet{yu_singing_2023,alonso_latent_2021, nercessian_differentiable_2023}.
Yu and Fazekas highlight the potential for their differentiable LPC method to be used for voice decomposition and analysis, while
\citet{nercessian_differentiable_2023} notes that even the DDSP autoencoder of \citet{engel_ddsp_2020} has limited explainability due to the learned mappings from audio features to synthesiser parameters, and proposes a differentiable implementation of the non-parametric WORLD feature analysis and synthesis model.

A further impetus for applying DDSP to SVS is audio quality, with two major challenges in neural vocoders being phase discontinuities and accurate pitch reconstruction. Reconstructing phase information from mel-spectrograms is difficult, and phase discontinuities can cause unnatural sound glitches and voice tremors \citep{wu_ddsp-based_2022}. Differentiable oscillators, like the sawtooth oscillator proposed by Wu et al., address this by enforcing phase continuity.

Accurate pitch control and reconstruction are known challenges for neural vocoders \citep{hono_periodnet_2021}. \citet{yoshimura_embedding_2023} argue that non-linear filtering operations in neural vocoders obscure the relationship between acoustic features and output, complicating accurate pitch control. They propose differentiable linear filters in a source-filter model to address this. \citet{nercessian_differentiable_2023} stress the importance of accurate pitch control and reconstruction for musical applications, a feature inherent to their differentiable WORLD synthesiser.

Computational efficiency is less emphasized in SVS literature compared to speech synthesis; however, GOLF \citet{yu_singing_2023} required less than 40\% GPU memory during training and provided nearly 10x faster inference speed than other DDSP SVS methods (which already supported real-time operation). This fast inference speed can facilitate downstream applications, including real-time musicking contexts or functioning in low-resource embedded devices.

\subsubsection{Singing Voice Conversion}

The task of singing voice conversion (SVC) aims to transform a recording of a source singer such that it sounds like it was sung by a given target singer.
This task, related to voice conversion, introduces further challenges \citep{huang_singing_2023}.
Specifically, there are a wider range of attributes to model, such as pitch contours, singing styles, and expressive variations.
Further, perceived pitch and timing must adhere to the source material while incorporating stylistic elements from the target.
Variations of this task include in-domain transfer, where target singing examples are available, and the more complex cross-domain transfer, where the target must be learned from speech samples \citep{huang_singing_2023} or from examples in a different language \citep{polyak_unsupervised_2020}.

Early methods used Gaussian mixture models (GMMs) to learn spectral envelopes for timbre conversion within a concatenative synthesis framework \citep{villavicencio_applying_2010} or applied directly to the source waveform \citep{kobayashi_statistical_2014}.
Later, GAN-based approaches were applied, as in SingGAN~\citep{sisman_singan_2019}.
However, a downside of these approaches was the need for datasets with aligned pairs of audio from source and target vocalists singing the same song. 
Thus, unsupervised training was explored \citep{nachmani_unsupervised_2019}, enabling training with unpaired data, and inference of the target vocalist from a different song.
Contemporary SVC models include diffusian-based methods such as DiffSinger \citep{liu_diffsinger_2022} and GAN-vocoder based methods like HiFi-SVC \citep{zhou_hifi-svc_2022}.

\citet{nercessian_differentiable_2023} argue that their differentiable implementation of the WORLD vocoder, paired with a deterministic WORLD encoder and a learned decoder, offers increased control over the pitch contour while ensuring phase coherence -- both of which are challenging for neural vocoders.
Further, the interpretable feature representation and extraction procedure allows for direct manipulation, as well as pitch and loudness conditioned timbre transfer, as described by~\citet{engel_ddsp_2020}.

The results of the 2023 SVC Challenge~\citep{huang_singing_2023} further suggest that DDSP may offer benefits to the task.
According to a subjective evaluation of naturalness, the top two performing models were both based on DSPGan~\citep{song_dspgan_2023}, which uses a pre-trained Neural Homomorphic Vocoder~\citep{liu_neural_2020} to generate mel-spectrograms for resynthesis using a differentiable source-filter based model. 
However, the organisers noted that it would be premature to conclude that DSPGan is unilaterally the best SVC model, given the small sample size.
Nonetheless, five other teams incorporated neural source-filter~\citep{wang_neural_2019-2} based components into HiFi-GAN~\citep{kong_hifi-gan_2020} to improve generalisation, implying that incorporation of domain knowledge via DDSP offers some benefit.

\section{Differentiable Digital Signal Processing}\label{sec:diffimp}

In this section we survey differentiable formulations of signal processing operations for audio synthesis.
Whilst many were first introduced for other tasks, their relevance to audio synthesis is often clear as many correspond to components of classical synthesis algorithms.
For this reason, we choose to include them here.

Our aim in this section is to provide an overview of the technical contributions that underpin DDSP in order to make clearer the connections between methods, as well as facilitate the identification of open directions for future research.
We also hope that this section will act as a technical entry point for those wishing to work with DDSP.
We thus do not intend to catalogue every \textit{application} of a given method, but instead endeavour to acknowledge technical contributions and any prominent variations.

\subsection{Filters}

\subsubsection{Infinite Impulse Response}\label{sec:iir}

\begin{table}
\centering
\resizebox{\textwidth}{!}{
\begin{tabular}{lllllll}
\toprule
\textbf{Type} & \textbf{Reference} & \textbf{Representation} & \textbf{Parameters} & \textbf{Training Algorithm} & \textbf{Stability Constraints} & \textbf{Time Varying} \\
\midrule
First Order & \cite{kuznetsov_differentiable_2020} & Direct & $b_0, b_1, a_1$ & TBPTT & $|a_1|<1$ & \xmark \\
 & \cite{bhattacharya_optimization_2020} & Shelving filter & Freq. ($f_c$), gain ($G$) & IBPTT & --- & \xmark \\
\midrule
Second Order & \cite{kuznetsov_differentiable_2020} & Direct & $b_0, b_1, b_2, a_1, a_2$& TBPTT (TDF-II) & \tiny \makecell[tl]{$|a_1|\leq0.5$\\$|a_2|<0.5$} & \xmark \\
 & \cite{kuznetsov_differentiable_2020} & State-variable filter & \makecell[tl]{Cutoff ($g$), damping ($R$),\\band gains ($c_\text{LP}, c_\text{HP}, c_\text{BP}$)} & TBPTT & \tiny \makecell[tl]{$g=1$\\$R=\frac{1}{\sqrt{2}}$} & \xmark \\
  & \cite{bhattacharya_optimization_2020} & Peak & \makecell[tl]{Freq. ($f_c$), bandwidth ($f_b$),\\gain ($G$)} & IBPTT (DF-II) & --- & \xmark \\
 & \cite{nercessian_neural_2020} & Low/high shelf, peak & \makecell[tl]{Freq. ($\omega_0$), gain ($A$),\\Q-factor ($q$)}  & Freq. Sampling & --- & \xmark \\
 & \cite{nercessian_lightweight_2021} & Direct &$b_0, b_1, b_2, a_1, a_2$ & Freq. Sampling & 
 \tiny
 \makecell[tl]{
$a_1 \leftarrow 2\tanh a_1$\\
    $a_2 \leftarrow \frac{1}{2}(|a_1| + (2 - |a_1|)\tanh a_2)$
    }
 & Via conditioning \\
 &  & Pole/zero & \tiny $p, q \in \mathbb{C}$ & Freq. Sampling & 
        \makecell[tl]{$p \leftarrow p \cdot \frac{\tanh|p|}{|p|}$}
 & Via conditioning \\
 &  & Low/high shelf, peak & \makecell[tl]{Freq. ($\omega_0$), gain ($A$),\\Q-factor ($q$)} & Freq. Sampling & \tiny \makecell[tl]{$Q \leftarrow \frac{Q_\text{max}}{1+e^{Q}}$} & Via conditioning \\
 & \cite{yu_singing_2023} & Direct (all-pole) & $a_1, a_2$ & TBPTT (DF-I) & 
 \tiny
 \makecell[tl]{
$a_1 \leftarrow 2\tanh a_1$\\
    $a_2 \leftarrow \frac{1}{2}(|a_1| + (2 - |a_1|)\tanh a_2)$
    }
 & Framewise \\
 & \cite{carson_differentiable_2023} & All-pass & \makecell[tl]{Break freq. ($\omega_b$), thru\\gain ($g_1$), feedback gain ($g_2$)}  & Freq. Sampling & --- & Framewise \\
\midrule
Arbitrary order & \cite{kuznetsov_differentiable_2020} & Linear state-space & Transition matrices ($\mathbf{A}, \mathbf{B},\mathbf{C},\mathbf{D})$ & TBPTT & $\mathbf{A}\sim\mathcal{U}_{n\times n}\left(-\frac{1}{n},\frac{1}{n}\right)$ & \xmark \\
 & \cite{mv_sfnet_2020} & LPC & Reflection Coefficients ($k_m$) & TBPTT & $k_m \in (-1, 1)$ & Framewise \\
 & \cite{subramani_end--end_2022} & LPC & Reflection Coefficients ($k_m$) & Autoregressive & $k_m \in (-1, 1)$ & Framewise \\
\bottomrule
\end{tabular}
}
\captionof{table}{Differentiable implementations of discrete time IIR filters with trainable parameters. Specific parameterisations, training algorithms, and stability constrains are presented. Recursive filter structure is also indicated where appropriate and when clear from the original manuscript.}%
\label{tab:iir_list}
\end{table}

A causal linear time-invariant (LTI) filter with impulse response $h(t)$ is said to have an infinite impulse response (IIR) if there does not exist a time $T$ such that $h(t)=0$ for all $t>T$.
In the case of digital filters, this property arises when the filter's difference equation includes a nonzero coefficient for a previous output.

We present the dominant methods for differentiable IIR filtering in Table~\ref{tab:iir_list}.

\paragraph{Recursive methods}
Optimising IIR filter coefficients by gradient descent is not a new topic.
Several algorithms for adaptive IIR filtering and online system identification rely on the computation of exact or approximate gradients \citep{shynk_adaptive_1989}.
Moreover, to facilitate the training of locally recurrent IIR multilayer perceptrons (IIR-MLP) \citep{back_fir_1991}, approximations to backpropogation-through-time (BPTT) have been proposed \citep{campolucci_-line_1995}. 
However, prior to widely available automatic differentiation, such methods required cumbersome manual gradient derivations, restricting the exploration of arbitrary filter parametrisations or topologies.

One such training algorithm, known as instantaneous backpropogation through time (IBPTT) \cite{back_fir_1991}\footnote{Note that the algorithm proposed by \cite{back_fir_1991} was not referred to as IBPTT in the original work. This name was given later by \cite{campolucci_-line_1995}.}
was applied by \cite{bhattacharya_optimization_2020} to a constrained parameterisation of IIR filters, namely peak and shelving filters such as those commonly found in audio equalisers.
This method was tested on a cascade of such filters and used to match the response of target head-related transfer functions (HRTFs).
However, the formulation of IBPTT precludes the use of most modern audio loss functions, somewhat hindering the applicability of this method.

\newcommand{\kuz}{Kuznetsov et al. }
\cite{kuznetsov_differentiable_2020} identified the close relationship between IIR filters and recurrent neural networks (RNNs).
In particular, they illustrated that in the case of a simple RNN known as the \textit{Elman network} \citep{elman_finding_1990}, the two are equivalent when the activations of the Elman network are element-wise linear maps, and the bias vectors are zero.
This is illustrated below:

\begin{equation}
\begin{array}{rlcrl}
    \mathbf{h}[n] &= \sigma_h\left(\mathbf{W}_h \mathbf{h}[n-1] + \mathbf{U}_h \mathbf{x}[n] + b_h\right), & & \mathbf{h}[n] &= \mathbf{W}_h \mathbf{h}[n-1] + \mathbf{U}_h \mathbf{x}[n] , \\
    \mathbf{y}[n] &= \sigma_y\left(\mathbf{W}_y \mathbf{h}[n] + b_y\right). & & \mathbf{y}[n] &= \mathbf{W}_y \mathbf{h}[n]. \\
    & \text{(Elman network)} & & & \text{(All-pole IIR filter)}
\end{array}
\end{equation}

Such RNNs are typically trained with backpropogation through time (BPTT), and more commonly its truncated variant (TBPTT) in which training sequences are split into shorter subsequences.\footnote{Note that the truncation in TBPTT, which is applied to input sequences, is distinct from that in CBPTT, which is applied to intermediate backpropogated errors.}
Based on this equivalence, \kuz directly applied TBPTT to IIR filters, effectively training various filter structures as linear recurrent networks.
To ensure filter stability, they propose simple constraints on parameter initialisation.

A challenge with recursive optimisation is the necessity of memory allocations for each ``unrolled'' time step.%
For long sequences, this can result in poor performance and high memory cost.
To address this and allow optimisation over high order IIR filters, \cite{yu_singing_2023} provided an efficient algorithm for applying BPTT to all-pole filters.
Specifically, for an all-pole filter with coefficients $\mathbf{a}\in\mathbb{R}^M$, they showed that the partial derivatives $\frac{\partial y[n]}{\partial a_i}$ and $\frac{\partial \mathcal{L}}{\partial x[n]}$ can be expressed as applications of the same all pass filter to $-y[n-i]$, and a filter with time-reversed coefficients to $\frac{\partial \mathcal{L}}{\partial y[M-n]}$, respectively. That is,

\begin{align}
    \frac{\partial y[n]}{\partial a_i}
    =
    -y[n-i]-\sum_{k=1}^M a_k \frac{\partial y[n-k]}{\partial a_i} 
    &\implies
    \mathcal{Z}\left\{
    \frac{\partial y[n]}{\partial a_i}
    \right\}
    =
    \frac{1}{1 + \sum_{k=1}^M a_k z^{-k}} \cdot
    Y\left(z\right)\cdot z^{-i},\label{eqn:yu_ya}\\
    \frac{\partial \mathcal{L}}{\partial x[n]}
    =
    \frac{\partial \mathcal{L}}{\partial y[n]}
    -\sum_{k=1}^M a_k \frac{\partial \mathcal{L}}{\partial x[n+k]} 
    &\implies
    \mathcal{Z}\left\{
    \frac{\partial \mathcal{L}}{\partial x[n]}
    \right\}
    =
    \frac{1}{1 + \sum_{k=0}^{M-1} a_{M-k} z^{-k}} \cdot
    \mathcal{Z}\left\{\frac{\partial \mathcal{L}}{\partial y[n]}\right\},\label{eqn:yu_la}
\end{align}

\noindent where $\mathcal{Z}\left\{\cdot\right\}$ denotes the z-transform operator.
It is clear from equations \ref{eqn:yu_ya} and \ref{eqn:yu_la} that these derivatives can be evaluated without the need for a computation graph of unrolled filter timesteps, enabling the use of efficient, recursive IIR implementations.
\footnote{This algorithm has been implemented by \cite{yu_singing_2023} and is available in the open source TorchAudio package. \url{https://github.com/pytorch/audio}. Accessed 21st July 2023.}.

\paragraph{Frequency sampling methods}\label{sec:freqsamp}

It is common, when working with higher order filters, to factorise the transfer function into a cascade of second order sections in order to ensure numerical stability.
It has been reported \citep{nercessian_lightweight_2021} that optimising differentiable cascades using BPTT/TBPTT limits the number of filters that can be practically learned in series, motivating an alternate algorithm for optimising the parameters of cascaded IIR filters. 

One such approach, proposed by \cite{nercessian_neural_2020}, circumvented the need for BPTT by defining a loss function in the spectral domain, with the desired filter magnitude response as the target.
This method was used to train a neural network to match a target magnitude response using a cascade of differentiable parametric shelving and peak filters.
To compute the frequency domain loss function, the underlying response of the filter must be sampled at some discrete set of complex frequencies, typically selected to be the $K^\text{th}$ roots of unity $e^{j\frac{k}{K}}, k = 0, 1, \dots, K-1$.

This procedure is equivalent to the frequency sampling method for finite-impulse response filter design.
That is, by sampling the filter's frequency response we are effectively optimising an FIR filter approximation to the underlying frequency response.
Naturally, this sampling operation results in time-domain aliasing of the filter impulse response, and the choice of $K$ thus represents a trade-off between accuracy and computational expense.

In subsequent work, \cite{nercessian_lightweight_2021} extended this frequency sampling approach to \textit{hyperconditioned} IIR filter cascades, to model an audio distortion effect.
Hyperconditioning refers to a hypernetwork-like \citep{ha_hypernetworks_2016} structure in which the hypernetwork introduces conditioning information to the main model by generating its parameters.
In this case, the hypernetwork's inputs are the user-facing controls of an audio effect, and the main model is a cascade of biquadratic IIR filters.

Whilst filter stability is less of a concern when training a model to produce fixed sets of filter coefficients, the hyperconditioned setting carries a greater risk.  
Both user error and erroneous model predictions may lead to a diverging impulse response at either inference or training time.
For this reason, Nercessian et al. tested three parameterisations of cascaded biquads (coefficient, conjugate pole-zero, and parametric EQ representations), and proposed stability enforcing activations for each.
Rather than directly optimising filter magnitude responses, the model was instead optimised using an audio reconstruction loss by applying the filters to an input signal.
During optimisation, this was approximated by complex multiplication of the frequency-sampled filter response in the discrete Fourier domain.

Subsequently, this constrained frequency sampling method was applied to the more general task of IIR filter design \citep{colonel_direct_2022-1}.
Here, a neural network was trained to parameterise a differentiable biquad cascade to match a given input magnitude response, using synthetic data sampled from the coefficient space of the differentiable filter.
However, the authors do note that training a model with higher order filters ($N \geq 64$) tended to lead to instability, suggesting that even the frequency sampling approach may be insufficient to solve the aforementioned challenges with cascade depth \citep{nercessian_lightweight_2021}.

\cite{diaz_rigid-body_2022} also applied the constrained frequency sampling method to hybrid parallel-cascade filter network structures, for the purpose of generating resonant filterbanks which match the modal frequencies of rigid objects.
In comparing filter structures, authors found that the best performance was achieved with a greater number of shallower cascades.
Paired with the insight from \cite{colonel_direct_2022-1}, this suggests that further research is necessary into the stability and convergence of the frequency sampling method under different filter network structures, as well as the effect of saturating nonlinearities on filter coefficients.

Some applications call for all-pass filters -- i.e. filters which leave frequency magnitudes unchanged, but which do alter their phases.
\cite{carson_differentiable_2023} optimised all-pass filter cascades directly using an audio reconstruction loss.
Due to the difficulty of optimising time-varying filter coefficients, the time-varying phaser filter is piecewise approximated by $M$ framewise time-invariant transfer functions $H^{(m)}(z)$, which are predicted and applied to windowed segments of the input signal.
Overlapping windows are then summed such that the constant overlap-add property (COLA) is satisfied.

In the case where an exact sinusoidal decomposition of the filter's input signal is known, such as in a synthesiser with additive oscillators, the filter's transfer function can be sampled at exactly the component frequencies of the input.
This is applied by \cite{masuda_synthesizer_2021} to the task of sound-matching with a differentiable subtractive synthesiser.

\paragraph{Source-filter models and linear predictive coding}\label{sec:sflpc}

The source-filter model of speech production describes speech in terms of a source signal, produced by the glottis, which is filtered by the vocal tract and lip radiation.
This is frequently approximated as the product of three LTI systems, i.e. $Y(z)=G(z)H(z)L(z)$ where $G(z)$ describes the glottal source, $H(z)$ describes the response of the vocal tract, and $L(z)$ is the lip radiation.
Often, $L(z)$ is assumed to be a first order differentiator, i.e. $L(z)=1-z^{-1}$, and the glottal flow derivative $G'(z)=G(z)L(z)$ is directly modelled.

Frequently, a local approximation to the time-varying filter $H(z)$ is obtained via linear prediction over a finite time window of length $M$:

\begin{equation}
    y[n] = e[n] + \sum_{m=1}^M a_m y[n-m],
\end{equation}

\noindent where $e[n]$ describes the excitation signal, which is equivalent to the linear prediction residual. The coefficients $a_m$ are exactly the coefficients of an $M^\text{th}$ order all-pole filter.
This representation is known as linear predictive coding (LPC)

To the best of our knowledge, \cite{juvela_gelp_2019} were the first to incorporate a differentiable synthesis filter into a neural network training pipeline.
In their method, a GAN was trained to generate excitation signals $e[n]$, while the synthesis filter coefficients were directly estimated from the acoustic feature input, i.e. non-differentiably.
Specifically, given a mel-spectrum input, $\mathbf{m}=\log\left(\mathbf{M}\hat{\mathbf{x}}\right)$, where $\mathbf{M}$ is the mel-frequency filterbank and $\hat{\mathbf{x}}$ is the discrete Fourier transform of a signal window, synthesis filter coefficients $a_m$ are estimated by approximating the discrete Fourier spectrum, $\hat{\mathbf{x}}\approx \max \left(\mathbf{M}^\dagger \exp(\mathbf{m}), \epsilon\right)$, from which the autocorrelation function is computed and the normal equations are solved for $a_m$.
In order to backpropogate error gradients through the synthesis filter to the excitation generator, the filter is approximated for training by truncating its impulse response, which is equivalent to the frequency sampling method discussed in section~\ref{sec:freqsamp}.

LPC has also been used to augment autoregressive neural audio synthesis models.
In LPCNet \citep{valin_lpcnet_2019}, linear prediction coefficients were non-differentiably computed from the input acoustic features, while the sample-level neural network predicted the excitation.
However, in subsequent work \citep{mv_sfnet_2020, subramani_end--end_2022} this estimation procedure was made differentiable, allowing synthesis filter coefficients to also be directly predicted by a neural network.
As LPC coefficients can easily be unstable, both \cite{subramani_end--end_2022} and \cite{mv_sfnet_2020} opted not to directly predict the LPC coefficients $a_m$, but instead described the system in terms of its \textit{reflection coefficients} $k_i$.
The fully differentiable Levinson recursion then allowed recovery of the coefficients $a_m$.
Specifically, where $a_m^{(i)}$ denotes the $m^\text{th}$ LPC coefficient in a filter of order $i$ (i.e. $M=i$), the recursion is defined:

\begin{equation}
    a_m^{(i)} = 
    \begin{cases}
    k_m & \text{if } m = i \\
    a_m^{(i-1)}+k_i a_{i-m}^{(i-1)} & \text{otherwise.}
    \end{cases}
\end{equation}

\noindent When $k_i\in(-1, 1)$ filter stability is thus guaranteed.
This is applied by \cite{subramani_end--end_2022} in an end-to-end differentiable extension of the LPCNet architecture, again with autoregressive prediction of the excitation signal.
Conversely, \cite{mv_sfnet_2020} employ a notably simpler source model, consisting of a mixture of an impulse train and filtered noise signal.
Synthesis filter coefficients were also estimated differentiably by~\citet{yoshimura_embedding_2023}, who used the truncated Maclaurin series of a mel-cepstral synthesis filter to enable FIR approximation.

An alternative approach to modelling the glottal source is offered by \cite{yu_singing_2023}, who use a one-parameter ($R_d$) formulation of the Liljencrants-Fant (LF) model of the glottal flow derivative $G'(z)$. 
The continuous time glottal source is sampled in both time and the $R_d$ dimension to create a two-dimensional wavetable.
To implement differentiable time-varying LPC filtering, the authors opt to use a locally stationary approximation and produce the full signal by overlap-add resynthesis.
Rather than indirect paramaterisation of the filter via reflection coefficients, the synthesis filter is factored to second order sections.
The coefficient representation and accompanying constraint to the biquad triangle proposed by \cite{nercessian_lightweight_2021} is then used to enforce stability.

\cite{sudholt_vocal_2023} apply a differentiable source-filter based model of speech production to the inverse task of recovering articulatory features from a reference recording.
To estimate articulatory parameters of speech production, S\"udholt et al. use a differentiable implementation of the Pink Trombone,\footnote{\url{https://dood.al/pinktrombone}. Accessed 3rd Aug 2023.} which approximates the geometry of the vocal tract as a series of cylindrical segments of varying cross-sectional area --- i.e. the Kelly-Lochbaum vocal tract model.
Instead of independently defining segment areas, these are parameterised by tongue position $t_p$, tongue diameter $t_d$, and some number of optional constrictions of the vocal tract.
To estimate these parameters, S\"udholt et al. derive the waveguide transfer function and perform gradient descent using the mean squared error between the log-scaled magnitude response and the vocal tract response estimated by inverse filtering.

\subsubsection{Finite Impulse Response}

A LTI filter is said to have a finite impulse response (FIR), if given impulse response $h(t)$ there exists a value of $T$ such that $h(t)=0$ for all $t>T$.
In discrete time, this is equivalent to a filter's difference equation being expressible as a sum of past inputs.
Due to the lack of recursion, FIR filters guarantee stability and are less susceptible to issues caused by the accumulation of numerical error.
However, this typically comes at the expense of ripple artifacts in the frequency response, or an increase in computational cost to reduce these artifacts.

As with IIR filters, the discrete time FIR filter is equivalent to a common building block of deep neural networks, nameley the \textit{convolutional layer}\footnote{Note that the convolution operation in neural networks is usually implemented as a cross-correlation operation. This is equivalent to convolution with a kernel reversed along the dimension(s) of convolution. Hence, we use these terms interchangeably to aid legibility.}.
Again, linear activations and zero bias yield the exact filter formulation:

\begin{equation}
\begin{array}{rlcrl}
    \mathbf{y}[n] &= \sigma\Bigl((\mathbf{W}\ast\mathbf{h})[n] + \mathbf{b}\Bigr), & & \mathbf{y}[n] &= (\mathbf{W} \ast\mathbf{h})[n]. \\
    & \text{(Convolutional layer)} & & & \text{(FIR filter)}
\end{array}
\end{equation}

In practice, we frequently wish to produce a time-varying frequency response in order to model the temporal dynamics of a signal. 
The stability guarantees and ease of parallel evaluation offered by FIR filters mean they are an appropriate choice for meeting these constraints in a deep learning context, but care must be taken to compensate for issues such as spectral leakage and phase distortion using filter design techniques.
Moreover, such compensation must be achieved differentiably.

To the best of our knowledge, the first such example of differentiable FIR filter design was proposed by \cite{wang_neural_2019}.
This work applied a pair of differentiable high- and low-pass FIR filters, with a time-varying cutoff frequency $f_c[m]$ predicted by a neural network conditioning module.\footnote{The specific parameterisations of $f_c$ proposed by \cite{wang_neural_2019} combine neural network predictions with domain knowledge about the behaviour of the maximum voice frequency during voiced and unvoiced signal regions. We omit these details here to focus on the differentiable filter implementation.}
The filters are then implemented as windowed sinc filters, with frame-wise impulse responses $\tilde{h}_\text{LP}[m,n]$ and $\tilde{h}_\text{HP}[m,n]$.

Whilst Wang et al. manually derive an the loss gradient with respect to $f_c[m]$, their implementation relies on automatic differentiation.

While closed form parameterisation of FIR filter families allows for relatively straightforward differentiable filter implementations, the complexity of the resulting frequency responses is limited.
However, filter design methods such as frequency sampling and least squares allow for FIR approximations to arbitrary responses to be created.
The first differentiable implementation of such a method was proposed by \cite{engel_ddsp_2020}, whose differentiable spectral-modelling synthesiser contained a framewise time-varying FIR filter applied to a white noise signal.
To avoid phase distortion and suppress frequency response ripple, Engel et al. proposed a differentiable implementation of the window design method for linear phase filter design.

Specifically, a frequency sampled framewise magnitude response $\mathbf{\hat{h}}[m] \in \mathbb{R}^N$ is output by the decoder, where $m$ denotes the frame index and $N$ is the length of the impulse response.\footnote{Note that here we adopt vector notation for the framewise impulse response (i.e. $\mathbf{\hat{h}}[m]_n = \hat{h}[m,n]$) to allow simplify the representation of operations.}
To recover the framewise symmetric impulse responses $\mathbf{h}[m]\in\mathbb{R}^N$, they take the inverse discrete Fourier transform, $\mathbf{h}[m]=\mathbf{W}_N^H \mathbf{\hat{h}}[m])$, for $N$-point DFT matrix $\mathbf{W}_N$.
The symmetry of the impulse responses a sufficient condition for a linear phase response as a function of frequency.
To mitigate spectral leakage, a window function $\mathbf{w}\in\mathbb{R}^N$ is applied to each symmetric impulse response. Finally, the filter is shifted to causal form such that it is centred at $\frac{N}{2}$ for a window length of $N$ samples.
The filters are then applied to the signal by circular convolution, achieved via complex multiplication in the frequency domain.

An alternative differentiable FIR parameterisation was suggested by \cite{liu_neural_2020}, who adopted complex cepstra as an internal representation.
This representation jointly describes the filter's magnitude response and group delay, resulting in a mixed phase filter response.
The authors note group delay exerts an influence over speech timbre, motivating such a design.
However, the performance of this method is not directly compared to the linear phase method described above.
The approximate framewise impulse response $\mathbf{h}[m]$ is recovered from the complex cepstrum $\mathbf{\hat{h}}[m]$ as follows:

\begin{equation}
    \mathbf{h}[m] = \mathbf{W}_N^H \exp \left( \mathbf{W}_N\mathbf{\hat{h}}[m]\right)
\end{equation}

\noindent where the exponential function is applied element-wise.

\citet{barahona-rios_noisebandnet_2023} applied an FIR filterbank to a white noise source for sound effect synthesis, but circumvented the need to implement the filters differentiably by pre-computing the filtered noise signal and predicting band gains.

\subsection{Additive Synthesis}

Many audio signals contain prominent oscillatory components as a direct consequence of the physical tendency of objects to produce periodic vibrations \citep{smith_physical_2010}.
A natural choice for modelling such signals, additive synthesis, thus also encodes such a preference for oscillation.
Motivated by the signal processing interpretation of Fourier's theorem, i.e. that a signal can be decomposed into sinusoidal components, additive synthesis describes a signal as a finite sum of sinusoidal components.
Unlike representation in the discrete Fourier basis, however, the frequency axis is not necessarily discretised, allowing for direct specification of component frequencies.
The general form for such a model in discrete time is thus:

\begin{equation}\label{eqn:gen_additive}
    y[n] = \sum_{k=1}^K \alpha_k[n] \sin \left(\phi_k + \sum_{m=0}^n \omega_k[m] \right),
\end{equation}

\noindent where $K$ is the number of sinusoidal components, $\mathbf{\alpha}[n]\in\mathbb{R}^K$ is a time series of component amplitudes, $\mathbf{\phi}\in\mathbb{R}^K$ is the component-wise initial phase, and $\mathbf{\omega}[n]$ is the time series of instantaneous component frequencies.
Often, parameters are somehow constrained or jointly parameterised, such as in the harmonic model where $\omega_k[n] = k \omega_0[n]$ for some fundamental frequency $\omega_0[n]$.

A prominent extension of additive synthesis is \textit{spectral modelling synthesis} \citep{serra_spectral_1990}.
In this approach, the \textit{residual} signal (i.e. the portion of the signal remaining after estimating sinusoidal model coefficients) is modelled stochastically, as a noise source processed by a time varying filter.
Such a model was implemented differentiably by \cite{engel_ddsp_2020}, using a bank of harmonic oscillators combined with a LTV-FIR filter applied to a noise source.
Specifically, the oscillator bank was parameterised as follows:

\begin{equation}\label{eqn:ddsp_harm}
    y[n] = A[n] \sum_{k=1}^K \hat{\alpha}_k[n] \sin\left(
    \sum_{m=0}^n k\omega_0[m]
    \right),
\end{equation}

\noindent where $A[n]$ is a global amplitude parameter, and $\hat{\mathbf{\alpha}}[n]$ is a normalised distribution over component amplitudes (i.e. $\sum_k \hat{\alpha}_k[n] = 1$ and $\hat{\alpha}_k[n] \geq 0$).
In practice, $A[n]$ and $\hat{\mathbf{\alpha}}[n]$ are predicted at lower sample rates, and interpolated before evaluating the final signal.
The fundamental frequency $\omega_0[n]$ is obtained by means of a pitch estimation algorithm -- in the paper, CREPE \citep{kim_crepe_2018} is used -- rather than by direct optimisation.
It is thus interesting to note that Eqn~\ref{eqn:ddsp_harm} is linear with respect to parameters $A[n]$ and $\hat{\mathbf{\alpha}}[n]$, and thus admits a convex optimisation problem for an appropriate choice of loss function.
This is, however, not the case with respect to $\omega_0[n]$, an issue discussed in detail in Section~\ref{sec:path}.

A more constrained version of the differentiable harmonic oscillator bank was applied by \cite{masuda_synthesizer_2021} for the purpose of synthesiser sound matching. Specifically, they produced anti-aliased approximations of square and sawtooth waveforms by summing harmonics at pre-determined amplitudes:

\begin{align}
     y_\text{saw}[n] &= \sum_{k=1}^K \frac{2}{\pi k} \sin\left( 2\pi k \sum_{m=0}^n \omega_0[m] \right) \\
     y_\text{square}[n] &= \sum_{k=1}^K \frac{4}{\pi\left(2k-1\right)} \sin\left( 2\pi (2k-1) \sum_{m=0}^n \omega_0[m] \right).
\end{align}

\noindent Masuda et al. also introduce a waveform interpolation parameter $p$, such that $y[n] = p y_\text{saw}[n] + (1-p) y_\text{square}[n]$, allowing for differentiable transformation between these fixed spectra.

\subsubsection{Wavetable Synthesis}

Historically, a major obstacle to the adoption of additive synthesis was simply the computational cost of evaluating potentially hundreds of sinusoidal components at every time step.
A practical solution was to pre-compute the values of a sinusoid at a finite number of time-steps and store them in a memory buffer.
This buffer can then be read periodically at varying ``frequencies'' by fractionally incrementing a circular read pointer and applying interpolation.
The buffer, referred to as a \textit{wavetable}, need not contain only samples from a sinusoidal function, however.
Instead, it can contain any values, allowing for the specification of arbitrary harmonic spectra (excluding the effect of interpolation error) when the wavetable is read periodically.
Wavetables can thus allow for efficient synthesis of a finite number of predetermined spectra, or even continuous interpolation between spectra through interpolation both between and within wavetables.

\cite{shan_differentiable_2022} introduced a differentiable implementation of wavetable synthesis, drawing comparison to a dictionary learning task.
In particular, their proposed model learns a finite collection of wavetables $D = \left\{w_k[n]\right\}_{k\in\left\{1\dots K\right\}}$.
The fractional read position of the wavetables is determined by integrating the $\omega_0$ parameter which, as in the harmonic oscillator bank of \cite{engel_ddsp_2020}, is provided by a separate model. 
To produce a particular timbre, the model predicts coefficients $c_k$ for a weighted sum over the wavetables, such that $\sum_{k=1}^K c_k = 1$ and $c_k \geq 0$.

Notably, Shan et al. found that this approach outperformed the DDSP additive model in terms of reconstruction error when using 20 wavetables, and performed almost equivalently with only 10 wavetables.
Further, this method provided a roughly $12\times$ improvement in inference speed, although as the authors note this is likely to be related to the 10-fold decrease in the number of parameter sequences that require interpolation..

\subsubsection{Unconstrained Additive Models}\label{sec:unconstrained-additive}

The implementations discussed thus far are all, in a sense, \textit{constrained} additive models.
This is because a harmonic relationship between sinusoidal component frequencies is enforced.
By contrast, the general model form illustrated in Eqn.~\ref{eqn:gen_additive} is \textit{unconstrained}, which is to say that its frequency parameters are independently specified.
In some circumstances, the greater freedom offered by such a model may be advantageous -- for example, many natural signals contain a degree of inharmonicity due to the geometry or stiffness of the resonating object.
However, vastly fewer examples of differentiable unconstrained models exist in the literature than of their constrained counterparts.

Nonetheless, \cite{engel_self-supervised_2020} introduced a differentiable unconstrained additive synthesiser, which they applied to the task of monophonic pitch estimation in an analysis-by-synthesis framework.
The differentiable unconstrained model follows the form in Eqn~\ref{eqn:gen_additive}, with the omission of the initial phase parameter.
Thus, unlike their differentiable harmonic model, there is no factorised global amplitude parameter -- instead each sinusoidal component is individually described by its amplitude envelope $\alpha_k[n]$ and frequency envelope $\omega_k[n]$.

Optimisation of this model, however, is not as straightforward as the constrained harmonic case where an estimate of $\omega_0[n]$ is provided \textit{a priori}.
The non-convexity of the sinusoidal oscillators with respect to their frequency parameters leads to a challenging optimisation problem, which does not appear to be directly solvable by straightforward gradient descent over an audio loss function.
This was experimentally explored by \citet{turian_im_2020} who found that, even in the single sinusoidal case, gradients for most loss functions were uninformative with respect to the ground truth frequency.
Thus, Engel et al. incorporated a self-supervised pre-training stage into their optimisation procedure.
Specifically, a dataset of synthetic signals from a harmonic-plus-noise synthesiser was generated, for which exact sinusoidal model parameters were know.
This was used to pretrain the sinusoidal parameter encoder network with a parameter regression loss (discussed further in section~\ref{sec:parameter}), which circumvents the non-convexity issue.

Subsequently, \cite{hayes_sinusoidal_2023} proposed an alternate formulation of the unconstrained differentiable sinusoidal model, which aimed to circumvent the issue of non-convexity with respect to the frequency parameters.
Specifically, they replaced the sinusoidal function with a complex surrogate, which produced a damped sinusoid:

\begin{equation}
    y[n] = \sum_{k=1}^K \mathfrak{Re}\left\{
    z_k^n
    \right\}
    = \sum_{k=1}^K \left|z_k\right|^n \cos n \angle z_k,
\end{equation}

\noindent where $\mathfrak{Re}$ denotes the real part of a complex variable, $|z|$ denotes the complex modulus, $\angle z$ denotes the complex phase, and $z_k$ are the complex parameters of the model which jointly encode both frequency and damping.
Note that as the sample index $n$ becomes the exponential parameter, this model does not directly accommodate time-varying frequency parameters.
However, at the time of writing, no published work exists applying this surrogate to enable differentiable unconstrained additive synthesis.

\subsection{Non-linear Synthesis}

Non-linear or distortion-based synthesis techniques emerged during the 60s, 70s and 80s, and became popular as alternatives to additive and subtractive methods due to the comparative efficiency with which they could produce varied complex spectra \citep{bilbao_numerical_2009}.
Broadly, these approaches involve evaluating a nonlinear function of either an audio signal or of synthesis parameters.
This typically results in the introduction of further frequency components, with frequencies and magnitudes determined by the specifics of the method.

\subsubsection{Waveshaping Synthesis}

Digital waveshaping synthesis \citep{le_brun_digital_1979} introduces frequency components through amplitude distortion of an audio signal.
Specifically, for a continuous time signal that can be exactly expressed as a sum of stationary sinusoids, i.e. $x(t) = \sum_{k=1}^K \alpha_k \sin f_k t $, the application of a nonlinear function $\sigma$ produces a signal that can be expressed as a potentially infinite sum of sinusoids at linear combinations of the input frequencies.%

In the original formulation of waveshaping synthesis, proposed by \cite{le_brun_digital_1979}, the input signal is a single sinusoid.
The nonlinear function $\sigma$, also referred to as the shaping function, is specified as a sum of Chebyshev polynomials of the first kind $T_k$, allowing.
These functions are defined such that the $k^\text{th}$ polynomial transforms a sinusoid to its $k^\text{th}$ harmonic: $T_k(\cos\theta)=\cos k\theta$.
In this way, a shaping function can be easily designed that produces any desired combination of harmonics.
Further efficiency gains can be achieved by storing this function in a memory buffer and applying it to a sinusoidal input by interpolated fractional lookups, in a manner similar to wavetable synthesis.
Timbral variations can be produced by altering the amplitude of the incoming signal and applying a compensatory normalisation after the shaping function.

A differentiable waveshaping synthesiser was proposed by \cite{hayes_neural_2021}.
This approach replaced the Chebyshev polynomial method for shaping function design with a small multilayer perceptron $\sigma_\theta$.
To allow time-varing control over timbral content, a separate network predicted affine transform parameters $\alpha_N, \beta_N, \alpha_a \beta_a$, applied to the signal before and after the shaper network, giving the formulation:

\begin{equation}
    y[n] = \alpha_N \sigma_\theta\left(\alpha_a x[n] + \beta_a\right) + \beta_N.
\end{equation}

\noindent In practice, the network is trained with a bank of multiple such waveshapers.

\paragraph{Neural Source-Filter Models}

As noted by \cite{engel_ddsp_2020}, the neural source-filter model introduced by \cite{wang_neural_2019-2} can be considered a form of waveshaping synthesiser.
This family of models, based on the classical source-filter model, replaces the linear synthesis filter $H(z)$ with a neural network $f_\theta$, which takes a source signal $x[n]$ as input, and also accepts a conditioning signal $\mathbf{z}[n]$ which alters its response.
Due to the nonlinearity of $f_\theta$, we can interpret its behaviour through the lens of waveshaping -- that is, it is able to introduce new frequency components to the signal.%
Thus, for a purely harmonic source signal, a harmonic output will be generated, excluding the effects of aliasing.

The proposed neural filter block follows a similar architecture to a non-autoregressive WaveNet \citep{oord_wavenet_2016}, featuring dilated convolutions, gated activations, and residual connections.
Wang et al. experimented with both a mixed sinusoid-plus-noise source signal processed by a single neural filter pathway, and separate harmonic sinusoidal and noise signals processed by individual neural filter pathways.
Later extensions of the technique included a hierarchical neural source-filter model, operating at increasing resolutions, for musical timre synthesis~\cite{michelashvili_hierarchical_2020}, and the introduction of quasi-periodic cyclic noise as an excitation source, allowing for more realistic voiced-unvoiced transitions~\citep{wang_using_2020}.

\subsubsection{Frequency Modulation}\label{sec:frequency-modulation}

Frequency modulation (FM) synthesis~\citep{chowning_synthesis_1973} produces rich spectra with complex timbral evolutions from a small number of simple oscillators.
Its parameters allow continuous transformation between entirely harmonic and discordantly inharmonic spectra.
It was applied in numerous commercially successful synthesisers in the 1980s, resulting in its widespread use in popular and electronic music.

A simple stationary formulation, consisting of one \textit{carrier} oscillator $(\alpha_c, \omega_c, \phi_c)$ and one \textit{modulator} $(\alpha_m, \omega_m, \phi_m)$ is given by:

\begin{equation}
    y[n] = \alpha_c \sin \left(
        \omega_c n + \phi_c + \alpha_m \sin \left(
            \omega_m n + \phi_m
        \right)
    \right).
\end{equation}

\noindent For a sinusoidal modulator, FM synthesis is equivalent to \textit{phase modulation} up to a phase shift.
Phase modulation is often preferred in practice, as it does not require integration of the modulation signal to produce an instantaneous phase value.

More complex synthesisers can be constructed by connecting multiple modulators and carriers, forming a modulation graph.\footnote{This graph is sometimes referred to in commercial synthesisers as an ``algorithm''.}
This graph may contain cycles, corresponding to oscillator feedback, which can be implemented in discrete time with single sample delays.
\citet{caspe_ddx7_2022} published the first differentiable FM synthesiser implementation, which was able to learn to control the modulation indices of modulation graphs taken from the Yamaha DX7, an influential FM synthesiser.
To increase the flexibility of the DDX7 approach, \citet{ye_nas-fm_2023} applied a neural architecture search algorithm over the modulation graph, with the intention of allowing optimal FM synthesiser structures to be inferred from target audio.

As the gradients of an FM synthesiser's output with respect to the majority of its parameters are oscillatory, optimisation of a differentiable implementation is challenging, due to the aforementioned issues with sinusoidal parameter gradients~\citep{turian_im_2020,hayes_sinusoidal_2023}.
For this reason, the differentiable DX7 of \citet{caspe_ddx7_2022} relied on fixed oscillator tuning ratios, specified \textit{a priori}, enforcing a harmonic spectral structure.
In this sense, FM synthesis continues to be an open challenge, representing a particularly difficult manifestation of the previously discussed issues with non-convex DDSP.

\subsection{Modulation Signals}

Automatic modulation of parameters over time is a crucial component of many modern audio synthesisers, especially those used for music and sound design.
Control signals for modulation are commonly realised through \textit{envelopes}, which typically take the value of a parametric function of time in response to a trigger event, and \textit{low frequency oscillators} (LFOs), which oscillate at sub-audible ($<20$Hz) frequencies.
Estimating the parameters of envelopes and LFOs is thus valuable for sound matching tasks.
This was addressed non-differentiably by \citet{mitcheltree_modulation_2023}, who estimated LFO shapes from mel spectrograms.

In the context of modelling an analog phaser pedal, \citet{carson_differentiable_2023} produced a differentiable LFO model using the complex sinusoidal surrogate method of \citet{hayes_sinusoidal_2023} in combination with a small waveshaper neural network.
Using this technique, they were able to directly approximate the shape of the LFO acting on the all-pass filter break frequency by gradient descent. 

\citet{masuda_improving_2023} introduced a differentiable attack-decay-sustain-release (ADSR) envelope\footnote{Far from being an arbitrary choice, this is perhaps the most commonly used parametric envelope generator in synthesisers.} for use in a sound matching task.
Specifically, they defined the envelope as follows:

\begin{equation}
    u(t) =
    \left|t\cdot\frac{v_\text{max}}{t_\text{atk}}\right|_{\left[0, v_\text{max}\right]} +
    \left|\left(t-t_\text{atk}\right)\cdot
    \frac{v_\text{sus}-v_\text{max}}{t_\text{dec}}\right|_{\left[v_\text{sus}-v_\text{max},0\right]} +
    \left|-\left(t-t_\text{off}\right)\cdot
    \frac{v_\text{sus}}{t_\text{rel}}\right|_{\left[-v_\text{sus},0\right]},
\end{equation}

\noindent where $t_\text{atk}, t_\text{dec}, t_\text{rel}, t_\text{off}$ represent the attack, decay, release, and note off times, respectively; $v_\text{sus}$ is the sustain amplitude and $v_\text{max}$ is the peak amplitude; and $|x|_{[a,b]}\triangleq \min(\max(x,a),b)$.
Through sound matching experiments and gradient visualisations, they demonstrate that their model is capable of learning to predict parameters for the differentiable envelope, though note that this constraint means that subtle variations in real sounds can not be entirely captured.

\section{Loss Functions \& Training Objectives}

A central benefit of DDSP is that the training loss function can be defined directly in the audio domain, allowing the design of losses which emphasise certain desirable signal characters, for example through phase invariance or perceptually-informed weighting of frequency bands.
Nonetheless, many works we reviewed combined audio losses with other forms of training objective, including parameter regression and adversarial training.
In this section, we review the most commonly used such loss functions.

We note that deep feature loss, sometimes referred to as ``perceptual loss'' --- that is, distances between intermediate activations of pretrained neural networks --- have been explored, including the use of CREPE~\citep{kim_crepe_2018} embeddings~\citep{engel_ddsp_2020, michelashvili_hierarchical_2020}. 
Engel et al. note that this loss during unsupervised learning of pitch. 
Additionally, \citet{turian_one_2021} evaluated a number of audio  representations (DSP-based and learned representations), comparing them on distance-based ranking tasks using synthesised sounds, and found that OpenL3 \citep{cramer_look_2019} performed well.
However, \cite{masuda_improving_2023} remarked that preliminary results using OpenL3 for sound matching worked poorly. 
Since this initial work on the topic, relatively little attention has been devoted to exploring such distances for DDSP training; however, they may be a promising direction for future work.

\subsection{Audio Loss Functions}

Audio loss functions compare a predicted audio signal $\hat{y}[n]$ to a ground truth signal $y[n]$.
The simplest such loss function is thus a direct distance between audio samples in the time domain:

\begin{equation}
    \mathcal{L}_\text{wav}(\mathbf{y}, \hat{\mathbf{y}}) = \sum_n \left\lVert y[n] - \hat{y}[n] \right\rVert_p
\end{equation}

\noindent where $\left\lVert \cdot \right\rVert_p$ is the $L^p$ norm.
\citet{engel_ddsp_2020} note that this loss is typically not ideal due to the weak correspondence between individual time-domain audio samples and auditory perception.
For example, a time-domain loss penalises imperceptible shifts in oscillator phase, which may not be desirable, depending on the behaviour of a particular synthesiser and target application \citep{engel_ddsp_2020, mv_sfnet_2020, liu_neural_2020}.
\citet{wang_neural_2019} applied a phase difference loss, but noted that despite the loss values falling during training, speech quality was not improved over a randomly initialized phase spectrum.
Conversely, \citet{webber_autovocoder_2023} explicitly modelled phase in the frequency domain, finding that an $L^2$ time-domain loss helped reduce audible artifacts.

The predominant approach to formulating an audio loss for DDSP tasks, however, is based on magnitude spectrograms.
These approaches are often referred to as \textit{spectral loss}.
While numerous variations on this approach, we found three to recur commonly in the literature:

\begin{enumerate}
    \item Spectral convergence loss~\citep{arik_fast_2019}
    \begin{equation}
        \mathcal{L}_\text{sc}(\mathbf{y}, \hat{\mathbf{y}}) = 
        \frac{\left\lVert \left| STFT(\mathbf{y}) \right| - \left| STFT(\hat{\mathbf{y}}) \right| \right\rVert_{F}}
        {\left\lVert \left| STFT(\mathbf{y}) \right| \right\rVert_{F}}
    \end{equation}

    \item Log magnitude spectral loss~\citep{arik_fast_2019}
    \begin{equation}
        \mathcal{L}_\text{log}(\mathbf{y}, \hat{\mathbf{y}}) = \left\lVert \log\left| STFT(\mathbf{y}) \right| - \log\left| STFT(\hat{\mathbf{y}}) \right| \right\rVert_{1}
    \end{equation}
    
    \item Linear magnitude spectral loss
    \begin{equation}
        \mathcal{L}_\text{lin}(\mathbf{y}, \hat{\mathbf{y}}) = \left\lVert \left| STFT(\mathbf{y}) \right| - \left| STFT(\hat{\mathbf{y}}) \right| \right\rVert_{1}
    \end{equation}
\end{enumerate}

\noindent where $\left\lVert \cdot \right\rVert_F$ and $\left\lVert \cdot \right\rVert_1$ denote the Frobenius and $L_1$ norms, respectively, and $\left|STFT(\cdot)\right|$ is the magnitude spectrogram from the short-time Fourier transform.
A weighting of $\frac{1}{N}$ is sometimes applied to $L_{log}$ and $L_{lin}$, where $N$ is the number of STFT bins \citep{yamamoto_parallel_2020}.
Perceptually motivated frequency scales, like the mel scale, can also be applied to spectrograms.
\citet{arik_fast_2019} note that the spectral convergence loss ``emphasises highly on large spectral components, which helps in early phases of training,'' whereas log magnitude spectral loss tend to help fit small amplitude components, which are to be more important later in training.

\citet{wang_neural_2019} proposed computing a spectral loss using multiple STFT window sizes and hop lengths, aggregating the outputs into a single loss value.\footnote{Wang et al. used a slightly different formulation of spectral loss they called \textit{spectral amplitude distance}}
This technique has since come to be known as the \textit{multi-resolution STFT} (MRSTFT) loss \citep{yamamoto_parallel_2020} or \textit{multi-scale spectral loss} (MSS) loss \citep{engel_ddsp_2020}.
The motivation behind this formulation is to compensate for the time-frequency resolution tradeoff inherent to the STFT.

A general form for MRSTFT losses is thus given by a weighted sum of the different spectral loss formulations at different resolutions:

\begin{equation}
    \mathcal{L}_\text{MRSTFT} = \sum_{k\in{K}}{\alpha_\text{sc}\mathcal{L}_{\text{sc},k} + \alpha_\text{log}\mathcal{L}_{\text{log},k} + \alpha_\text{lin}\mathcal{L}_{\text{lin},k}}
\end{equation}

\noindent where $K$ is the set of STFT configurations, $\mathcal{L}_{\cdot,k}$ is a spectral loss computed with a particular configuration, and $\alpha$ is the weighting for a loss term.\footnote{We point readers to the \textit{auroloss} python package \citep{steinmetz_auraloss_2020} for implementations of these methods and more auditory loss functions not mentioned here. Available on GitHub \url{https://github.com/csteinmetz1/auraloss}. Accessed 25th August 2023.}
A consensus on the best spectral loss configuration has not emerged, suggesting that tuning of such losses is highly task-dependent.

Mel scaled spectral losses have also been used in a number of works~\citep{mv_sfnet_2020, choi_nansy_2022, watts_puffin_2023, kaneko_istftnet_2022, song_dspgan_2023,diaz_rigid-body_2022}, after \citet{fabbro_speech_2020} introduced the multi-resolution formulation and \citet{kong_diffwave_2020} demonstrated their suitability for using in conjunction with adversarial objectives.

A large number of different multi-resolution configurations have been explored. 
\citet{wu_ddsp-based_2022} suggested just four resolutions was sufficient for satisfactory results. 
\citet{liu_neural_2020}, on the other hand, used 12 different configurations noting that increasing this number resulted in fewer artefacts.
\citep{barahona-rios_noisebandnet_2023} showed that a very small hop-size of 8 samples, with a window length of 32 samples, allowed good transient reconstruction.
This is congruent with the findings of \citet{kumar_high-fidelity_2023}, who noted a small hop-size improved transient reconstruction in their neural audio codec.

\citet{martinez_ramirez_differentiable_2021} noted that filtering can introduce frequency-dependent delays and phase inversions, which can cause problems for auditory loss functions.
They proposed a delay invariant loss to address this, which computes an optimal delay between $y[n]$ and $\hat{y}[n]$ using a cross-correlation function, and evaluates loss functions on time-aligned waveforms. 

\citet{wang_using_2020} observed issues learning a stable pitch with the introduction of cyclic noise and proposed a masked spectral loss as a solution, which evaluates loss only in frequency bins containing harmonics of the known fundamental frequency.
This is intended to penalise only harmonic mismatch, instead of accounting for the full spectral envelope.
\citet{wu_ddsp-based_2022} trained their parameter estimation model to predict $f_0$ from a mel spectrogram and used an explicit $f_0$ regression loss where both the ground truth and target $f_0$ were extracted using the WORLD vocoder~\citep{morise_world_2016}, noting that MRSTFT alone was not sufficient to learn to reconstruct singing voices in their case while jointly learning $f_0$.

\subsection{Parameter Loss \& Self-supervision}\label{sec:parameter}

Historically, parameter loss was commonly used in sound matching tasks involving black box or non-differentiable synthesisers~\citep{yee-king_automatic_2018, barkan_inversynth_2019}.
This was, however, identified as a sub-optimal training objective \citep{esling_flow_2020}, as synthesisers are ill-conditioned with respect to their parameters --- that is, small changes in parameter space may yield large changes in the output.

However, with a differentiable synthesiser, parameter loss can be combined with auditory loss functions as a form of self-supervision, seemingly helping to avoid convergence on bad minima during training \citep{engel_self-supervised_2020,masuda_improving_2023}.
For most parameters, loss is computed directly between estimated and ground truth parameter values,
where ground truth parameters are randomly sampled to form a synthetic dataset of audio-parameter pairs.

\citet{engel_self-supervised_2020} used a parameter regression pretraining phase over a large dataset of synthetic audio signals with complete parameter annotations.
This enabled them to the fine-tune their network with an unconstrained differentiable sinusoidal model in conjunction with several other DDSP components for self-supervised pitch estimation.
They additionally introduced a \textit{sinusoidal consistency loss}, which is a permutation invariant parameter loss inspired by the two-way mismatch algorithm, to measure the error between sets of parameters for sinusoids representing partials of a target sound.

In a sound matching task with a differentiable subtractive synthesiser, \citet{masuda_synthesizer_2021} observed that training only with spectral loss was ineffective, speculating that there was not a clear relationship between the loss and subtractive synthesis parameters.
In subsequent work \citep{masuda_improving_2023}, they used a combination of parameter loss, with a synthetic dataset, and various methods for introducing \textit{out-of-domain} audio during training with a spectral loss. 
Through this procedure, they noted that certain parameters, such as the frequency of oscillators and chorus delay, were poorly optimised by a spectral loss.

Despite their tenuous perceptual correspondence, an advantage of parameter losses is their relative efficiency, particularly when parameters are predicted globally, or below audio sample rate.
\citet{han_perceptual-neural-physical_2023} proposed a method for reweighting the contributions of individual parameters to provide the best quadratic approximation of a given ``perceptual'' loss --- i.e. a differentiable audio loss with desirable perceptual qualities, such as MRSTFT or joint time-frequency scattering~\citep{muradeli_differentiable_2022}.
The proposed technique requires that loss gradients with respect to ground truth to parameters be evaluated once before training, limiting the technique to synthetic datasets, but the advantage is that online backpropogation through the differentiable synthesiser can be effectively avoided.

\subsection{Adversarial training}

Generative adversarial networks~\citep{goodfellow_generative_2014} consist of two components: a generator, which produces synthetic examples, and a discriminator, which attempts to classify generated examples from real ones.
These components are trained to optimise a \textit{minimax} game.
In later work, this adversarial training formulation was combined with a reconstruction loss \citep{isola_image--image_2017} for image generation, a technique which has since become popular in audio generation \citep{kong_hifi-gan_2020}.

From the perspective of reconstruction, the main motivation for adversarial training is that it tends to improve the naturalness and perceived quality of results \citep{michelashvili_hierarchical_2020, choi_nansy_2022} and enables learning fine temporal structures \citep{liu_neural_2020}, particularly when a multi-resolution discriminator is used~\citep{you_gan_2021}.
Further, despite using a phase invariant reconstruction loss, both \citet{liu_neural_2020} and \citet{watts_puffin_2023} observed that adversarial training improved phase reconstruction and reduce phase-related audio artefacts.
However, these benefits come at the expense of a more complex training setup.

Several variations on adversarial training for DDSP synthesis have been explored.
The HiFi-GAN methodology \citep{kong_hifi-gan_2020} has been particularly popular \citep{choi_nansy_2022, watts_puffin_2023, webber_autovocoder_2023, kaneko_istftnet_2022}. 
This involves multiple discriminators operating at different periods and scales in a least-squares GAN setup, and includes a \textit{feature matching loss} \citet{kumar_melgan_2019}, which involves using distances between discriminator activations as an auxiliary loss.
Others have used a hinge loss \citep{liu_neural_2020, caillon_rave_2021, nercessian_differentiable_2023} and a Wasserstein GAN \citep{juvela_gelp_2019}.
\citet{caillon_rave_2021} and \citet{watts_puffin_2023} both train in two stages, first optimising for reconstruction, then introducing adversarial training to fine-tune the model.

\section{Discussion}\label{sec:discussion}

In this section we discuss both the strengths and limitations of DDSP for audio synthesis.
For those new to the topic, we hope to assist in evaluating the suitability of DDSP for applications and research directions of interest.
With more experienced practitioners in mind, we seek to highlight promising directions for future work and present open research questions that we argue hinder wider applicability and adoption of DDSP for audio synthesis.

\subsection{DDSP \& Inductive Bias}

Underpinning DDSP, and the related field of differentiable rendering~\cite{kato_differentiable_2020}, is the notion of incorporating a domain-appropriate \textit{inductive bias}.
Imposing strong assumptions on the form a model's outputs should take --- for example, producing a signal using only an additive synthesiser --- limits model flexibility, but in return can improve the appropriateness of outputs.
When such a bias is appropriate to the task and data, this can be highly beneficial.
However, this bias also limits the broader applicability of the model, and may cause issues with generalisation.

In reviewing the literature, we found that authors most commonly referred to the following strengths of DDSP methods:

\begin{enumerate}
    \item \textbf{Audio quality}: differentiable oscillators have helped reduce artefacts caused by phase discontinuities~\citep{engel_ddsp_2020} and pitch errors \citep{nercessian_differentiable_2023}, and have enabled SOTA results when incorporated into hybrid models \citep{choi_nansy_2022, song_dspgan_2023};
    \item \textbf{Data efficiency}: an appropriately specified differentiable signal processor seems to reduce the data burden, with good results achievable using only minutes of audio \citep{engel_ddsp_2020,michelashvili_hierarchical_2020};
    \item \textbf{Computational efficiency}: offloading signal generation to efficient synthesis algorithms carries the further benefit of faster inference \citep{shan_differentiable_2022, hayes_neural_2021, carney_tone_2021}.
    \item \textbf{Interpretability}: framing model outputs in terms of signal processor parameters allows for post hoc interpretation. Differentiable articulatory models can provide insights into vocal production \citep{sudholt_vocal_2023}, while common audio synthesiser designs can be used to enable interpretable decomposition of target sounds \citep{caspe_ddx7_2022, masuda_improving_2023};
    \item \textbf{Control/creative affordances}: explicit controls based on perceptual attributes such as pitch and loudness have enabled creative applications such as real-time timbre transfer \citep{carney_tone_2021}, expressive musical performance rendering \citep{wu_midi-ddsp_2022}, and voice designing \citep{choi_nansy_2022}.
\end{enumerate}

Furthermore, differentiable audio synthesis has enabled new techniques in tasks beyond the realm of speech and music synthesis.
For example, \citet{schulze-forster_unsupervised_2023} applied differentiable source-filter models to perform unsupervised source separation, while \citet{engel_self-supervised_2020} used an analysis-by-synthesis framework to predict pitch without explicit supervision from ground truth labels.
Further, DDSP-based synthesisers and audio effects have been used for tasks such as data amplification~\citep{wu_generating_2022} and data augmentation~\citep{guo_dent-ddsp_2022}.

However, the majority of these benefits have been realised within the context of synthesising monophonic audio with a predominantly harmonic spectral structure, where it is possible to explicitly provide a fundamental frequency annotation.
This includes solo monophonic instruments, singing, and speech.
This limitation is necessitated by the choices of synthesis model that have made up the majority of DDSP synthesis research --- typically, these encode a strong bias towards the generation of harmonic signals, while the reliance on accurate $f_0$ estimates renders polyphony significantly more challenging.
In this sense, the trade-off is a lack of straightforward generalisation to other classes of sound --- producing drum sounds with a differentiable harmonic-plus-noise synthesiser conditioned on loudness and $f_0$ is unlikely to produce usable results, for example.
This is not an inherently negative characteristic of the methodology ---
excluding possibilities from the solution space (e.g., non-harmonic sounds or phase discontinuities) has been instrumental in realising the above benefits such as data-efficient training and improved audio quality.

The application of DDSP to a broader range of sounds has consequently been slow.
Many synthesis techniques exist which are capable of generating polyphonic, inharmonic, or transient-dense audio, but to date there has been limited exploration of these in the literature. 
This may, in part, be due to the difficulty inherent in optimising their parameters by gradient descent, as noted by \citet{turian_im_2020} and \citet{hayes_sinusoidal_2023}.
Work on differentiable FM synthesis~\citep{caspe_ddx7_2022,ye_nas-fm_2023}, for example, may eventually lead to the modelling of more varied sound sources due to its ability to produce complex inharmonic spectra, but as noted by~\citet{caspe_ddx7_2022}, optimisation of carrier and modulator frequencies is currently not possible due to loss surface ripple. 

In summary, the inductive bias inherent to DDSP represents a trade-off.
It has enabled in certain tasks, through constraint of solution spaces, improved audio quality, data efficiency, computational efficiency, interpretability, and control affordances.
In exchange, these strong assumptions narrowly constrain these models to their respective domains of application, and limit their generalisation to many real world scenarios where perfect harmonicity or isolation of monophic sources can't be guaranteed.
Hence, we argue that a deeper understanding of the trade-offs induced by specific differentiable synthesisers would be a valuable future research direction for the audio synthesis community.

\subsection{Applying DDSP}

The process of taking a DSP operation and making it differentiable is, at a high level, relatively accessible due to the robust automatic differentiation available in major deep learning libraries.\footnote{PyTorch \url{https://pytorch.org/} and TensorFlow \url{https://www.tensorflow.org/} are two examples that are well supported and that have been used extensively in previous DDSP work. URLs accessed August 27th 2023.} 
Such libraries expose APIs containing a variety of mathematical and numerical functions, with the corresponding CUDA kernels and backward pass implementations.
This allows DSP algorithms to be expressed directly using these libraries and gradients to be calculated directly.
Additionally, with the growing interest in audio machine learning research and DDSP, several specialized libraries have emerged.\footnote{These include the original DDSP library \url{https://github.com/magenta/ddsp} introduced by \citep{engel_ddsp_2020}, a PyTorch port of the DDSP library \url{https://github.com/acids-ircam/ddsp_pytorch}, TorchAudio \url{https://pytorch.org/audio/stable/index.html}, and torchsynth \url{https://github.com/torchsynth/torchsynth} introduced by ~\citep{turian_one_2021}, accessed August 27th 2023.}

\subsubsection{Differentiable Implementations}
Implementing DSP algorithms using these libraries is generally straightforward using the existing API.
However, as we have seen, certain techniques may be non-trivial to implement in a way that allows straightforward optimisation without workarounds or specialised algorithms. 
These include ADSR envelopes \citep{masuda_improving_2023}, quantisation \citep{subramani_end--end_2022}, IIR filters~\citep{kuznetsov_differentiable_2020,nercessian_lightweight_2021}, sinusoidal models~\citep{hayes_sinusoidal_2023}, and more.

Rewriting DSP algorithms in an automatic differentiation library can, itself, be time consuming and introduces an additional burden to the research pipeline.
Furthermore, the programming language of a particular library -- most commonly Python -- may not be the same language used in an end application.
However, recent efforts have sought to support the translation of DSP code into differentiable implementations\footnote{\citep{braun_dawdreamer_2021} recently introduced the ability to transpile DSP code written in Faust\url{https://faust.grame.fr/} to Jax code \citep{bradbury_jax_2018} into the DawDreamer library. Released on GitHub, v0.6.14 \url{https://github.com/DBraun/DawDreamer/releases/tag/v0.6.14}. URLs accessed 27th August 2023.} and support the deployment of audio code written in machine learning libraries into audio plugins\footnote{\url{https://neutone.space/} accessed August 27th 2023}.
Future efforts could explore how fast inference libraries focused on real-time audio applications could be integrated with DDSP audio synthesis~\citep{chowdhury_rtneural_2021}.

\subsubsection{Alternatives to Automatic Differentiation}
In certain situations, manual implementation of DSP algorithms in automatic differentiation software is not possible (e.g., when using a black-box like a VST audio plugin or physical hardware audio processor) or is undesirable given the additional challenge and engineering overhead. Three alternative methods to manually implementing DSP operations differentiably have been explored in previous audio research, although have not been extensively applied to synthesis at the time of writing: (i) neural proxies \citep{steinmetz_automatic_2021}; (ii) hybrid neural proxies \citep{steinmetz_style_2022}; and (iii), numerical gradient approximation methods \citep{martinez_ramirez_differentiable_2021}.

Neural proxies seek to train a deep learning network to mimic the behaviour of an audio processor, including the effect of parameter changes.
Hybrid neural proxies use a neural proxy only during training, replacing the audio processing component of the proxy with the original DSP during inference.
\citet{steinmetz_style_2022} introduced the concept of hybrid neural proxies for audio research and further distinguished between half hybrid approaches which use a neural proxy for both forward and backward optimization passes and full hybrid approaches that only use a proxy for the backward pass.
Numerical gradient approximation methods, on the other hand, don't require any component of the audio processing chain to be differentiable and instead use a numerial method, such as simultaneous permutation stochastic approximation (SPSA) \citep{spall_overview_1998}, to estimate gradients during the backward pass.
\citep{martinez_ramirez_differentiable_2021} used SPSA to estimate gradients for audio plugin-based audio effects.

\subsection{Looking Ahead}

We wish, finally, to discuss the opportunities, risks, and open challenges in future research into DDSP for audio synthesis.

\subsubsection{Hybrid Methods}

In this review, we undertook a cross-domain survey of DDSP-based audio synthesis encompassing both speech and music.
In doing so, we note that the progression of DDSP methods in speech synthesis commenced with mixed approaches, integrating DSP components to neural audio synthesisers and benefiting from the strengths of both.
Early examples included the incorporation of LPC synthesis filters~\citep{juvela_gelp_2019,valin_lpcnet_2019}, effectively ``offloading'' part of the synthesis task.
Conversely, the dominant applications of DDSP to music tend to offload \textit{all} signal generation to differentiable DSP components.
We note that there may be opportunities for both fields to benefit from one another's findings here, exploring further intermediate \textit{hybrid} approaches.

Hybrid methods, in general, combine the aforementioned strengths of DDSP with more general deep learning models.
This is visible in the use of pre- and post-nets in speech and singing synthesis, for example~\citep{nercessian_end--end_2021,choi_nansy_2022,nercessian_differentiable_2023}; in the integration of a differentiable filtered noise synthesiser to RAVE~\citep{caillon_rave_2021}; or in the results of the recent SVC challenge, in which the top two results used a pre-trained DDSP synthesiser to condition a GAN \citep{huang_singing_2023}.
We thus expect hybrid methods to be a fruitful future research direction, where DSP domain knowledge can help guide and constrain more general neural audio synthesisers, and neural networks can help generalise narrow DDSP solutions.  

\subsubsection{Implementations, Efficiency, \& Stability}

A major challenge in DDSP research is ensuring that it is computationally and numerically feasible to even perform optimisation. 
While many DSP operations are deeply connected to neural network components, the transition into gradient descent over DDSP models is not inherently straightforward.
IIR filters operating on audio, for instance, are likely to be applied over many more time steps than a conventional RNN, leading to specific challenges such as the memory cost of unrolled operations, and filter stability. 
As a result, efficient training algorithms have received considerable attention.
Further, recent work by \citet{yu_singing_2023} presented an efficient GPU implementation of all-pole IIR filters, evaluated exactly and recursively, with efficient backpropogation based on a simplified algorithm for evaluating the gradient.
Similarly, accompanying their original contribution, \citet{engel_ddsp_2020} included an ``angular cumsum'' CUDA kernel, enabling phase accumulation without numerical issues.
The development of open-source and efficient implementations of differentiable signal processing operations or their constituent parts is of clear benefit to future work, and thus we argue that this is a valuable area in which to direct future work.

\subsubsection{Interdisciplinary Research and Participatory Design}
At a higher level, consideration of the intended real-world application is also crucial, and can provide guidance to future experimental enquiry.  
Interdisciplinary research, and projects that involve end-users through practices such as participatory design could be of particular value to the field.
In the case of music, for example, \citep{carney_tone_2021} conducted usability studies with musicians, informing the development of an in-browser DDSP timbre transfer application.
Future work could seek to collaborate not only with end users, but also digital instrument makers.
Developing pipelines for integration with and deployment to hardware platforms such as Bela~\citep{pelinski_pipeline_2023} could enable much more straightforward involvement of such key stakeholder groups, while also paving the way to more immediately tangible research outcomes.

We note, in the research into DDSP for speech synthesis we surveyed, that the use of subjective evaluations such as the mean opinion score (MOS) is popular, but found limited examples of stakeholder or end user involvement in the reserach process.
Nonetheless, we do note that such research does tend to have a strong end-user focus.
Some, in particular, is particularly targeted --- \citep{watts_puffin_2023} consider the specific low-resource computation constraints of Augmentative and alternative communication (AAC) \citep{murray_augmentative_2009}, which focuses on supporting or replacing speech for individuals whose needs are not met by natural speech.

Given that the use of DDSP is so frequently motivated in the literature by considerations of efficiency, control, quality, etc., whose definitions arguably vary with application area and end user needs, we suggest that the involvement of end users through qualitative research methods would be a valuable practice in prioritising future technical reserach. 

\subsubsection{Open Questions}

Through our review we noted a small number of recurring themes relating to specific challenges in working with DDSP, often mentioned as a corollary to the main results, or to motivate an apparent workaround.
We also observed that certain methods have received attention in one application domain but not yet been applied to another, or have simply received only a small amount of attention.
In this section, we briefly compile these observations in the hopes that they might help direct future research.
These open questions include:
\begin{enumerate}
    \item The difficulty estimating oscillator frequencies by gradient descent, discussed in detail by~\citet{turian_im_2020} and \citet{hayes_sinusoidal_2023} (see section \ref{sec:unconstrained-additive})
    \begin{itemize}
        \item A related issue is the tuning of modulator frequencies in FM synthesis~\citep{caspe_ddx7_2022} (see section \ref{sec:frequency-modulation})
    \end{itemize}
    \item The invariance of some signal processors under permutations of their parameters appears to be relevant to optimisation, as noted by~\citet{nercessian_neural_2020}, \citet{engel_self-supervised_2020}, and \citet{masuda_improving_2023}, but there has been no specific investigation as to how this impacts training.
    \item Estimating global parameters like ADSR segment times appears to be challenging, especially when these lead to complex interactions~\citep{masuda_improving_2023}. This is important for modelling commercial synthesisers differentiably.
    \item Estimating delay parameters and compensating for delays poses specific challenges, which will most likely require specialised loss functions.~\citep{masuda_improving_2023, martinez_ramirez_differentiable_2021}
    \item Neural proxy~\citep{steinmetz_style_2022} and gradient approximation~\citep{martinez_ramirez_differentiable_2021} techniques have been applied in automatic mixing and intelligent audio production, but not yet explored for audio synthesis. Could these allow black-box software instruments to be used pseudo-differentiably? 
    \item Directed acyclic audio signal processing graphs have been estimated blindly \citep{lee_blind_2023}, while FM routing has been optimised through neural architecture search \citep{ye_nas-fm_2023}. Can these methods be generalised to allow estimation of arbitrary synthesiser topologies? 
\end{enumerate}

\section{Conclusion}

In this article, we have surveyed the literature on differentiable digital signal processing (DDSP) across the domains of speech, music, and singing voice synthesis.
We provided a detailed overview of the major tasks and application areas where DDSP has been used, identifying recurring motivations for its adoption.
Additionally, we summarised major technical contributions and discussed various approaches to optimising models that incorporate DDSP components.
In our discussion, we noted that the purported advantages of DDSP techniques are typically gained at the expense of broad applicability and generalisation.
Finally, we identified promising avenues for future research, such as the development of hybrid models and the exploration of interdisciplinary research opportunities, and concluded our review by highlighting several knowledge gaps that warrant attention in future work.

\section*{Conflict of Interest Statement}

The authors declare that the research was conducted in the absence of any commercial or financial relationships that could be construed as a potential conflict of interest.

\section*{Funding}
This work was supported by UK Research and Innovation [grant number EP/S022694/1].

\bibliographystyle{Frontiers-Harvard} %
\bibliography{diffref}

\begin{thebibliography}{183}
\providecommand{\natexlab}[1]{#1}
\expandafter\ifx\csname urlstyle\endcsname\relax
  \providecommand{\doi}[1]{doi:\discretionary{}{}{}#1}\else
  \providecommand{\doi}{doi:\discretionary{}{}{}\begingroup
  \urlstyle{rm}\Url}\fi
\providecommand{\selectlanguage}[1]{\relax}
\providecommand{\bibAnnoteFile}[1]{%
  \IfFileExists{#1}{\begin{quotation}\noindent\textsc{Key:} #1\\
  \textsc{Annotation:}\ \input{#1}\end{quotation}}{}}
\providecommand{\bibAnnote}[2]{%
  \begin{quotation}\noindent\textsc{Key:} #1\\
  \textsc{Annotation:}\ #2\end{quotation}}

\bibitem[{Alonso and Erkut(2021)}]{alonso_latent_2021}
Alonso, J. and Erkut, C. (2021).
\newblock Latent {{Space Explorations}} of {{Singing Voice Synthesis}} using
  {{DDSP}}
\bibAnnoteFile{alonso_latent_2021}

\bibitem[{ANSI(1960)}]{ansi_psychoacoustic_1960}
ANSI (1960).
\newblock Psychoacoustic {{Terminology}}: {{Timbre}}
\bibAnnoteFile{ansi_psychoacoustic_1960}

\bibitem[{Ar{\i}k et~al.(2019)Ar{\i}k, Jun, and Diamos}]{arik_fast_2019}
Ar{\i}k, S.~{\"O}., Jun, H., and Diamos, G. (2019).
\newblock Fast {{Spectrogram Inversion Using Multi-Head Convolutional Neural
  Networks}}.
\newblock \emph{IEEE Signal Processing Letters} 26, 94--98.
\newblock \doi{10.1109/LSP.2018.2880284}
\bibAnnoteFile{arik_fast_2019}

\bibitem[{Atal and Hanauer(1971)}]{atal_speech_1971}
Atal, B.~S. and Hanauer, S.~L. (1971).
\newblock Speech {{Analysis}} and {{Synthesis}} by {{Linear Prediction}} of the
  {{Speech Wave}}.
\newblock \emph{The Journal of the Acoustical Society of America} 50, 637--655.
\newblock \doi{10.1121/1.1912679}
\bibAnnoteFile{atal_speech_1971}

\bibitem[{Back and Tsoi(1991)}]{back_fir_1991}
Back, A.~D. and Tsoi, A.~C. (1991).
\newblock {{FIR}} and {{IIR Synapses}}, a {{New Neural Network Architecture}}
  for {{Time Series Modeling}}.
\newblock \emph{Neural Computation} 3, 375--385.
\newblock \doi{10.1162/neco.1991.3.3.375}
\bibAnnoteFile{back_fir_1991}

\bibitem[{Bakhturina et~al.(2021)Bakhturina, Lavrukhin, Ginsburg, and
  Zhang}]{bakhturina_hi-fi_2021}
Bakhturina, E., Lavrukhin, V., Ginsburg, B., and Zhang, Y. (2021).
\newblock Hi-{{Fi Multi-Speaker English TTS Dataset}}
\bibAnnoteFile{bakhturina_hi-fi_2021}

\bibitem[{{Barahona-R{\'i}os} and
  Collins(2023)}]{barahona-rios_noisebandnet_2023}
{Barahona-R{\'i}os}, A. and Collins, T. (2023).
\newblock {{NoiseBandNet}}: {{Controllable Time-Varying Neural Synthesis}} of
  {{Sound Effects Using Filterbanks}}.
\newblock \doi{10.48550/arXiv.2307.08007}
\bibAnnoteFile{barahona-rios_noisebandnet_2023}

\bibitem[{Barkan et~al.(2019)Barkan, Tsiris, Katz, and
  Koenigstein}]{barkan_inversynth_2019}
Barkan, O., Tsiris, D., Katz, O., and Koenigstein, N. (2019).
\newblock {{InverSynth}}: {{Deep Estimation}} of {{Synthesizer Parameter
  Configurations}} from {{Audio Signals}}.
\newblock \emph{IEEE/ACM Transactions on Audio, Speech, and Language
  Processing} 27, 2385--2396.
\newblock \doi{10.1109/TASLP.2019.2944568}
\bibAnnoteFile{barkan_inversynth_2019}

\bibitem[{Bhattacharya et~al.(2020)Bhattacharya, Nowak, and
  Z{\"o}lzer}]{bhattacharya_optimization_2020}
Bhattacharya, P., Nowak, P., and Z{\"o}lzer, U. (2020).
\newblock Optimization of cascaded parametric peak and shelving filters with
  backpropagation algorithm.
\newblock In \emph{Proc. {{Digit}}. {{Audio}} Effects}. 101--108
\bibAnnoteFile{bhattacharya_optimization_2020}

\bibitem[{Bilbao(2009)}]{bilbao_numerical_2009}
Bilbao, S. (2009).
\newblock \emph{Numerical {{Sound Synthesis}}: {{Finite Difference Schemes}}
  and {{Simulation}} in {{Musical Acoustics}}} ({John Wiley \& Sons})
\bibAnnoteFile{bilbao_numerical_2009}

\bibitem[{Birkholz(2013)}]{birkholz_modeling_2013}
Birkholz, P. (2013).
\newblock Modeling {{Consonant-Vowel Coarticulation}} for {{Articulatory Speech
  Synthesis}}.
\newblock \emph{PLOS ONE} 8, e60603.
\newblock \doi{10.1371/journal.pone.0060603}
\bibAnnoteFile{birkholz_modeling_2013}

\bibitem[{Bitton et~al.(2018)Bitton, Esling, and
  {Chemla-Romeu-Santos}}]{bitton_modulated_2018}
Bitton, A., Esling, P., and {Chemla-Romeu-Santos}, A. (2018).
\newblock Modulated {{Variational}} auto-{{Encoders}} for many-to-many musical
  timbre transfer
\bibAnnoteFile{bitton_modulated_2018}

\bibitem[{Blaauw and Bonada(2017)}]{blaauw_neural_2017-1}
Blaauw, M. and Bonada, J. (2017).
\newblock A {{Neural Parametric Singing Synthesizer}}
\bibAnnoteFile{blaauw_neural_2017-1}

\bibitem[{Bradbury et~al.(2018)Bradbury, Frostig, Hawkins, Johnson, Leary,
  Maclaurin et~al.}]{bradbury_jax_2018}
Bradbury, J., Frostig, R., Hawkins, P., Johnson, M.~J., Leary, C., Maclaurin,
  D., et~al. (2018).
\newblock {{JAX}}: Composable transformations of {{Python}}+{{NumPy}} programs
\bibAnnoteFile{bradbury_jax_2018}

\bibitem[{Braun(2021)}]{braun_dawdreamer_2021}
Braun, D. (2021).
\newblock {{DawDreamer}}: {{Bridging}} the {{Gap Between Digital Audio
  Workstations}} and {{Python Interfaces}}.
\newblock \doi{10.48550/arXiv.2111.09931}
\bibAnnoteFile{braun_dawdreamer_2021}

\bibitem[{Cahill(1897)}]{cahill_art_1897}
Cahill, T. (1897).
\newblock Art of and apparatus for generating and distributing music
  electrically
\bibAnnoteFile{cahill_art_1897}

\bibitem[{Caillon and Esling(2021)}]{caillon_rave_2021}
Caillon, A. and Esling, P. (2021).
\newblock {{RAVE}}: {{A}} variational autoencoder for fast and high-quality
  neural audio synthesis.
\newblock \emph{arXiv:2111.05011 [cs, eess]}
\bibAnnoteFile{caillon_rave_2021}

\bibitem[{{\c C}ak{\i}r and Virtanen(2018)}]{cakir_musical_2018}
{\c C}ak{\i}r, E. and Virtanen, T. (2018).
\newblock Musical {{Instrument Synthesis}} and {{Morphing}} in
  {{Multidimensional Latent Space Using Variational}}, {{Convolutional
  Recurrent Autoencoders}}.
\newblock In \emph{Proceedings of the {{Audio Engineerings Society}} 145th
  {{Convention}}}
\bibAnnoteFile{cakir_musical_2018}

\bibitem[{Campolucci et~al.(1995)Campolucci, Piazza, and
  Uncini}]{campolucci_-line_1995}
Campolucci, P., Piazza, F., and Uncini, A. (1995).
\newblock On-line learning algorithms for neural networks with {{IIR}}
  synapses.
\newblock In \emph{Proceedings of {{ICNN}}'95 - {{International Conference}} on
  {{Neural Networks}}} ({Perth, WA, Australia}: {IEEE}), vol.~2, 865--870.
\newblock \doi{10.1109/ICNN.1995.487532}
\bibAnnoteFile{campolucci_-line_1995}

\bibitem[{Carney et~al.(2021)Carney, Li, Toh, Yu, and Engel}]{carney_tone_2021}
Carney, M., Li, C., Toh, E., Yu, P., and Engel, J. (2021).
\newblock Tone {{Transfer}}: {{In-Browser Interactive Neural Audio Synthesis}}.
\newblock In \emph{{{IUI}}}
\bibAnnoteFile{carney_tone_2021}

\bibitem[{Carson et~al.(2023)Carson, {Valentini-Botinhao}, King, and
  Bilbao}]{carson_differentiable_2023}
Carson, A., {Valentini-Botinhao}, C., King, S., and Bilbao, S. (2023).
\newblock Differentiable {{Grey-box Modelling}} of {{Phaser Effects}} using
  {{Frame-based Spectral Processing}}.
\newblock \doi{10.48550/arXiv.2306.01332}
\bibAnnoteFile{carson_differentiable_2023}

\bibitem[{Caspe et~al.(2022)Caspe, McPherson, and Sandler}]{caspe_ddx7_2022}
Caspe, F., McPherson, A., and Sandler, M. (2022).
\newblock {{DDX7}}: {{Differentiable FM Synthesis}} of {{Musical Instrument
  Sounds}}.
\newblock \doi{10.48550/arXiv.2208.06169}
\bibAnnoteFile{caspe_ddx7_2022}

\bibitem[{Castellon et~al.(2020)Castellon, Donahue, and
  Liang}]{castellon_towards_2020}
Castellon, R., Donahue, C., and Liang, P. (2020).
\newblock Towards realistic {{MIDI}} instrument synthesizers.
\newblock In \emph{4th {{Workshop}} on {{Machine Learning}} for {{Creativity}}
  and {{Design}}} ({Vancouver, Canada})
\bibAnnoteFile{castellon_towards_2020}

\bibitem[{Chen et~al.(2020{\natexlab{a}})Chen, Tan, Luan, Qin, and
  Liu}]{chen_hifisinger_2020}
Chen, J., Tan, X., Luan, J., Qin, T., and Liu, T.-Y. (2020{\natexlab{a}}).
\newblock {{HiFiSinger}}: {{Towards High-Fidelity Neural Singing Voice
  Synthesis}}
\bibAnnoteFile{chen_hifisinger_2020}

\bibitem[{Chen et~al.(2020{\natexlab{b}})Chen, Zhang, Zen, Weiss, Norouzi, and
  Chan}]{chen_wavegrad_2020}
Chen, N., Zhang, Y., Zen, H., Weiss, R.~J., Norouzi, M., and Chan, W.
  (2020{\natexlab{b}}).
\newblock {WaveGrad}: {Estimating} {Gradients} for {Waveform} {Generation}.
\newblock \emph{arXiv:2009.00713 [cs, eess, stat]} ArXiv: 2009.00713
\bibAnnoteFile{chen_wavegrad_2020}

\bibitem[{Childers et~al.(1985)Childers, Yegnanarayana, and
  Wu}]{childers_voice_1985}
Childers, D., Yegnanarayana, B., and Wu, K. (1985).
\newblock Voice conversion: {{Factors}} responsible for quality.
\newblock In \emph{{{ICASSP}} '85. {{IEEE International Conference}} on
  {{Acoustics}}, {{Speech}}, and {{Signal Processing}}}. vol.~10, 748--751.
\newblock \doi{10.1109/ICASSP.1985.1168479}
\bibAnnoteFile{childers_voice_1985}

\bibitem[{Cho et~al.(2021)Cho, Yang, Chang, Cheng, Wang, and
  Liu}]{cho_survey_2021}
Cho, Y.-P., Yang, F.-R., Chang, Y.-C., Cheng, C.-T., Wang, X.-H., and Liu,
  Y.-W. (2021).
\newblock A {{Survey}} on {{Recent Deep Learning-driven Singing Voice Synthesis
  Systems}}.
\newblock In \emph{2021 {{IEEE International Conference}} on {{Artificial
  Intelligence}} and {{Virtual Reality}} ({{AIVR}})}. 319--323.
\newblock \doi{10.1109/AIVR52153.2021.00067}
\bibAnnoteFile{cho_survey_2021}

\bibitem[{Choi et~al.(2022)Choi, Yang, Lee, and Kim}]{choi_nansy_2022}
Choi, H.-S., Yang, J., Lee, J., and Kim, H. (2022).
\newblock {{NANSY}}++: {{Unified Voice Synthesis}} with {{Neural Analysis}} and
  {{Synthesis}}
\bibAnnoteFile{choi_nansy_2022}

\bibitem[{Chowdhury(2021)}]{chowdhury_rtneural_2021}
Chowdhury, J. (2021).
\newblock {{RTNeural}}: {{Fast}} neural inferencing for real-time systems.
\newblock \emph{arXiv preprint arXiv:2106.03037}
\bibAnnoteFile{chowdhury_rtneural_2021}

\bibitem[{Chowning(1973)}]{chowning_synthesis_1973}
Chowning, J.~M. (1973).
\newblock The synthesis of complex audio spectra by means of frequency
  modulation.
\newblock \emph{J. Audio Eng. Soc} 21, 526--534
\bibAnnoteFile{chowning_synthesis_1973}

\bibitem[{Colonel et~al.(2022)Colonel, Steinmetz, Michelen, and
  Reiss}]{colonel_direct_2022-1}
Colonel, J.~T., Steinmetz, C.~J., Michelen, M., and Reiss, J.~D. (2022).
\newblock Direct design of biquad filter cascades with deep learning by
  sampling random polynomials.
\newblock \doi{10.48550/arXiv.2110.03691}.
\newblock ArXiv:2110.03691 [cs, eess]
\bibAnnoteFile{colonel_direct_2022-1}

\bibitem[{Cook(1996)}]{cook_singing_1996}
Cook, P.~R. (1996).
\newblock Singing {{Voice Synthesis}}: {{History}}, {{Current Work}}, and
  {{Future Directions}}.
\newblock \emph{Computer Music Journal} 20, 38--46.
\newblock \doi{10.2307/3680822}
\bibAnnoteFile{cook_singing_1996}

\bibitem[{Cooley and Tukey(1965)}]{cooley_algorithm_1965}
Cooley, J.~W. and Tukey, J.~W. (1965).
\newblock An algorithm for the machine calculation of complex {{Fourier}}
  series.
\newblock \emph{Mathematics of computation} 19, 297--301.
\newblock \doi{10.1090/S0025-5718-1965-0178586-1}
\bibAnnoteFile{cooley_algorithm_1965}

\bibitem[{Cramer et~al.(2019)Cramer, Wu, Salamon, and Bello}]{cramer_look_2019}
Cramer, A.~L., Wu, H.-H., Salamon, J., and Bello, J.~P. (2019).
\newblock Look, {{Listen}}, and {{Learn More}}: {{Design Choices}} for {{Deep
  Audio Embeddings}}.
\newblock In \emph{{{ICASSP}} 2019 - 2019 {{IEEE International Conference}} on
  {{Acoustics}}, {{Speech}} and {{Signal Processing}} ({{ICASSP}})}.
  3852--3856.
\newblock \doi{10.1109/ICASSP.2019.8682475}
\bibAnnoteFile{cramer_look_2019}

\bibitem[{Dai et~al.(2018)Dai, Zhang, and Xia}]{dai_music_2018-1}
Dai, S., Zhang, Z., and Xia, G.~G. (2018).
\newblock Music {{Style Transfer}}: {{A Position Paper}}
\bibAnnoteFile{dai_music_2018-1}

\bibitem[{De~Man et~al.(2019)De~Man, Stables, and
  Reiss}]{de_man_intelligent_2019}
De~Man, B., Stables, R., and Reiss, J.~D. (2019).
\newblock \emph{Intelligent Music Production} ({Routledge})
\bibAnnoteFile{de_man_intelligent_2019}

\bibitem[{D{\'e}fossez et~al.(2018)D{\'e}fossez, Zeghidour, Usunier, Bottou,
  and Bach}]{defossez_sing_2018}
D{\'e}fossez, A., Zeghidour, N., Usunier, N., Bottou, L., and Bach, F. (2018).
\newblock {{SING}}: {{Symbol-to-Instrument Neural Generator}}.
\newblock \emph{arXiv:1810.09785 [cs, eess, stat]}
\bibAnnoteFile{defossez_sing_2018}

\bibitem[{Devis et~al.(2023)Devis, Demerlé, Nabi, Genova, and
  Esling}]{devis_continuous_2023}
Devis, N., Demerlé, N., Nabi, S., Genova, D., and Esling, P. (2023).
\newblock Continuous {Descriptor}-{Based} {Control} for {Deep} {Audio}
  {Synthesis}.
\newblock In \emph{{ICASSP} 2023 - 2023 {IEEE} {International} {Conference} on
  {Acoustics}, {Speech} and {Signal} {Processing} ({ICASSP})} (Rhodes Island,
  Greece: IEEE), 1--5.
\newblock \doi{10.1109/ICASSP49357.2023.10096670}
\bibAnnoteFile{devis_continuous_2023}

\bibitem[{Diaz et~al.(2022)Diaz, Hayes, Saitis, Fazekas, and
  Sandler}]{diaz_rigid-body_2022}
Diaz, R., Hayes, B., Saitis, C., Fazekas, G., and Sandler, M. (2022).
\newblock Rigid-{{Body Sound Synthesis}} with {{Differentiable Modal
  Resonators}}.
\newblock \doi{10.48550/arXiv.2210.15306}
\bibAnnoteFile{diaz_rigid-body_2022}

\bibitem[{Donahue et~al.(2019)Donahue, McAuley, and
  Puckette}]{donahue_adversarial_2019}
Donahue, C., McAuley, J., and Puckette, M. (2019).
\newblock {{ADVERSARIAL AUDIO SYNTHESIS}}
\bibAnnoteFile{donahue_adversarial_2019}

\bibitem[{Dudley and Tarnoczy(1950)}]{dudley_speaking_1950}
Dudley, H. and Tarnoczy, T.~H. (1950).
\newblock The {{Speaking Machine}} of {{Wolfgang}} von {{Kempelen}}.
\newblock \emph{The Journal of the Acoustical Society of America} 22, 151--166.
\newblock \doi{10.1121/1.1906583}
\bibAnnoteFile{dudley_speaking_1950}

\bibitem[{Dudley(1939)}]{dudley_vocoder_1939}
Dudley, {\relax WH}. (1939).
\newblock The vocoder 18, 122
\bibAnnoteFile{dudley_vocoder_1939}

\bibitem[{Dupre et~al.(2021)Dupre, Denjean, Aramaki, and
  {Kronland-Martinet}}]{dupre_spatial_2021}
Dupre, T., Denjean, S., Aramaki, M., and {Kronland-Martinet}, R. (2021).
\newblock Spatial {{Sound Design}} in a {{Car Cockpit}}: {{Challenges}} and
  {{Perspectives}}.
\newblock In \emph{2021 {{Immersive}} and {{3D Audio}}: From {{Architecture}}
  to {{Automotive}} ({{I3DA}})} ({Bologna, Italy}: {IEEE}), 1--5.
\newblock \doi{10.1109/I3DA48870.2021.9610910}
\bibAnnoteFile{dupre_spatial_2021}

\bibitem[{Elman(1990)}]{elman_finding_1990}
Elman, J.~L. (1990).
\newblock Finding {{Structure}} in {{Time}}.
\newblock \emph{Cognitive Science} 14, 179--211.
\newblock \doi{10.1207/s15516709cog1402_1}
\bibAnnoteFile{elman_finding_1990}

\bibitem[{Engel et~al.(2018)Engel, Agrawal, Chen, Gulrajani, Donahue, and
  Roberts}]{engel_gansynth_2018}
Engel, J., Agrawal, K.~K., Chen, S., Gulrajani, I., Donahue, C., and Roberts,
  A. (2018).
\newblock {{GANSynth}}: {{Adversarial Neural Audio Synthesis}}.
\newblock In \emph{International {{Conference}} on {{Learning
  Representations}}}
\bibAnnoteFile{engel_gansynth_2018}

\bibitem[{Engel et~al.(2019)Engel, Agrawal, Chen, Gulrajani, Donahue, and
  Roberts}]{engel_gansynth_2019}
Engel, J., Agrawal, K.~K., Chen, S., Gulrajani, I., Donahue, C., and Roberts,
  A. (2019).
\newblock {GANSynth}: {Adversarial} {Neural} {Audio} {Synthesis}.
\newblock In \emph{7th {International} {Conference} on {Learning}
  {Representations}} (New Orleans, LA, USA), 17
\bibAnnoteFile{engel_gansynth_2019}

\bibitem[{Engel et~al.(2020{\natexlab{a}})Engel, Hantrakul, Gu, and
  Roberts}]{engel_ddsp_2020}
Engel, J., Hantrakul, L.~H., Gu, C., and Roberts, A. (2020{\natexlab{a}}).
\newblock {{DDSP}}: {{Differentiable Digital Signal Processing}}.
\newblock In \emph{8th {{International Conference}} on {{Learning
  Representations}}} ({Addis Ababa, Ethiopia})
\bibAnnoteFile{engel_ddsp_2020}

\bibitem[{Engel et~al.(2017)Engel, Resnick, Roberts, Dieleman, Norouzi, Eck
  et~al.}]{engel_neural_2017}
Engel, J., Resnick, C., Roberts, A., Dieleman, S., Norouzi, M., Eck, D., et~al.
  (2017).
\newblock Neural audio synthesis of musical notes with {{WaveNet}}
  autoencoders.
\newblock In \emph{Proceedings of the 34th {{International Conference}} on
  {{Machine Learning}} - {{Volume}} 70} ({Sydney, Australia}), 1068--1077
\bibAnnoteFile{engel_neural_2017}

\bibitem[{Engel et~al.(2020{\natexlab{b}})Engel, Swavely, and
  Roberts}]{engel_self-supervised_2020}
Engel, J., Swavely, R., and Roberts, A. (2020{\natexlab{b}}).
\newblock Self-supervised {{Pitch Detection}} by {{Inverse Audio Synthesis}}.
\newblock In \emph{Proceedings of the {{International Conference}} on {{Machine
  Learning}}}. 9
\bibAnnoteFile{engel_self-supervised_2020}

\bibitem[{Esling et~al.(2018)Esling, Chemla, and Bitton}]{esling_bridging_2018}
Esling, P., Chemla, A., and Bitton, A. (2018).
\newblock Bridging audio analysis, perception and synthesis with
  perceptually-regularized variational timbre spaces.
\newblock In \emph{Proceedings of the 19th {{International Society}} for
  {{Music Information Retrieval Conference}}} ({Paris, France}), 175--181
\bibAnnoteFile{esling_bridging_2018}

\bibitem[{Esling et~al.(2020)Esling, Masuda, Bardet, Despres, and
  {Chemla-Romeu-Santos}}]{esling_flow_2020}
Esling, P., Masuda, N., Bardet, A., Despres, R., and {Chemla-Romeu-Santos}, A.
  (2020).
\newblock Flow {{Synthesizer}}: {{Universal Audio Synthesizer Control}} with
  {{Normalizing Flows}}.
\newblock \emph{Applied Sciences} 10, 302.
\newblock \doi{10.3390/app10010302}
\bibAnnoteFile{esling_flow_2020}

\bibitem[{Fabbro et~al.(2020)Fabbro, Golkov, Kemp, and
  Cremers}]{fabbro_speech_2020}
Fabbro, G., Golkov, V., Kemp, T., and Cremers, D. (2020).
\newblock Speech {{Synthesis}} and {{Control Using Differentiable DSP}}
\bibAnnoteFile{fabbro_speech_2020}

\bibitem[{Gatys et~al.(2015)Gatys, Ecker, and Bethge}]{gatys_neural_2015}
Gatys, L.~A., Ecker, A.~S., and Bethge, M. (2015).
\newblock A {{Neural Algorithm}} of {{Artistic Style}}
\bibAnnoteFile{gatys_neural_2015}

\bibitem[{G{\'o}mez et~al.(2018)G{\'o}mez, Blaauw, Bonada, Chandna, and
  Cuesta}]{gomez_deep_2018}
G{\'o}mez, E., Blaauw, M., Bonada, J., Chandna, P., and Cuesta, H. (2018).
\newblock Deep {{Learning}} for {{Singing Processing}}: {{Achievements}},
  {{Challenges}} and {{Impact}} on {{Singers}} and {{Listeners}}
\bibAnnoteFile{gomez_deep_2018}

\bibitem[{Goodfellow et~al.(2014)Goodfellow, Pouget-Abadie, Mirza, Xu,
  Warde-Farley, Ozair et~al.}]{goodfellow_generative_2014}
Goodfellow, I.~J., Pouget-Abadie, J., Mirza, M., Xu, B., Warde-Farley, D.,
  Ozair, S., et~al. (2014).
\newblock Generative {Adversarial} {Nets}.
\newblock In \emph{Advances in {Neural} {Information} {Processing} {Systems}}
  (Montreal, Canada: Curran Associates, Inc.), vol.~27 of \emph{{NIPS}'14},
  2672--2680.
\newblock Number of pages: 9
\bibAnnoteFile{goodfellow_generative_2014}

\bibitem[{Guo et~al.(2022{\natexlab{a}})Guo, Zhou, Meng, and
  Liu}]{guo_improving_2022}
Guo, H., Zhou, Z., Meng, F., and Liu, K. (2022{\natexlab{a}}).
\newblock Improving {{Adversarial Waveform Generation Based Singing Voice
  Conversion}} with {{Harmonic Signals}}.
\newblock In \emph{{{ICASSP}} 2022 - 2022 {{IEEE International Conference}} on
  {{Acoustics}}, {{Speech}} and {{Signal Processing}} ({{ICASSP}})}.
  6657--6661.
\newblock \doi{10.1109/ICASSP43922.2022.9746709}
\bibAnnoteFile{guo_improving_2022}

\bibitem[{Guo et~al.(2022{\natexlab{b}})Guo, Chen, and
  Chng}]{guo_dent-ddsp_2022}
Guo, Z., Chen, C., and Chng, E.~S. (2022{\natexlab{b}}).
\newblock {{DENT-DDSP}}: {{Data-efficient}} noisy speech generator using
  differentiable digital signal processors for explicit distortion modelling
  and noise-robust speech recognition
\bibAnnoteFile{guo_dent-ddsp_2022}

\bibitem[{Ha et~al.(2016)Ha, Dai, and Le}]{ha_hypernetworks_2016}
Ha, D., Dai, A., and Le, Q.~V. (2016).
\newblock {HyperNetworks}.
\newblock In \emph{International {Conference} on {Learning} {Representations}}.
\newblock ArXiv: 1609.09106
\bibAnnoteFile{ha_hypernetworks_2016}

\bibitem[{Hagiwara et~al.(2022)Hagiwara, Cusimano, and
  Liu}]{hagiwara_modeling_2022}
Hagiwara, M., Cusimano, M., and Liu, J.-Y. (2022).
\newblock Modeling {{Animal Vocalizations}} through {{Synthesizers}}
\bibAnnoteFile{hagiwara_modeling_2022}

\bibitem[{Han et~al.(2023)Han, Lostanlen, and
  Lagrange}]{han_perceptual-neural-physical_2023}
Han, H., Lostanlen, V., and Lagrange, M. (2023).
\newblock Perceptual-{{Neural-Physical Sound Matching}}
\bibAnnoteFile{han_perceptual-neural-physical_2023}

\bibitem[{Hantrakul et~al.(2019)Hantrakul, Engel, Roberts, and
  Gu}]{hantrakul_fast_2019}
Hantrakul, L., Engel, J., Roberts, A., and Gu, C. (2019).
\newblock Fast and {{Flexible Neural Audio Synthesis}}.
\newblock In \emph{Proceedings of the 20th {{International Society}} for
  {{Music Information Retrieval Conference}}} ({Delft, The Netherlands}),
  524--530
\bibAnnoteFile{hantrakul_fast_2019}

\bibitem[{Hawthorne et~al.(2022)Hawthorne, Simon, Roberts, Zeghidour, Gardner,
  Manilow et~al.}]{hawthorne_multi-instrument_2022}
Hawthorne, C., Simon, I., Roberts, A., Zeghidour, N., Gardner, J., Manilow, E.,
  et~al. (2022).
\newblock Multi-instrument {{Music Synthesis}} with {{Spectrogram Diffusion}}
\bibAnnoteFile{hawthorne_multi-instrument_2022}

\bibitem[{Hayes et~al.(2021)Hayes, Saitis, and Fazekas}]{hayes_neural_2021}
Hayes, B., Saitis, C., and Fazekas, G. (2021).
\newblock Neural {{Waveshaping Synthesis}}.
\newblock In \emph{Proceedings of the 22nd {{International Society}} for
  {{Music Information Retrieval Conference}}} ({Online})
\bibAnnoteFile{hayes_neural_2021}

\bibitem[{Hayes et~al.(2023)Hayes, Saitis, and Fazekas}]{hayes_sinusoidal_2023}
Hayes, B., Saitis, C., and Fazekas, G. (2023).
\newblock Sinusoidal {Frequency} {Estimation} by {Gradient} {Descent}.
\newblock In \emph{{ICASSP} 2023 - 2023 {IEEE} {International} {Conference} on
  {Acoustics}, {Speech} and {Signal} {Processing} ({ICASSP})}. 1--5.
\newblock \doi{10.1109/ICASSP49357.2023.10095188}
\bibAnnoteFile{hayes_sinusoidal_2023}

\bibitem[{Holmes(2008)}]{holmes_electronic_2008}
Holmes, T. (2008).
\newblock \emph{Electronic and experimental music: technology, music, and
  culture} (New York: Routledge), 3rd ed edn.
\bibAnnoteFile{holmes_electronic_2008}

\bibitem[{Hono et~al.(2021)Hono, Takaki, Hashimoto, Oura, Nankaku, and
  Tokuda}]{hono_periodnet_2021}
Hono, Y., Takaki, S., Hashimoto, K., Oura, K., Nankaku, Y., and Tokuda, K.
  (2021).
\newblock Periodnet: {{A Non-Autoregressive Waveform Generation Model}} with a
  {{Structure Separating Periodic}} and {{Aperiodic Components}}.
\newblock In \emph{{{ICASSP}} 2021 - 2021 {{IEEE International Conference}} on
  {{Acoustics}}, {{Speech}} and {{Signal Processing}} ({{ICASSP}})} ({Toronto,
  ON, Canada}: {IEEE}), 6049--6053.
\newblock \doi{10.1109/ICASSP39728.2021.9414401}
\bibAnnoteFile{hono_periodnet_2021}

\bibitem[{Horner et~al.(1993)Horner, Beauchamp, and
  Haken}]{horner_machine_1993}
Horner, A., Beauchamp, J., and Haken, L. (1993).
\newblock Machine tongues {{XVI}}. {{Genetic}} algorithms and their application
  to {{FM}} matching synthesis.
\newblock \emph{Computer Music Journal} 17, 17--29.
\newblock \doi{10.2307/3680541}
\bibAnnoteFile{horner_machine_1993}

\bibitem[{Huang et~al.(2019)Huang, Li, Anil, Bao, Oore, and
  Grosse}]{huang_timbretron_2019}
Huang, S., Li, Q., Anil, C., Bao, X., Oore, S., and Grosse, R.~B. (2019).
\newblock {{TimbreTron}}: {{A WaveNet}}({{CycleGAN}}({{CQT}}({{Audio}})))
  {{Pipeline}} for {{Musical Timbre Transfer}}
\bibAnnoteFile{huang_timbretron_2019}

\bibitem[{Huang et~al.(2023)Huang, Violeta, Liu, Shi, and
  Toda}]{huang_singing_2023}
Huang, W.-C., Violeta, L.~P., Liu, S., Shi, J., and Toda, T. (2023).
\newblock The {{Singing Voice Conversion Challenge}} 2023
\bibAnnoteFile{huang_singing_2023}

\bibitem[{Hunt and Black(1996)}]{hunt_unit_1996}
Hunt, A. and Black, A. (1996).
\newblock Unit selection in a concatenative speech synthesis system using a
  large speech database.
\newblock In \emph{1996 {{IEEE International Conference}} on {{Acoustics}},
  {{Speech}}, and {{Signal Processing Conference Proceedings}}}. vol.~1,
  373--376 vol. 1.
\newblock \doi{10.1109/ICASSP.1996.541110}
\bibAnnoteFile{hunt_unit_1996}

\bibitem[{Huzaifah and Wyse(2020)}]{huzaifah_deep_2020}
Huzaifah, M. and Wyse, L. (2020).
\newblock Deep generative models for musical audio synthesis.
\newblock \emph{arXiv:2006.06426 [cs, eess, stat]} ArXiv: 2006.06426
\bibAnnoteFile{huzaifah_deep_2020}

\bibitem[{Isola et~al.(2017)Isola, Zhu, Zhou, and
  Efros}]{isola_image--image_2017}
Isola, P., Zhu, J.-Y., Zhou, T., and Efros, A.~A. (2017).
\newblock Image-to-{{Image Translation}} with {{Conditional Adversarial
  Networks}}.
\newblock In \emph{2017 {{IEEE Conference}} on {{Computer Vision}} and
  {{Pattern Recognition}} ({{CVPR}})} ({Honolulu, HI}: {IEEE}), 5967--5976.
\newblock \doi{10.1109/CVPR.2017.632}
\bibAnnoteFile{isola_image--image_2017}

\bibitem[{Ito and Johnson(2017)}]{ito_lj_2017}
Ito, K. and Johnson, L. (2017).
\newblock The {{LJ}} speech dataset
\bibAnnoteFile{ito_lj_2017}

\bibitem[{Jack et~al.(2018)Jack, Mehrabi, Stockman, and
  McPherson}]{jack_action-sound_2018}
Jack, R.~H., Mehrabi, A., Stockman, T., and McPherson, A. (2018).
\newblock Action-sound {Latency} and the {Perceived} {Quality} of {Digital}
  {Musical} {Instruments}.
\newblock \emph{Music Perception} 36, 109--128.
\newblock \doi{10.1525/mp.2018.36.1.109}
\bibAnnoteFile{jack_action-sound_2018}

\bibitem[{Jain et~al.(2020)Jain, Kumar, Cai, Singhal, and
  Kumar}]{jain_att_2020}
Jain, D.~K., Kumar, A., Cai, L., Singhal, S., and Kumar, V. (2020).
\newblock {ATT}: {Attention}-based {Timbre} {Transfer}.
\newblock In \emph{2020 {International} {Joint} {Conference} on {Neural}
  {Networks} ({IJCNN})} (Glasgow, United Kingdom: IEEE), 1--6.
\newblock \doi{10.1109/IJCNN48605.2020.9207146}
\bibAnnoteFile{jain_att_2020}

\bibitem[{Jin et~al.(2018)Jin, Finkelstein, Mysore, and Lu}]{jin_fftnet_2018}
Jin, Z., Finkelstein, A., Mysore, G.~J., and Lu, J. (2018).
\newblock Fftnet: {{A Real-Time Speaker-Dependent Neural Vocoder}}.
\newblock In \emph{2018 {{IEEE International Conference}} on {{Acoustics}},
  {{Speech}} and {{Signal Processing}} ({{ICASSP}})}. 2251--2255.
\newblock \doi{10.1109/ICASSP.2018.8462431}
\bibAnnoteFile{jin_fftnet_2018}

\bibitem[{Jonason et~al.(2020)Jonason, Sturm, and
  Thome}]{jonason_control-synthesis_2020}
Jonason, N., Sturm, B. L.~T., and Thome, C. (2020).
\newblock The control-synthesis approach for making expressive and controllable
  neural music synthesizers.
\newblock In \emph{Proceedings of the 2020 {{AI Music Creativity Conference}}}.
  9
\bibAnnoteFile{jonason_control-synthesis_2020}

\bibitem[{Juvela et~al.(2019)Juvela, Bollepalli, Yamagishi, and
  Alku}]{juvela_gelp_2019}
Juvela, L., Bollepalli, B., Yamagishi, J., and Alku, P. (2019).
\newblock {{GELP}}: {{GAN-Excited Linear Prediction}} for {{Speech Synthesis}}
  from {{Mel-Spectrogram}}.
\newblock In \emph{Interspeech 2019} ({ISCA}), 694--698.
\newblock \doi{10.21437/Interspeech.2019-2008}
\bibAnnoteFile{juvela_gelp_2019}

\bibitem[{Kalchbrenner et~al.(2018)Kalchbrenner, Elsen, Simonyan, Noury,
  Casagrande, Lockhart et~al.}]{kalchbrenner_efficient_2018}
Kalchbrenner, N., Elsen, E., Simonyan, K., Noury, S., Casagrande, N., Lockhart,
  E., et~al. (2018).
\newblock Efficient {{Neural Audio Synthesis}}.
\newblock In \emph{Proceedings of the 35th {{International Conference}} on
  {{Machine Learning}}} ({PMLR}), 2410--2419
\bibAnnoteFile{kalchbrenner_efficient_2018}

\bibitem[{Kaneko et~al.(2022)Kaneko, Tanaka, Kameoka, and
  Seki}]{kaneko_istftnet_2022}
Kaneko, T., Tanaka, K., Kameoka, H., and Seki, S. (2022).
\newblock {{ISTFTNET}}: {{Fast}} and {{Lightweight Mel-Spectrogram Vocoder
  Incorporating Inverse Short-Time Fourier Transform}}.
\newblock In \emph{{{ICASSP}} 2022 - 2022 {{IEEE International Conference}} on
  {{Acoustics}}, {{Speech}} and {{Signal Processing}} ({{ICASSP}})}.
  6207--6211.
\newblock \doi{10.1109/ICASSP43922.2022.9746713}
\bibAnnoteFile{kaneko_istftnet_2022}

\bibitem[{Kato et~al.(2020)Kato, Beker, Morariu, Ando, Matsuoka, Kehl
  et~al.}]{kato_differentiable_2020}
Kato, H., Beker, D., Morariu, M., Ando, T., Matsuoka, T., Kehl, W., et~al.
  (2020).
\newblock Differentiable {Rendering}: {A} {Survey}.
\newblock \doi{10.48550/arXiv.2006.12057}.
\newblock ArXiv:2006.12057 [cs]
\bibAnnoteFile{kato_differentiable_2020}

\bibitem[{Kawamura et~al.(2022)Kawamura, Nakamura, Kitamura, Saruwatari,
  Takahashi, and Kondo}]{kawamura_differentiable_2022}
Kawamura, M., Nakamura, T., Kitamura, D., Saruwatari, H., Takahashi, Y., and
  Kondo, K. (2022).
\newblock Differentiable {{Digital Signal Processing Mixture Model}} for
  {{Synthesis Parameter Extraction}} from {{Mixture}} of {{Harmonic Sounds}}.
\newblock In \emph{{{ICASSP}} 2022 - 2022 {{IEEE International Conference}} on
  {{Acoustics}}, {{Speech}} and {{Signal Processing}} ({{ICASSP}})}
  ({Singapore, Singapore}: {IEEE}), 941--945.
\newblock \doi{10.1109/ICASSP43922.2022.9746399}
\bibAnnoteFile{kawamura_differentiable_2022}

\bibitem[{Keller(1994)}]{keller_fundamentals_1994}
Keller, E. (ed.) (1994).
\newblock \emph{Fundamentals of Speech Synthesis and Speech Recognition: Basic
  Concepts, State of the Art, and Future Challenges} ({Chichester [England] ;
  New York}: {Wiley})
\bibAnnoteFile{keller_fundamentals_1994}

\bibitem[{Khan and Chitode(2016)}]{khan_concatenative_2016}
Khan, R.~A. and Chitode, J.~S. (2016).
\newblock Concatenative speech synthesis: {{A}} review.
\newblock \emph{International Journal of Computer Applications} 136, 1--6.
\newblock \doi{10.5120/ijca2016907992}
\bibAnnoteFile{khan_concatenative_2016}

\bibitem[{Kim et~al.(2018)Kim, Salamon, Li, and Bello}]{kim_crepe_2018}
Kim, J.~W., Salamon, J., Li, P., and Bello, J.~P. (2018).
\newblock Crepe: {A} {Convolutional} {Representation} for {Pitch} {Estimation}.
\newblock In \emph{2018 {IEEE} {International} {Conference} on {Acoustics},
  {Speech}, and {Signal} {Processing}, {ICASSP} 2018 - {Proceedings}}
  (Institute of Electrical and Electronics Engineers Inc.), 161--165.
\newblock \doi{10.1109/ICASSP.2018.8461329}
\bibAnnoteFile{kim_crepe_2018}

\bibitem[{Kobayashi et~al.(2014)Kobayashi, Toda, Neubig, Sakti, and
  Nakamura}]{kobayashi_statistical_2014}
Kobayashi, K., Toda, T., Neubig, G., Sakti, S., and Nakamura, S. (2014).
\newblock Statistical singing voice conversion with direct waveform
  modification based on the spectrum differential.
\newblock In \emph{Interspeech 2014} ({ISCA}), 2514--2518.
\newblock \doi{10.21437/Interspeech.2014-539}
\bibAnnoteFile{kobayashi_statistical_2014}

\bibitem[{Kong et~al.(2020{\natexlab{a}})Kong, Kim, and
  Bae}]{kong_hifi-gan_2020}
Kong, J., Kim, J., and Bae, J. (2020{\natexlab{a}}).
\newblock {{HiFi-GAN}}: {{Generative Adversarial Networks}} for {{Efficient}}
  and {{High Fidelity Speech Synthesis}}.
\newblock In \emph{Advances in {{Neural Information Processing Systems}}}
  ({Curran Associates, Inc.}), vol.~33, 17022--17033
\bibAnnoteFile{kong_hifi-gan_2020}

\bibitem[{Kong et~al.(2020{\natexlab{b}})Kong, Ping, Huang, Zhao, and
  Catanzaro}]{kong_diffwave_2020}
Kong, Z., Ping, W., Huang, J., Zhao, K., and Catanzaro, B.
  (2020{\natexlab{b}}).
\newblock {DiffWave}: {A} {Versatile} {Diffusion} {Model} for {Audio}
  {Synthesis}.
\newblock \emph{arXiv:2009.09761 [cs, eess, stat]} ArXiv: 2009.09761
\bibAnnoteFile{kong_diffwave_2020}

\bibitem[{Kong et~al.(2021)Kong, Ping, Huang, Zhao, and
  Catanzaro}]{kong_diffwave_2021}
Kong, Z., Ping, W., Huang, J., Zhao, K., and Catanzaro, B. (2021).
\newblock {{DiffWave}}: {{A Versatile Diffusion Model}} for {{Audio Synthesis}}
\bibAnnoteFile{kong_diffwave_2021}

\bibitem[{Krumhansl(1989)}]{krumhansl_why_1989}
Krumhansl, C.~L. (1989).
\newblock Why is musical timbre so hard to understand.
\newblock \emph{Structure and perception of electroacoustic sound and music} 9,
  43--55
\bibAnnoteFile{krumhansl_why_1989}

\bibitem[{Kumar et~al.(2019)Kumar, Kumar, {de Boissiere}, Gestin, Teoh, Sotelo
  et~al.}]{kumar_melgan_2019}
Kumar, K., Kumar, R., {de Boissiere}, T., Gestin, L., Teoh, W.~Z., Sotelo, J.,
  et~al. (2019).
\newblock {{MelGAN}}: {{Generative Adversarial Networks}} for {{Conditional
  Waveform Synthesis}}.
\newblock In \emph{Advances in {{Neural Information Processing Systems}}}
  ({Curran Associates, Inc.}), vol.~32
\bibAnnoteFile{kumar_melgan_2019}

\bibitem[{Kumar et~al.(2023)Kumar, Seetharaman, Luebs, Kumar, and
  Kumar}]{kumar_high-fidelity_2023}
Kumar, R., Seetharaman, P., Luebs, A., Kumar, I., and Kumar, K. (2023).
\newblock High-{{Fidelity Audio Compression}} with {{Improved RVQGAN}}
\bibAnnoteFile{kumar_high-fidelity_2023}

\bibitem[{Kuznetsov et~al.(2020)Kuznetsov, Parker, and
  Esqueda}]{kuznetsov_differentiable_2020}
Kuznetsov, B., Parker, J.~D., and Esqueda, F. (2020).
\newblock Differentiable {{IIR}} filters for machine learning applications.
\newblock In \emph{{{DaFX}}}
\bibAnnoteFile{kuznetsov_differentiable_2020}

\bibitem[{Le~Brun(1979)}]{le_brun_digital_1979}
Le~Brun, M. (1979).
\newblock Digital {Waveshaping} {Synthesis}.
\newblock \emph{Journal of the Audio Engineering Society} 27, 250--266
\bibAnnoteFile{le_brun_digital_1979}

\bibitem[{Lee et~al.(2022)Lee, Choi, and Lee}]{lee_differentiable_2022}
Lee, S., Choi, H.-S., and Lee, K. (2022).
\newblock Differentiable {{Artificial Reverberation}}.
\newblock \emph{IEEE/ACM Transactions on Audio, Speech, and Language
  Processing} 30, 2541--2556.
\newblock \doi{10.1109/TASLP.2022.3193298}
\bibAnnoteFile{lee_differentiable_2022}

\bibitem[{Lee et~al.(2023)Lee, Park, Paik, and Lee}]{lee_blind_2023}
Lee, S., Park, J., Paik, S., and Lee, K. (2023).
\newblock Blind {Estimation} of {Audio} {Processing} {Graph}.
\newblock In \emph{{ICASSP} 2023 - 2023 {IEEE} {International} {Conference} on
  {Acoustics}, {Speech} and {Signal} {Processing} ({ICASSP})} (Rhodes Island,
  Greece: IEEE), 1--5.
\newblock \doi{10.1109/ICASSP49357.2023.10096581}
\bibAnnoteFile{lee_blind_2023}

\bibitem[{Liu et~al.(2022)Liu, Li, Ren, Chen, and Zhao}]{liu_diffsinger_2022}
Liu, J., Li, C., Ren, Y., Chen, F., and Zhao, Z. (2022).
\newblock {{DiffSinger}}: {{Singing Voice Synthesis}} via {{Shallow Diffusion
  Mechanism}}
\bibAnnoteFile{liu_diffsinger_2022}

\bibitem[{Liu et~al.(2020)Liu, Chen, and Yu}]{liu_neural_2020}
Liu, Z., Chen, K., and Yu, K. (2020).
\newblock Neural {{Homomorphic Vocoder}}.
\newblock In \emph{Interspeech 2020} ({ISCA}), 240--244.
\newblock \doi{10.21437/Interspeech.2020-3188}
\bibAnnoteFile{liu_neural_2020}

\bibitem[{Martinez~Ramirez et~al.(2021)Martinez~Ramirez, Wang, Smaragdis, and
  Bryan}]{martinez_ramirez_differentiable_2021}
Martinez~Ramirez, M.~A., Wang, O., Smaragdis, P., and Bryan, N.~J. (2021).
\newblock Differentiable {{Signal Processing With Black-Box Audio Effects}}.
\newblock In \emph{{{ICASSP}} 2021 - 2021 {{IEEE International Conference}} on
  {{Acoustics}}, {{Speech}} and {{Signal Processing}} ({{ICASSP}})} ({Toronto,
  ON, Canada}: {IEEE}), 66--70.
\newblock \doi{10.1109/ICASSP39728.2021.9415103}
\bibAnnoteFile{martinez_ramirez_differentiable_2021}

\bibitem[{Masuda and Saito(2021)}]{masuda_synthesizer_2021}
Masuda, N. and Saito, D. (2021).
\newblock Synthesizer {{Sound Matching}} with {{Differentiable DSP}}.
\newblock \doi{10.5281/zenodo.5624609}
\bibAnnoteFile{masuda_synthesizer_2021}

\bibitem[{Masuda and Saito(2023)}]{masuda_improving_2023}
Masuda, N. and Saito, D. (2023).
\newblock Improving {{Semi-Supervised Differentiable Synthesizer Sound
  Matching}} for {{Practical Applications}}.
\newblock \emph{IEEE/ACM Transactions on Audio, Speech, and Language
  Processing} 31, 863--875.
\newblock \doi{10.1109/TASLP.2023.3237161}
\bibAnnoteFile{masuda_improving_2023}

\bibitem[{Matsubara et~al.(2022)Matsubara, Okamoto, Takashima, Takiguchi, Toda,
  and Kawai}]{matsubara_comparison_2022}
Matsubara, K., Okamoto, T., Takashima, R., Takiguchi, T., Toda, T., and Kawai,
  H. (2022).
\newblock Comparison of real-time multi-speaker neural vocoders on {{CPUs}}.
\newblock \emph{Acoustical Science and Technology} 43, 121--124.
\newblock \doi{10.1250/ast.43.121}
\bibAnnoteFile{matsubara_comparison_2022}

\bibitem[{McAdams(2019)}]{siedenburg_perceptual_2019}
McAdams, S. (2019).
\newblock The {{Perceptual Representation}} of {{Timbre}}.
\newblock In \emph{Timbre: {{Acoustics}}, {{Perception}}, and {{Cognition}}},
  eds. K.~Siedenburg, C.~Saitis, S.~McAdams, A.~N. Popper, and R.~R. Fay
  ({Cham}: {Springer International Publishing}), vol.~69. 23--57.
\newblock \doi{10.1007/978-3-030-14832-4_2}
\bibAnnoteFile{siedenburg_perceptual_2019}

\bibitem[{Mehri et~al.(2017)Mehri, Kumar, Gulrajani, Kumar, Jain, Sotelo
  et~al.}]{mehri_samplernn_2017}
Mehri, S., Kumar, K., Gulrajani, I., Kumar, R., Jain, S., Sotelo, J., et~al.
  (2017).
\newblock {{SampleRNN}}: {{An Unconditional End-to-End Neural Audio Generation
  Model}}
\bibAnnoteFile{mehri_samplernn_2017}

\bibitem[{Michelashvili and Wolf(2020)}]{michelashvili_hierarchical_2020}
Michelashvili, M.~M. and Wolf, L. (2020).
\newblock Hierarchical {{Timbre-painting}} and {{Articulation Generation}}.
\newblock In \emph{Proceedings of the 21th {{International Society}} for
  {{Music Information Retrieval Conference}}}
\bibAnnoteFile{michelashvili_hierarchical_2020}

\bibitem[{Mitcheltree et~al.(2023)Mitcheltree, Steinmetz, Comunit{\`a}, and
  Reiss}]{mitcheltree_modulation_2023}
Mitcheltree, C., Steinmetz, C.~J., Comunit{\`a}, M., and Reiss, J.~D. (2023).
\newblock Modulation {{Extraction}} for {{LFO-driven Audio Effects}}
\bibAnnoteFile{mitcheltree_modulation_2023}

\bibitem[{Moffat and Sandler(2019)}]{moffat_approaches_2019}
Moffat, D. and Sandler, M.~B. (2019).
\newblock Approaches in {{Intelligent Music Production}}.
\newblock \emph{Arts} , 1--13\doi{10.3390/arts8040125}
\bibAnnoteFile{moffat_approaches_2019}

\bibitem[{Mohammadi and Kain(2017)}]{mohammadi_overview_2017}
Mohammadi, S.~H. and Kain, A. (2017).
\newblock An overview of voice conversion systems.
\newblock \emph{Speech Communication} 88, 65--82.
\newblock \doi{10.1016/j.specom.2017.01.008}
\bibAnnoteFile{mohammadi_overview_2017}

\bibitem[{Morise et~al.(2016)Morise, Yokomori, and Ozawa}]{morise_world_2016}
Morise, M., Yokomori, F., and Ozawa, K. (2016).
\newblock {{WORLD}}: {{A Vocoder-Based High-Quality Speech Synthesis System}}
  for {{Real-Time Applications}}.
\newblock \emph{IEICE Transactions on Information and Systems} E99.D,
  1877--1884.
\newblock \doi{10.1587/transinf.2015EDP7457}
\bibAnnoteFile{morise_world_2016}

\bibitem[{Muradeli et~al.(2022)Muradeli, Vahidi, Wang, Han, Lostanlen, Lagrange
  et~al.}]{muradeli_differentiable_2022}
Muradeli, J., Vahidi, C., Wang, C., Han, H., Lostanlen, V., Lagrange, M.,
  et~al. (2022).
\newblock Differentiable {Time}–frequency {Scattering} on {GPU}.
\newblock ISSN: 2413-6689
\bibAnnoteFile{muradeli_differentiable_2022}

\bibitem[{Murray and Goldbart(2009)}]{murray_augmentative_2009}
Murray, J. and Goldbart, J. (2009).
\newblock Augmentative and alternative communication: A review of current
  issues.
\newblock \emph{Paediatrics and Child Health} 19, 464--468.
\newblock \doi{10.1016/j.paed.2009.05.003}
\bibAnnoteFile{murray_augmentative_2009}

\bibitem[{Mv and Ghosh(2020)}]{mv_sfnet_2020}
Mv, A.~R. and Ghosh, P.~K. (2020).
\newblock {{SFNet}}: {{A Computationally Efficient Source Filter Model Based
  Neural Speech Synthesis}}.
\newblock \emph{IEEE Signal Processing Letters} 27, 1170--1174.
\newblock \doi{10.1109/LSP.2020.3005031}
\bibAnnoteFile{mv_sfnet_2020}

\bibitem[{Nachmani and Wolf(2019)}]{nachmani_unsupervised_2019}
Nachmani, E. and Wolf, L. (2019).
\newblock Unsupervised {{Singing Voice Conversion}}
\bibAnnoteFile{nachmani_unsupervised_2019}

\bibitem[{Nercessian(2020)}]{nercessian_neural_2020}
Nercessian, S. (2020).
\newblock Neural {{Parametric Equalizer Matching Using Differentiable
  Biquads}}.
\newblock In \emph{Proceedings of the 23rd {{International Conference}} on
  {{Digital Audio Effects}}} ({Vienna, Austria}), 8
\bibAnnoteFile{nercessian_neural_2020}

\bibitem[{Nercessian(2021)}]{nercessian_end--end_2021}
Nercessian, S. (2021).
\newblock End-to-{{End Zero-Shot Voice Conversion Using}} a {{DDSP Vocoder}}.
\newblock In \emph{2021 {{IEEE Workshop}} on {{Applications}} of {{Signal
  Processing}} to {{Audio}} and {{Acoustics}} ({{WASPAA}})}. 1--5.
\newblock \doi{10.1109/WASPAA52581.2021.9632754}
\bibAnnoteFile{nercessian_end--end_2021}

\bibitem[{Nercessian(2023)}]{nercessian_differentiable_2023}
Nercessian, S. (2023).
\newblock Differentiable {{WORLD Synthesizer-Based Neural Vocoder With
  Application To End-To-End Audio Style Transfer}}.
\newblock In \emph{Audio {{Engineering Society Convention}} 154} ({Audio
  Engineering Society})
\bibAnnoteFile{nercessian_differentiable_2023}

\bibitem[{Nercessian et~al.(2021)Nercessian, Sarroff, and
  Werner}]{nercessian_lightweight_2021}
Nercessian, S., Sarroff, A., and Werner, K.~J. (2021).
\newblock Lightweight and {{Interpretable Neural Modeling}} of an {{Audio
  Distortion Effect Using Hyperconditioned Differentiable Biquads}}.
\newblock In \emph{{{ICASSP}} 2021 - 2021 {{IEEE International Conference}} on
  {{Acoustics}}, {{Speech}} and {{Signal Processing}} ({{ICASSP}})} ({Toronto,
  ON, Canada}: {IEEE}), 890--894.
\newblock \doi{10.1109/ICASSP39728.2021.9413996}
\bibAnnoteFile{nercessian_lightweight_2021}

\bibitem[{Nishimura et~al.(2016)Nishimura, Hashimoto, Oura, Nankaku, and
  Tokuda}]{nishimura_singing_2016}
Nishimura, M., Hashimoto, K., Oura, K., Nankaku, Y., and Tokuda, K. (2016).
\newblock Singing {{Voice Synthesis Based}} on {{Deep Neural Networks}}.
\newblock In \emph{Interspeech 2016} ({ISCA}), 2478--2482.
\newblock \doi{10.21437/Interspeech.2016-1027}
\bibAnnoteFile{nishimura_singing_2016}

\bibitem[{Oord et~al.(2018)Oord, Li, Babuschkin, Simonyan, Vinyals, Kavukcuoglu
  et~al.}]{oord_parallel_2018}
Oord, A., Li, Y., Babuschkin, I., Simonyan, K., Vinyals, O., Kavukcuoglu, K.,
  et~al. (2018).
\newblock Parallel {{WaveNet}}: {{Fast High-Fidelity Speech Synthesis}}.
\newblock In \emph{Proceedings of the 35th {{International Conference}} on
  {{Machine Learning}}} ({PMLR}), 3918--3926
\bibAnnoteFile{oord_parallel_2018}

\bibitem[{Pelinski et~al.(2023)Pelinski, Diaz, Temprano, and
  McPherson}]{pelinski_pipeline_2023}
Pelinski, T., Diaz, R., Temprano, A. L.~B., and McPherson, A. (2023).
\newblock Pipeline for recording datasets and running neural networks on the
  {{Bela}} embedded hardware platform
\bibAnnoteFile{pelinski_pipeline_2023}

\bibitem[{Polyak et~al.(2020)Polyak, Wolf, Adi, and
  Taigman}]{polyak_unsupervised_2020}
Polyak, A., Wolf, L., Adi, Y., and Taigman, Y. (2020).
\newblock Unsupervised {{Cross-Domain Singing Voice Conversion}}
\bibAnnoteFile{polyak_unsupervised_2020}

\bibitem[{Pons et~al.(2021)Pons, Pascual, Cengarle, and
  Serra}]{pons_upsampling_2021}
Pons, J., Pascual, S., Cengarle, G., and Serra, J. (2021).
\newblock Upsampling {Artifacts} in {Neural} {Audio} {Synthesis}.
\newblock In \emph{{ICASSP} 2021 - 2021 {IEEE} {International} {Conference} on
  {Acoustics}, {Speech} and {Signal} {Processing} ({ICASSP})} (Toronto, ON,
  Canada: IEEE), 3005--3009.
\newblock \doi{10.1109/ICASSP39728.2021.9414913}
\bibAnnoteFile{pons_upsampling_2021}

\bibitem[{Prenger et~al.(2019)Prenger, Valle, and
  Catanzaro}]{prenger_waveglow_2019}
Prenger, R., Valle, R., and Catanzaro, B. (2019).
\newblock Waveglow: {{A Flow-based Generative Network}} for {{Speech
  Synthesis}}.
\newblock In \emph{{{ICASSP}} 2019 - 2019 {{IEEE International Conference}} on
  {{Acoustics}}, {{Speech}} and {{Signal Processing}} ({{ICASSP}})}.
  3617--3621.
\newblock \doi{10.1109/ICASSP.2019.8683143}
\bibAnnoteFile{prenger_waveglow_2019}

\bibitem[{Ramachandran et~al.(2017)Ramachandran, Paine, Khorrami, Babaeizadeh,
  Chang, Zhang et~al.}]{ramachandran_fast_2017}
Ramachandran, P., Paine, T.~L., Khorrami, P., Babaeizadeh, M., Chang, S.,
  Zhang, Y., et~al. (2017).
\newblock Fast {{Generation}} for {{Convolutional Autoregressive Models}}
\bibAnnoteFile{ramachandran_fast_2017}

\bibitem[{Ram{\'i}rez et~al.(2020)Ram{\'i}rez, Benetos, and
  Reiss}]{ramirez_deep_2020}
Ram{\'i}rez, M.~A., Benetos, E., and Reiss, J.~D. (2020).
\newblock Deep learning for black-box modeling of audio effects.
\newblock \emph{Applied Sciences (Switzerland)} 10.
\newblock \doi{10.3390/app10020638}
\bibAnnoteFile{ramirez_deep_2020}

\bibitem[{Ren et~al.(2022)Ren, Xiao, Chang, Huang, Li, Chen
  et~al.}]{ren_comprehensive_2022}
Ren, P., Xiao, Y., Chang, X., Huang, P.-y., Li, Z., Chen, X., et~al. (2022).
\newblock A {{Comprehensive Survey}} of {{Neural Architecture Search}}:
  {{Challenges}} and {{Solutions}}.
\newblock \emph{ACM Computing Surveys} 54, 1--34.
\newblock \doi{10.1145/3447582}
\bibAnnoteFile{ren_comprehensive_2022}

\bibitem[{Renault et~al.(2022)Renault, Mignot, and
  Roebel}]{renault_differentiable_2022}
Renault, L., Mignot, R., and Roebel, A. (2022).
\newblock Differentiable {{Piano Model}} for {{Midi-to-Audio Performance
  Synthesis}}.
\newblock In \emph{Proceedings of the 25th {{International Conference}} on
  {{Digital Audio Effects}}} ({Vienna, Austria}), 8
\bibAnnoteFile{renault_differentiable_2022}

\bibitem[{Rodet(2002)}]{rodet_synthesis_2002}
Rodet, X. (2002).
\newblock Synthesis and processing of the singing voice.
\newblock In \emph{Proc. 1st {{IEEE}} Benelux Workshop on Model Based
  Processing and Coding of Audio ({{MPCA-2002}})}. 15--31
\bibAnnoteFile{rodet_synthesis_2002}

\bibitem[{Rouard and Hadjeres(2021)}]{rouard_crash_2021}
Rouard, S. and Hadjeres, G. (2021).
\newblock {{CRASH}}: {{Raw Audio Score-based Generative Modeling}} for
  {{Controllable High-resolution Drum Sound Synthesis}}.
\newblock \emph{arXiv:2106.07431 [cs, eess]}
\bibAnnoteFile{rouard_crash_2021}

\bibitem[{Saino et~al.(2006)Saino, Zen, Nankaku, Lee, and
  Tokuda}]{saino_hmm-based_2006}
Saino, K., Zen, H., Nankaku, Y., Lee, A., and Tokuda, K. (2006).
\newblock An {{HMM-based}} singing voice synthesis system.
\newblock In \emph{Ninth International Conference on Spoken Language
  Processing}.
\newblock \doi{10.21437/Interspeech.2006-584}
\bibAnnoteFile{saino_hmm-based_2006}

\bibitem[{{Schulze-Forster} et~al.(2023){Schulze-Forster}, Richard, Kelley,
  Doire, and Badeau}]{schulze-forster_unsupervised_2023}
{Schulze-Forster}, K., Richard, G., Kelley, L., Doire, C. S.~J., and Badeau, R.
  (2023).
\newblock Unsupervised {{Music Source Separation Using Differentiable
  Parametric Source Models}}.
\newblock \emph{IEEE/ACM Transactions on Audio, Speech, and Language
  Processing} 31, 1276--1289.
\newblock \doi{10.1109/TASLP.2023.3252272}
\bibAnnoteFile{schulze-forster_unsupervised_2023}

\bibitem[{Schwarz(2006)}]{schwarz_concatenative_2006}
Schwarz, D. (2006).
\newblock Concatenative sound synthesis: {{The}} early years.
\newblock \emph{Journal of New Music Research} 35, 3--22.
\newblock \doi{10.1080/09298210600696857}
\bibAnnoteFile{schwarz_concatenative_2006}

\bibitem[{Schwarz(2007)}]{schwarz_corpus-based_2007}
Schwarz, D. (2007).
\newblock Corpus-{{Based Concatenative Synthesis}}.
\newblock \emph{IEEE Signal Processing Magazine} 24, 92--104.
\newblock \doi{10.1109/MSP.2007.323274}
\bibAnnoteFile{schwarz_corpus-based_2007}

\bibitem[{Seeviour et~al.(1976)Seeviour, Holmes, and
  Judd}]{seeviour_automatic_1976}
Seeviour, P., Holmes, J., and Judd, M. (1976).
\newblock Automatic generation of control signals for a parallel formant speech
  synthesizer.
\newblock In \emph{{{ICASSP}} '76. {{IEEE International Conference}} on
  {{Acoustics}}, {{Speech}}, and {{Signal Processing}}}. vol.~1, 690--693.
\newblock \doi{10.1109/ICASSP.1976.1169987}
\bibAnnoteFile{seeviour_automatic_1976}

\bibitem[{Serra and Smith(1990)}]{serra_spectral_1990}
Serra, X. and Smith, J. (1990).
\newblock Spectral {{Modeling Synthesis}}: {{A Sound Analysis}}/{{Synthesis
  System Based}} on a {{Deterministic Plus Stochastic Decomposition}}.
\newblock \emph{Computer Music Journal} 14, 12--24.
\newblock \doi{10.2307/3680788}
\bibAnnoteFile{serra_spectral_1990}

\bibitem[{Shadle and Damper(2001)}]{shadle_prospects_2001}
Shadle, C.~H. and Damper, R.~I. (2001).
\newblock Prospects for articulatory synthesis: {{A}} position paper.
\newblock In \emph{4th {{ISCA}} Tutorial and Research Workshop ({{ITRW}}) on
  Speech Synthesis}
\bibAnnoteFile{shadle_prospects_2001}

\bibitem[{Shan et~al.(2022)Shan, Hantrakul, Chen, Avent, and
  Trevelyan}]{shan_differentiable_2022}
Shan, S., Hantrakul, L., Chen, J., Avent, M., and Trevelyan, D. (2022).
\newblock Differentiable {{Wavetable Synthesis}}.
\newblock In \emph{{{ICASSP}} 2022 - 2022 {{IEEE International Conference}} on
  {{Acoustics}}, {{Speech}} and {{Signal Processing}} ({{ICASSP}})}.
  4598--4602.
\newblock \doi{10.1109/ICASSP43922.2022.9746940}
\bibAnnoteFile{shan_differentiable_2022}

\bibitem[{Shynk(1989)}]{shynk_adaptive_1989}
Shynk, J. (1989).
\newblock Adaptive {{IIR}} filtering.
\newblock \emph{IEEE ASSP Magazine} 6, 4--21.
\newblock \doi{10.1109/53.29644}
\bibAnnoteFile{shynk_adaptive_1989}

\bibitem[{Siedenburg and McAdams(2017)}]{siedenburg_four_2017}
Siedenburg, K. and McAdams, S. (2017).
\newblock Four {{Distinctions}} for the {{Auditory}} ``{{Wastebasket}}'' of
  {{Timbre}}.
\newblock \emph{Frontiers in Psychology} 8, 1747.
\newblock \doi{10.3389/fpsyg.2017.01747}
\bibAnnoteFile{siedenburg_four_2017}

\bibitem[{Sisman et~al.(2019)Sisman, Vijayan, Dong, and
  Li}]{sisman_singan_2019}
Sisman, B., Vijayan, K., Dong, M., and Li, H. (2019).
\newblock {{SINGAN}}: {{Singing Voice Conversion}} with {{Generative
  Adversarial Networks}}.
\newblock In \emph{2019 {{Asia-Pacific Signal}} and {{Information Processing
  Association Annual Summit}} and {{Conference}} ({{APSIPA ASC}})}. 112--118.
\newblock \doi{10.1109/APSIPAASC47483.2019.9023162}
\bibAnnoteFile{sisman_singan_2019}

\bibitem[{Sisman et~al.(2021)Sisman, Yamagishi, King, and
  Li}]{sisman_overview_2021}
Sisman, B., Yamagishi, J., King, S., and Li, H. (2021).
\newblock An {{Overview}} of {{Voice Conversion}} and {{Its Challenges}}:
  {{From Statistical Modeling}} to {{Deep Learning}}.
\newblock \emph{IEEE/ACM Transactions on Audio, Speech, and Language
  Processing} 29, 132--157.
\newblock \doi{10.1109/TASLP.2020.3038524}
\bibAnnoteFile{sisman_overview_2021}

\bibitem[{Smith(1992)}]{smith_physical_1992}
Smith, J.~O. (1992).
\newblock Physical {{Modeling Using Digital Waveguides}}.
\newblock \emph{Computer Music Journal} 16, 74.
\newblock \doi{10.2307/3680470}
\bibAnnoteFile{smith_physical_1992}

\bibitem[{Smith(2010)}]{smith_physical_2010}
Smith, J.~O. (2010).
\newblock \emph{Physical {{Audio Signal Processing}}: For {{Virtual Musical
  Instruments}} and {{Audio Effects}}} ({Stanford, Calif}: {Stanford
  University, CCRMA})
\bibAnnoteFile{smith_physical_2010}

\bibitem[{Song et~al.(2023)Song, Zhang, Lei, Cong, Li, Xie
  et~al.}]{song_dspgan_2023}
Song, K., Zhang, Y., Lei, Y., Cong, J., Li, H., Xie, L., et~al. (2023).
\newblock {{DSPGAN}}: {{A Gan-Based Universal Vocoder}} for {{High-Fidelity
  TTS}} by {{Time-Frequency Domain Supervision}} from {{DSP}}.
\newblock In \emph{{{ICASSP}} 2023 - 2023 {{IEEE International Conference}} on
  {{Acoustics}}, {{Speech}} and {{Signal Processing}} ({{ICASSP}})}. 1--5.
\newblock \doi{10.1109/ICASSP49357.2023.10095105}
\bibAnnoteFile{song_dspgan_2023}

\bibitem[{Spall(1998)}]{spall_overview_1998}
Spall, J.~C. (1998).
\newblock An overview of the simultaneous perturbation method for efficient
  optimization.
\newblock \emph{Johns Hopkins apl technical digest} 19, 482--492
\bibAnnoteFile{spall_overview_1998}

\bibitem[{Ssergejewitsch(1928)}]{ssergejewitsch_method_1928}
Ssergejewitsch, T.~L. (1928).
\newblock Method of and apparatus for the generation of sounds
\bibAnnoteFile{ssergejewitsch_method_1928}

\bibitem[{Stanton et~al.(2022)Stanton, Shannon, Mariooryad, {Skerry-Ryan},
  Battenberg, Bagby et~al.}]{stanton_speaker_2022}
Stanton, D., Shannon, M., Mariooryad, S., {Skerry-Ryan}, {\relax RJ}.,
  Battenberg, E., Bagby, T., et~al. (2022).
\newblock Speaker {{Generation}}.
\newblock In \emph{{{ICASSP}} 2022 - 2022 {{IEEE International Conference}} on
  {{Acoustics}}, {{Speech}} and {{Signal Processing}} ({{ICASSP}})}.
  7897--7901.
\newblock \doi{10.1109/ICASSP43922.2022.9747345}
\bibAnnoteFile{stanton_speaker_2022}

\bibitem[{Steinmetz et~al.(2022{\natexlab{a}})Steinmetz, Bryan, and
  Reiss}]{steinmetz_style_2022}
Steinmetz, C.~J., Bryan, N.~J., and Reiss, J.~D. (2022{\natexlab{a}}).
\newblock Style {{Transfer}} of {{Audio Effects}} with {{Differentiable Signal
  Processing}}.
\newblock \emph{J. Audio Eng. Soc.} 70, 14.
\newblock \doi{10.17743/jaes.2022.0025}
\bibAnnoteFile{steinmetz_style_2022}

\bibitem[{Steinmetz et~al.(2021)Steinmetz, Pons, Pascual, and
  Serr{\`a}}]{steinmetz_automatic_2021}
Steinmetz, C.~J., Pons, J., Pascual, S., and Serr{\`a}, J. (2021).
\newblock Automatic {{Multitrack Mixing With A Differentiable Mixing Console Of
  Neural Audio Effects}}.
\newblock In \emph{{{ICASSP}} 2021 - 2021 {{IEEE International Conference}} on
  {{Acoustics}}, {{Speech}} and {{Signal Processing}} ({{ICASSP}})}. 71--75.
\newblock \doi{10.1109/ICASSP39728.2021.9414364}
\bibAnnoteFile{steinmetz_automatic_2021}

\bibitem[{Steinmetz and Reiss(2020)}]{steinmetz_auraloss_2020}
Steinmetz, C.~J. and Reiss, J.~D. (2020).
\newblock auraloss: {Audio} focused loss functions in {PyTorch}.
\newblock In \emph{Digital music research network one-day workshop ({DMRN}+15)}
\bibAnnoteFile{steinmetz_auraloss_2020}

\bibitem[{Steinmetz et~al.(2022{\natexlab{b}})Steinmetz, Sai~Vanka, Bromham,
  and Martinez~Ramirez}]{steinmetz_deep_2022}
Steinmetz, C.~J., Sai~Vanka, S., Bromham, G., and Martinez~Ramirez, M.~A.
  (2022{\natexlab{b}}).
\newblock Deep learning for automatic mixing
\bibAnnoteFile{steinmetz_deep_2022}

\bibitem[{Stylianou(2009)}]{stylianou_voice_2009}
Stylianou, Y. (2009).
\newblock Voice {{Transformation}}: {{A}} survey.
\newblock In \emph{2009 {{IEEE International Conference}} on {{Acoustics}},
  {{Speech}} and {{Signal Processing}}}. 3585--3588.
\newblock \doi{10.1109/ICASSP.2009.4960401}
\bibAnnoteFile{stylianou_voice_2009}

\bibitem[{Subramani et~al.(2022)Subramani, Valin, Isik, Smaragdis, and
  Krishnaswamy}]{subramani_end--end_2022}
Subramani, K., Valin, J.-M., Isik, U., Smaragdis, P., and Krishnaswamy, A.
  (2022).
\newblock End-to-end {{LPCNet}}: {{A Neural Vocoder With Fully-Differentiable
  LPC Estimation}}.
\newblock In \emph{Interspeech 2022} ({ISCA}), 818--822.
\newblock \doi{10.21437/Interspeech.2022-912}
\bibAnnoteFile{subramani_end--end_2022}

\bibitem[{S{\"u}dholt et~al.(2023)S{\"u}dholt, C{\'a}mara, Xu, and
  Reiss}]{sudholt_vocal_2023}
S{\"u}dholt, D., C{\'a}mara, M., Xu, Z., and Reiss, J.~D. (2023).
\newblock Vocal {{Tract Area Estimation}} by {{Gradient Descent}}
\bibAnnoteFile{sudholt_vocal_2023}

\bibitem[{Tamamori et~al.(2017)Tamamori, Hayashi, Kobayashi, Takeda, and
  Toda}]{tamamori_speaker-dependent_2017}
Tamamori, A., Hayashi, T., Kobayashi, K., Takeda, K., and Toda, T. (2017).
\newblock Speaker-dependent wavenet vocoder.
\newblock In \emph{Interspeech}. vol. 2017, 1118--1122.
\newblock \doi{10.21437/Interspeech.2017-314}
\bibAnnoteFile{tamamori_speaker-dependent_2017}

\bibitem[{Tan et~al.(2021)Tan, Qin, Soong, and Liu}]{tan_survey_2021}
Tan, X., Qin, T., Soong, F., and Liu, T.-Y. (2021).
\newblock A {{Survey}} on {{Neural Speech Synthesis}}
\bibAnnoteFile{tan_survey_2021}

\bibitem[{Tian et~al.(2020)Tian, Zhang, Lu, Chen, and
  Liu}]{tian_featherwave_2020}
Tian, Q., Zhang, Z., Lu, H., Chen, L.-H., and Liu, S. (2020).
\newblock {{FeatherWave}}: {{An}} efficient high-fidelity neural vocoder with
  multi-band linear prediction
\bibAnnoteFile{tian_featherwave_2020}

\bibitem[{Turian and Henry(2020)}]{turian_im_2020}
Turian, J. and Henry, M. (2020).
\newblock I'm {{Sorry}} for {{Your Loss}}: {{Spectrally-Based Audio Distances
  Are Bad}} at {{Pitch}}.
\newblock \emph{arXiv:2012.04572 [cs, eess]}
\bibAnnoteFile{turian_im_2020}

\bibitem[{Turian et~al.(2021)Turian, Shier, Tzanetakis, McNally, and
  Henry}]{turian_one_2021}
Turian, J., Shier, J., Tzanetakis, G., McNally, K., and Henry, M. (2021).
\newblock One {{Billion Audio Sounds}} from {{GPU-enabled Modular Synthesis}}.
\newblock \emph{arXiv:2104.12922 [cs, eess]}
\bibAnnoteFile{turian_one_2021}

\bibitem[{Valin and Skoglund(2019)}]{valin_lpcnet_2019}
Valin, J.-M. and Skoglund, J. (2019).
\newblock {{LPCNET}}: {{Improving Neural Speech Synthesis}} through {{Linear
  Prediction}}.
\newblock In \emph{{{ICASSP}} 2019 - 2019 {{IEEE International Conference}} on
  {{Acoustics}}, {{Speech}} and {{Signal Processing}} ({{ICASSP}})}.
  5891--5895.
\newblock \doi{10.1109/ICASSP.2019.8682804}
\bibAnnoteFile{valin_lpcnet_2019}

\bibitem[{van~den Oord et~al.(2016)van~den Oord, Dieleman, Zen, Simonyan,
  Vinyals, Graves et~al.}]{oord_wavenet_2016}
van~den Oord, A., Dieleman, S., Zen, H., Simonyan, K., Vinyals, O., Graves, A.,
  et~al. (2016).
\newblock {{WaveNet}}: {{A Generative Model}} for {{Raw Audio}}
\bibAnnoteFile{oord_wavenet_2016}

\bibitem[{Villavicencio and Bonada(2010)}]{villavicencio_applying_2010}
Villavicencio, F. and Bonada, J. (2010).
\newblock Applying voice conversion to concatenative singing-voice synthesis.
\newblock In \emph{Interspeech 2010} ({ISCA}), 2162--2165.
\newblock \doi{10.21437/Interspeech.2010-596}
\bibAnnoteFile{villavicencio_applying_2010}

\bibitem[{Vipperla et~al.(2020)Vipperla, Park, Choo, Ishtiaq, Min, Bhattacharya
  et~al.}]{vipperla_bunched_2020}
Vipperla, R., Park, S., Choo, K., Ishtiaq, S., Min, K., Bhattacharya, S.,
  et~al. (2020).
\newblock Bunched {{LPCNet}} : {{Vocoder}} for {{Low-cost Neural Text-To-Speech
  Systems}}
\bibAnnoteFile{vipperla_bunched_2020}

\bibitem[{Wang et~al.(2019{\natexlab{a}})Wang, Takaki, and
  Yamagishi}]{wang_neural_2019-1}
Wang, X., Takaki, S., and Yamagishi, J. (2019{\natexlab{a}}).
\newblock Neural {{Source-filter-based Waveform Model}} for {{Statistical
  Parametric Speech Synthesis}}.
\newblock In \emph{{{ICASSP}} 2019 - 2019 {{IEEE International Conference}} on
  {{Acoustics}}, {{Speech}} and {{Signal Processing}} ({{ICASSP}})} ({Brighton,
  United Kingdom}: {IEEE}), 5916--5920.
\newblock \doi{10.1109/ICASSP.2019.8682298}
\bibAnnoteFile{wang_neural_2019-1}

\bibitem[{Wang et~al.(2019{\natexlab{b}})Wang, Takaki, and
  Yamagishi}]{wang_neural_2019-2}
Wang, X., Takaki, S., and Yamagishi, J. (2019{\natexlab{b}}).
\newblock Neural {{Source-Filter Waveform Models}} for {{Statistical Parametric
  Speech Synthesis}}.
\newblock \emph{IEEE/ACM Transactions on Audio, Speech, and Language
  Processing} 28, 402--415.
\newblock \doi{10.1109/TASLP.2019.2956145}
\bibAnnoteFile{wang_neural_2019-2}

\bibitem[{Wang and Yamagishi(2019)}]{wang_neural_2019}
Wang, X. and Yamagishi, J. (2019).
\newblock Neural {{Harmonic-plus-Noise Waveform Model}} with {{Trainable
  Maximum Voice Frequency}} for {{Text-to-Speech Synthesis}}.
\newblock In \emph{10th {{ISCA Workshop}} on {{Speech Synthesis}} ({{SSW}} 10)}
  ({ISCA}), 1--6.
\newblock \doi{10.21437/SSW.2019-1}
\bibAnnoteFile{wang_neural_2019}

\bibitem[{Wang and Yamagishi(2020)}]{wang_using_2020}
Wang, X. and Yamagishi, J. (2020).
\newblock Using {{Cyclic Noise}} as the {{Source Signal}} for {{Neural
  Source-Filter-based Speech Waveform Model}}
\bibAnnoteFile{wang_using_2020}

\bibitem[{Wang et~al.(2022)Wang, Wang, Zhu, Wu, Li, Xue
  et~al.}]{wang_opencpop_2022}
Wang, Y., Wang, X., Zhu, P., Wu, J., Li, H., Xue, H., et~al. (2022).
\newblock Opencpop: {{A High-Quality Open Source Chinese Popular Song Corpus}}
  for {{Singing Voice Synthesis}}.
\newblock \doi{10.48550/arXiv.2201.07429}
\bibAnnoteFile{wang_opencpop_2022}

\bibitem[{Watts et~al.(2023)Watts, Wihlborg, and
  {Valentini-Botinhao}}]{watts_puffin_2023}
Watts, O., Wihlborg, L., and {Valentini-Botinhao}, C. (2023).
\newblock {{PUFFIN}}: {{Pitch-Synchronous Neural Waveform Generation}} for
  {{Fullband Speech}} on {{Modest Devices}}.
\newblock In \emph{{{ICASSP}} 2023 - 2023 {{IEEE International Conference}} on
  {{Acoustics}}, {{Speech}} and {{Signal Processing}} ({{ICASSP}})} ({Rhodes
  Island, Greece}: {IEEE}), 1--5.
\newblock \doi{10.1109/ICASSP49357.2023.10094729}
\bibAnnoteFile{watts_puffin_2023}

\bibitem[{Webber et~al.(2023)Webber, {Valentini-Botinhao}, Williams, Henter,
  and King}]{webber_autovocoder_2023}
Webber, J.~J., {Valentini-Botinhao}, C., Williams, E., Henter, G.~E., and King,
  S. (2023).
\newblock Autovocoder: {{Fast Waveform Generation}} from a {{Learned Speech
  Representation Using Differentiable Digital Signal Processing}}.
\newblock In \emph{{{ICASSP}} 2023 - 2023 {{IEEE International Conference}} on
  {{Acoustics}}, {{Speech}} and {{Signal Processing}} ({{ICASSP}})}. 1--5.
\newblock \doi{10.1109/ICASSP49357.2023.10095729}
\bibAnnoteFile{webber_autovocoder_2023}

\bibitem[{Wu et~al.(2022{\natexlab{a}})Wu, Hsiao, Yang, Friedman, Jackson,
  Bruzenak et~al.}]{wu_ddsp-based_2022}
Wu, D.-Y., Hsiao, W.-Y., Yang, F.-R., Friedman, O., Jackson, W., Bruzenak, S.,
  et~al. (2022{\natexlab{a}}).
\newblock {{DDSP-based Singing Vocoders}}: {{A New Subtractive-based
  Synthesizer}} and {{A Comprehensive Evaluation}}.
\newblock \doi{10.48550/arXiv.2208.04756}
\bibAnnoteFile{wu_ddsp-based_2022}

\bibitem[{Wu et~al.(2022{\natexlab{b}})Wu, Gardner, Manilow, Simon, Hawthorne,
  and Engel}]{wu_generating_2022}
Wu, Y., Gardner, J., Manilow, E., Simon, I., Hawthorne, C., and Engel, J.
  (2022{\natexlab{b}}).
\newblock Generating {{Detailed Music Datasets}} with {{Neural Audio
  Synthesis}}.
\newblock In \emph{Proceedings of the 39th {{International Conference}} on
  {{Machine Learning}}} ({Baltimore, Maryland, USA}), vol. 162
\bibAnnoteFile{wu_generating_2022}

\bibitem[{Wu et~al.(2022{\natexlab{c}})Wu, Manilow, Deng, Swavely, Kastner,
  Cooijmans et~al.}]{wu_midi-ddsp_2022}
Wu, Y., Manilow, E., Deng, Y., Swavely, R., Kastner, K., Cooijmans, T., et~al.
  (2022{\natexlab{c}}).
\newblock {{MIDI-DDSP}}: {{Detailed}} control of musical performance via
  hierarchical modeling.
\newblock In \emph{International Conference on Learning Representations}
\bibAnnoteFile{wu_midi-ddsp_2022}

\bibitem[{Yamamoto et~al.(2020)Yamamoto, Song, and
  Kim}]{yamamoto_parallel_2020}
Yamamoto, R., Song, E., and Kim, J.-M. (2020).
\newblock Parallel {{Wavegan}}: {{A Fast Waveform Generation Model Based}} on
  {{Generative Adversarial Networks}} with {{Multi-Resolution Spectrogram}}.
\newblock In \emph{{{ICASSP}} 2020 - 2020 {{IEEE International Conference}} on
  {{Acoustics}}, {{Speech}} and {{Signal Processing}} ({{ICASSP}})}.
  6199--6203.
\newblock \doi{10.1109/ICASSP40776.2020.9053795}
\bibAnnoteFile{yamamoto_parallel_2020}

\bibitem[{Ye et~al.(2023)Ye, Xue, Tan, Liu, and Guo}]{ye_nas-fm_2023}
Ye, Z., Xue, W., Tan, X., Liu, Q., and Guo, Y. (2023).
\newblock {{NAS-FM}}: {{Neural Architecture Search}} for {{Tunable}} and
  {{Interpretable Sound Synthesis}} based on {{Frequency Modulation}}.
\newblock \doi{10.48550/arXiv.2305.12868}
\bibAnnoteFile{ye_nas-fm_2023}

\bibitem[{{Yee-King} et~al.(2018){Yee-King}, Fedden, and
  {d'Inverno}}]{yee-king_automatic_2018}
{Yee-King}, M.~J., Fedden, L., and {d'Inverno}, M. (2018).
\newblock Automatic {{Programming}} of {{VST Sound Synthesizers Using Deep
  Networks}} and {{Other Techniques}}.
\newblock \emph{IEEE Transactions on Emerging Topics in Computational
  Intelligence} 2, 150--159.
\newblock \doi{10.1109/TETCI.2017.2783885}
\bibAnnoteFile{yee-king_automatic_2018}

\bibitem[{Yoshimura et~al.(2023)Yoshimura, Takaki, Nakamura, Oura, Hono,
  Hashimoto et~al.}]{yoshimura_embedding_2023}
Yoshimura, T., Takaki, S., Nakamura, K., Oura, K., Hono, Y., Hashimoto, K.,
  et~al. (2023).
\newblock Embedding a {{Differentiable Mel-Cepstral Synthesis Filter}} to a
  {{Neural Speech Synthesis System}}.
\newblock In \emph{{{ICASSP}} 2023 - 2023 {{IEEE International Conference}} on
  {{Acoustics}}, {{Speech}} and {{Signal Processing}} ({{ICASSP}})}. 1--5.
\newblock \doi{10.1109/ICASSP49357.2023.10094872}
\bibAnnoteFile{yoshimura_embedding_2023}

\bibitem[{You et~al.(2021)You, Kim, Nam, Hwang, and Chae}]{you_gan_2021}
You, J., Kim, D., Nam, G., Hwang, G., and Chae, G. (2021).
\newblock {GAN} vocoder: {Multi}-resolution discriminator is all you need.
\newblock In \emph{Interspeech}.
\newblock \doi{10.21437/Interspeech.2021-41}
\bibAnnoteFile{you_gan_2021}

\bibitem[{Yu et~al.(2020)Yu, Lu, Hu, Yu, Weng, Xu et~al.}]{yu_durian_2020}
Yu, C., Lu, H., Hu, N., Yu, M., Weng, C., Xu, K., et~al. (2020).
\newblock {{DurIAN}}: {{Duration Informed Attention Network}} for {{Speech
  Synthesis}}.
\newblock In \emph{Interspeech 2020} ({ISCA}), 2027--2031.
\newblock \doi{10.21437/Interspeech.2020-2968}
\bibAnnoteFile{yu_durian_2020}

\bibitem[{Yu and Fazekas(2023)}]{yu_singing_2023}
Yu, C.-Y. and Fazekas, G. (2023).
\newblock Singing {{Voice Synthesis Using Differentiable LPC}} and
  {{Glottal-Flow-Inspired Wavetables}}
\bibAnnoteFile{yu_singing_2023}

\bibitem[{Zen et~al.(2009)Zen, Tokuda, and Black}]{zen_statistical_2009}
Zen, H., Tokuda, K., and Black, A.~W. (2009).
\newblock Statistical parametric speech synthesis.
\newblock \emph{Speech Communication} 51, 1039--1064.
\newblock \doi{10.1016/j.specom.2009.04.004}
\bibAnnoteFile{zen_statistical_2009}

\bibitem[{Zhao et~al.(2020)Zhao, Wang, Juvela, and
  Yamagishi}]{zhao_transferring_2020}
Zhao, Y., Wang, X., Juvela, L., and Yamagishi, J. (2020).
\newblock Transferring {Neural} {Speech} {Waveform} {Synthesizers} to {Musical}
  {Instrument} {Sounds} {Generation}.
\newblock In \emph{{ICASSP} 2020 - 2020 {IEEE} {International} {Conference} on
  {Acoustics}, {Speech} and {Signal} {Processing} ({ICASSP})} (Barcelona,
  Spain: IEEE), 6269--6273.
\newblock \doi{10.1109/ICASSP40776.2020.9053047}
\bibAnnoteFile{zhao_transferring_2020}

\bibitem[{Zhou and Lu(2022)}]{zhou_hifi-svc_2022}
Zhou, Y. and Lu, X. (2022).
\newblock {{HiFi-SVC}}: {{Fast High Fidelity Cross-Domain Singing Voice
  Conversion}}.
\newblock In \emph{{{ICASSP}} 2022 - 2022 {{IEEE International Conference}} on
  {{Acoustics}}, {{Speech}} and {{Signal Processing}} ({{ICASSP}})}.
  6667--6671.
\newblock \doi{10.1109/ICASSP43922.2022.9746812}
\bibAnnoteFile{zhou_hifi-svc_2022}

\end{thebibliography}

\end{document}